\documentclass[a4paper,12pt]{article}
\pdfoutput=1 

\usepackage{jheppub,amsmath} 
\usepackage{graphicx}
\usepackage{dcolumn}
\usepackage{bm}
\usepackage{amsmath,amsfonts,amssymb}
\usepackage{slashed}

\DeclareMathOperator{\sech}{sech}
\usepackage{braket,xcolor}
\usepackage{verbatim}

\usepackage{caption}
\usepackage{subcaption}
\usepackage{multirow}
\usepackage{amsfonts}
\usepackage[utf8]{inputenc}
\usepackage{hyperref}
\usepackage{braket}
\usepackage[T1]{fontenc} 
\usepackage{cancel}

\usepackage{tikz} 

\title{\boldmath Krylov Complexity in Free and Interacting Scalar Field Theories with Bounded Power Spectrum}

\author[a]{Hugo A. Camargo,}
\author[a]{Viktor Jahnke,}
\author[a,b]{Keun-Young Kim}
\author[c]{and Mitsuhiro Nishida}
\affiliation[a]{Department of Physics and Photon Science, Gwangju Institute of Science and Technology,\\123 Cheomdan-gwagiro, Gwangju 61005, Korea}
\affiliation[b]{Research Center for Photon Science Technology, Gwangju Institute of Science and Technology, 123 Cheomdan-gwagiro, Gwangju 61005, Korea}
\affiliation[c]{Department of Physics, Pohang University of Science and Technology, Pohang 37673, Korea}

\emailAdd{hugo.camargo@gist.ac.kr}
\emailAdd{viktorjahnke@gist.ac.kr}
\emailAdd{fortoe@gist.ac.kr}
\emailAdd{nishida124@postech.ac.kr}

\abstract{We study a notion of operator growth known as Krylov complexity in free and interacting massive scalar quantum field theories in $d$-dimensions at finite temperature. We consider the effects of mass, one-loop self-energy due to perturbative interactions, and finite ultraviolet cutoffs in continuous momentum space. These deformations change the behavior of Lanczos coefficients and Krylov complexity and induce effects such as the ``staggering'' of the former into two families, a decrease in the exponential growth rate of the latter, and transitions in their asymptotic behavior. We also discuss the relation between the existence of a mass gap and the property of staggering, and the relation between our ultraviolet cutoffs in continuous theories and lattice theories.}

\begin{document} 
\maketitle
\flushbottom

\section{Introduction}
\label{sec:introduction}

The study of chaos in quantum many-body systems has received a significant amount of attention in recent years in various fields ranging from condensed matter physics to quantum gravity. During this time, deep connections have been uncovered between thermalization in closed quantum systems~\cite{Deutsch1991,Srednicki:1994mfb, Deutsch_2018}, the emergence of ergodicity and the universality of spectral statistics in random-matrix theory (RMT)~\cite{Cotler:2016fpe}. To this end, several probes of chaotic behavior in quantum many-body systems have been introduced such as out-of-time-order correlators (OTOCs)~\cite{Larkin1969, Shenker:2013pqa, Shenker:2013yza}, spectral form factors (SFF)~\cite{Gaikwad:2017odv,Liu:2018hlr} as well as different notions of operator entanglement and growth~\cite{Prosen2007, Prosen2009, Alba:2019okd, MacCormack:2020auw}. Within the latter class, there has been recent interest in studying a particular notion which quantifies the growth of operators and which has the potential of bridging the gap between quantum dynamics and the study of the black hole interior in anti-De Sitter (AdS) gravity. This notion is known as \emph{Krylov complexity}, or simply as \emph{K-complexity}.

K-complexity $K_{\mathcal{O}}(t)$ measures the growth in time, or spread, of an operator $\mathcal{O}$ in \emph{Krylov space}, a local Hilbert space $\mathcal{H}_{\mathcal{O}}$ defined as the span of nested commutators $\mathrm{Span}\lbrace \left[H,\ldots,\left[H,\mathcal{O}\right]\right] \rbrace$\footnote{In~\cite{Pastawski:2020LOSCH} authors performed experimental measurements of the growth rate of other probes of chaotic behavior such as OTOCs and the Loschmidt Echo~\cite{Wisniacki_2012}. It was argued that the evolution of an operator through the commutation with the Hamiltonian gives rise to scrambling, pointing at a link between these measures and K-complexity.} where $H$ is the Hamiltonian of the system. In \cite{Parker:2018yvk}, authors conjectured that exponential growth of the K-complexity $K_{\mathcal{O}}(t)\propto e^{\lambda_{K}t}$ could be interpreted as a signature of chaotic behavior in local quantum many-body systems with finite-dimensional Hilbert spaces, such as non-integrable spin chains or the Sachdev--Ye--Kitaev (SYK) model~\cite{KitaevTalks,SachdevYeModel}. The authors furthermore showed that for systems at infinite temperature $1/\beta\rightarrow \infty$, $\lambda_{K}$ is an upper bound for the growth rate of a large class of operator growth measures, including OTOCs. In particular, they conjectured a sharp bound on \emph{Lyapunov exponents}
\begin{equation}
\label{eq:BoundLyapKrylov}
    \lambda_{L}\leq \lambda_{K}~,
\end{equation}
which could represent a tighter bound than the universal Maldacena--Shenker--Stanford (MSS) chaos bound~\cite{Maldacena:2015waa}
\begin{equation}
\label{eq:MSS}
    \lambda_{L}\leq 2\pi /\beta~,
\end{equation} 
since they conjectured that the relation~\eqref{eq:BoundLyapKrylov} could be valid at finite temperature $0<1/\beta<\infty$, where $\lambda_{K}\leq2\pi/\beta$. As a consequence, this observation led to a wealth of studies of K-complexity in different settings, ranging from quantum many-body systems~\cite{Barbon:2019wsy,Avdoshkin:2019trj,Dymarsky:2019elm,Rabinovici:2020ryf,Cao:2020zls,Kim:2021okd,Rabinovici:2021qqt,Trigueros:2021rwj,Balasubramanian:2022tpr,Fan:2022xaa,Heveling:2022hth,Bhattacharjee:2022vlt,Caputa:2022eye,Muck:2022xfc,Rabinovici:2022beu,He:2022ryk,Hornedal:2022pkc,Alishahiha:2022nhe,Alishahiha:2022anw}, gauge theories~\cite{Magan:2020iac}, holographic models~\cite{Jian:2020qpp}, conformal field theories (CFT)~\cite{Dymarsky:2021bjq,Caputa:2021ori}, Lie groups~\cite{Caputa:2021sib,Patramanis:2021lkx} and open quantum systems~\cite{Bhattacharya:2022gbz,Liu:2022god,Bhattacharjee:2022lzy}. 

At the same time, the notion of quantum complexity~\cite{NielsenChuangBook} has taken a center stage in the study of the black hole interior through the AdS/CFT correspondence~\cite{Maldacena:1997re}. Borrowed from quantum computing, quantum complexity can be broadly thought of as a measure of how difficult it is to achieve a certain task with finite resources, e.g. producing a particular quantum state or a given operator using a finite set of operations or \emph{gates}. Intuitions related to the relevance of quantum complexity in decoding the information contained in Hawking radiation~\cite{Harlow:2013tf,Maldacena:2013xja} were sharpened in a series of works~\cite{Susskind:2014rva,Stanford:2014jda,Brown:2015bva,Brown:2015lvg} giving rise to holographic complexity proposals, which have been the subject of recent efforts to understand the interior of AdS black holes. See~\cite{Chapman:2021jbh} for a review. Nevertheless, despite significant progress, an outstanding complication that remains is the absence of a proper definition of quantum complexity beyond Gaussian states in free quantum field theories (QFTs)~\cite{Jefferson:2017sdb,Chapman:2017rqy,Khan:2018rzm,Hackl:2018ptj}, weakly-interacting theories~\cite{Bhattacharyya:2018bbv} and CFTs~\cite{Caputa:2017yrh,Caputa:2018kdj,Flory:2020eot,Chagnet:2021uvi,Camargo:2022wkd}. 

In most but certainly not all cases, the notion of quantum complexity is taken to be that of circuit complexity, which was generalized from spin chains to quantum many-body systems by implementing a geometrization approach developed in~\cite{Nielsen2006AGA,NielsenQuantumCompasGrav,NielsenDowning}. This notion, usually referred to as Nielsen complexity, can be thought of as a family of circuit complexity measures that have a number of ambiguities (e.g. choice of reference and target states --or equivalently desired operator--, cost function, universal gate set, and penalty factors) which lack a precise understanding from the gravitational perspective despite a qualitative agreement with holographic proposals\footnote{One should remark, nonetheless, that recent developments~\cite{Belin:2021bga, Belin:2022xmt} posit the existence of a family of notions of holographic complexity that generalize the previously conjectured ones. It is an active subject of research to understand whether the features of this family of holographic complexity measures have a field-theoretic counterpart, e.g. within the family of circuit complexity measures of the Nielsen-type.}. In contrast, K-complexity depends only on the choice of the operator $\mathcal{O}$ (or quantum state) in a given quantum system with Hamiltonian $H$. In~\cite{Balasubramanian:2022tpr}, authors also showed that it also exhibits universal features analogous to SFF~\cite{Cotler:2016fpe}\footnote{Readers are also referred to~\cite{He:2022ryk} for a comparison of the SFF, OTOCs and K-complexity in the context of the SYK model and some of its deformations.}. Furthermore, in~\cite{Barbon:2019wsy,Jian:2020qpp,Rabinovici:2020ryf,Caputa:2021sib} K-complexity was found to have a qualitative agreement with holographic expectations, particularly in the context of the so-called ``complexity=volume'' (CV) proposal~\cite{Stanford:2014jda}. Nevertheless, despite these qualitative similarities with holographic proposals, it is not clear whether K-complexity can be regarded as the field-theoretic counterpart to a holographic complexity notion, or conversely, what could be the holographic counterpart of K-complexity. 

Given the broad interest in K-complexity and lack of results beyond quantum many-body systems, CFTs, and systems with a high degree of symmetry, it is important to pave the way toward an understanding of this quantity in more general QFTs. An outstanding question is whether K-complexity can probe chaotic features of QFTs in a similar fashion to quantum many-body systems. A conundrum arose in~\cite{Dymarsky:2021bjq}, where it was shown that K-complexity generically grows exponentially in both chaotic (holographic) and non-chaotic (rational) CFTs and therefore this feature should not be na\"{i}vely regarded as an indication of chaos in QFTs with unbounded power spectrum. Dymarsky and his collaborators~\cite{DymarskyTalk} have given an interpretation of this issue by considering a discretized lattice model as follows: in order to capture signatures of quantum chaos, one needs to introduce a UV cutoff, such as a lattice spacing, which may produce a bound on the power spectrum, and study the growth of K-complexity for times after the UV cutoff scale. Inspired by their argument, in this work we focus on a similar but different way to introduce the UV cutoff. Instead of considering discretized lattice models, we consider a cutoff for the momentum integrals in QFTs at finite temperature $1/\beta$ in different spacetime dimensions $d$. Our approach consisting of the introduction of a momentum cutoff can be considered complementary to the analysis of discretized lattice models.

This paper is structured as follows: In Section~\ref{sec:Lanczos} we review the basics of the so-called \emph{Lanczos algorithm}~\cite{Lanczos:1950zz}, a procedure which allows us to compute important quantities in the study of operator growth, namely the so-called \emph{Lanczos coefficients} $b_{n}$ and ``probability amplitudes'' $\varphi_{n}(t)$. We also show how these quantities are used to compute the Krylov complexity $K_{\mathcal{O}}(t)$. Then, in Sec.~\ref{subsec:LanczosChaos} we discuss in detail the connection between the exponential growth of K-complexity and features of quantum chaos and the MSS chaos bound~\eqref{eq:MSS}, while in Sec.~\ref{subsec:KrylovSpectral} we discuss the effects that we expect to observe in K-complexity from considering a bounded power spectrum (i.e. the existence of IR and UV cutoffs).

Afterwards, in Section~\ref{sec:FreeMassScalar} we apply the Lanczos algorithm to the free massive scalar theory in $d$-dimensions. Throughout this section we consider the heavy scalar limit, $\beta m \gg 1$, where analytic formulas for most quantities can be found. In Sec.~\ref{subsec:MassScalarMoments} we carefully analyze the Wightman power spectrum $f^{W}(\omega)$ and the associated thermal Wightman $2$-point function $\Pi^{W}(t)$, also known as the auto-correlation function. In Sec.~\ref{subsec:MassScalarHeavyLimitNoUVCutoff} we focus on the case where there is no upper bound in the power spectrum (i.e. no UV cutoff). We compute the moments $\mu_{2n}$ from the power spectrum $f^{W}(\omega)$ which allows us to compute the Lanczos coefficients and probability amplitudes $\varphi_{n}(t)$. With these, we obtain an expression for the K-complexity $K(t)$ which we analyze in different odd dimensions $d$. We study its early and late-time behavior and we obtain an expectation for its growth rate as a function of $\beta m$.

In Sec.~\ref{subsec:MassScalarHeavyLimitHardUVCutoff} we introduce a hard UV cutoff $\Lambda$ in momentum space and study the effects that it has on both the Lanczos coefficients and the K-complexity, following an analogous procedure to the preceding section. We complement this analysis in Sec.~\ref{subsec:MassScalarHeavyLimitSmoothUVCutoff}, where we discuss the case where we modify the large-frequency behavior of the power spectrum $f^{W}(\omega)$ by introducing a smooth cutoff. Then, in Sec.~\ref{sec:InterMassScalar} we study the behavior of the Lanczos coefficients for an interacting scalar theory in $d=4$ spacetime dimensions. We first study the case where we have a marginally irrelevant deformation in Sec.~\ref{subsec:MargIrrelDefo}. This analysis is contrasted with the case where we consider a relevant deformation, which we study in Sec.~\ref{subsec:RelDefo}. Finally, in Sec.~\ref{sec:DiscussionConclusion} we discuss our results and offer some concluding remarks and possible future directions.

\paragraph{Note added}
On the same day that this work was announced on arXiv, Ref.~\cite{Avdoshkin:2022xuw} regarding the talk~\cite{DymarskyTalk} also appeared. There is a conceptual overlap between our works, particularly on the effects that the mass and UV cutoff have on K-complexity; although the way in which we introduce the UV cutoff is different. Nevertheless, our conclusions are consistent with each other.

\section{The Lanczos Algorithm}
\label{sec:Lanczos}

In this section we provide an outline of the recursion method, which is the basis for the Lanczos algorithm. The reader can refer to~\cite{RecursionBook,Parker:2018yvk} for further details. The basic idea is to study the Heisenberg evolution of an operator $\mathcal{O}$ in a given quantum system governed by a Hamiltonian $H$. The time evolution of $\mathcal{O}$ is determined by the Heisenberg equation
\begin{equation}
    \label{eq:HeisenbergEq}
     \partial_{t}\mathcal{O}(t)= i~[H,\mathcal{O}(t)]~,
\end{equation}
whose formal solution for a time-independent Hamiltonian $H$ can be written as
\begin{equation}
    \label{eq:HeisenbergSol}
     \mathcal{O}(t)= e^{i\,tH}~\mathcal{O}(0)~e^{-i\,tH}~.
\end{equation}
One may consider an expansion of $\mathcal{O}(t)$ as a power series in the following way
\begin{equation}
    \label{eq:HeisenbergSolSeries}
     \mathcal{O}(t)= \sum_{n=0}^{\infty}\frac{(i\, t)^{n}}{n!}\tilde{\mathcal{O}}_{n}~,
\end{equation}
where the $\tilde{\mathcal{O}}_{n}$ are nested commutators of $\mathcal{O}(0)$ with $H$. This expression intuitively shows how a ``simple'' initial operator $\mathcal{O}(0)$ may become increasingly ``complex'' at each time step as it evolves through its commutation with $H$. However, it is rare that one can exactly solve the operator dynamic for generic physical systems in this way. Instead, a method that is better suited for this study lies in the definition of a super-operator called the \emph{Liouvillian}
\begin{equation}
    \label{eq:Liouvillian}
     \mathcal{L}:=[H,\cdot~]~,
\end{equation}
which acts linearly on the space of operators and commutes them with the Hamiltonian. This allows us to write the formal solution to the Heisenberg evolution of $\mathcal{O}(t)$ simply as $\mathcal{O}(t)= e^{i\mathcal{L}t}~\mathcal{O}(0)$ and to re-interpret the operators $\tilde{\mathcal{O}}_{n}$ in~\eqref{eq:HeisenbergSolSeries} as $\tilde{\mathcal{O}}_{n}=\mathcal{L}^{n}~\mathcal{O}(0)$. This bears resemblance to the time-evolution of states in the Schr\"{o}dinger picture and suggests re-interpreting~\eqref{eq:HeisenbergSolSeries} as the operator $\mathcal{O}$'s ``wavefunction'' expanded in some local basis of states $\tilde{\mathcal{O}}_{n}$ belonging to a local Hilbert space $\mathcal{H}_{\mathcal{O}}$, known as Krylov space. Formally, $\mathcal{H}_{\mathcal{O}}$ is the Gelfand--Naimark--Segal (GNS) Hilbert space $\mathcal{H}_{\textrm{GNS}}$ constructed from the quantum system's operator algebra $\mathcal{A}$. As mentioned in Sec.~\ref{sec:introduction}, $\mathcal{H}_{\mathcal{O}}$ is spanned by the $\lbrace \vert \tilde{\mathcal{O}}_{n})\rbrace$, which we interpret as a local basis of states in this abstract Hilbert space. 

In order to complete the analogy with the Schr\"{o}dinger picture we require a notion of inner product in $\mathcal{H}_{\mathcal{O}}$. Since we are interested in considering finite temperature $1/\beta$ effects, we will work with the Wightman inner product induced by the thermal expectation value 
\begin{equation}
    \label{eq:WightInnProd}
     (A \vert B) := \langle e^{\beta H/2}A^{\dagger}e^{-\beta H/2}B\rangle_{\beta}\equiv\frac{1}{\mathcal{Z}_{\beta}}\textrm{Tr}(e^{-\beta H/2}A^{\dagger}e^{-\beta H/2}B)~,
\end{equation}
where $\mathcal{Z}_{\beta}:=$Tr$\left(e^{-\beta H}\right)$ is the thermal partition function. This choice will allow us to connect our results with the study of Lyapunov exponents and the MSS chaos bound~\eqref{eq:MSS}.
Having introduced the inner product, we apply the recursion method~\cite{RecursionBook} to study the dynamics of the operator $\mathcal{O}$. Starting from $\lbrace \vert \tilde{\mathcal{O}}_{n})\rbrace$ we construct an orthonormal basis $\lbrace \vert \mathcal{O}_{n})\rbrace$ of $\mathcal{H}_{\mathcal{O}}$ by using the Gram--Schmidt orthogonalization procedure. This orthonormal basis is known as the \emph{Krylov basis}, and in a sense is an optimal choice of basis in Krylov space. In the context of quantum state complexity, the Krylov basis has been shown to minimize the \emph{spread complexity} ~\cite{Balasubramanian:2022tpr,Caputa:2022eye,Caputa:2022yju}.

Starting with the first two operators in $\lbrace \vert \tilde{\mathcal{O}}_{n})\rbrace$, which are orthogonal to each other with respect to the Wightman inner product~\eqref{eq:WightInnProd}, we construct the first two elements of the Krylov basis
\begin{equation}
    \label{eq:KrylovBasis0and1}
     \vert \mathcal{O}_{0}) := \vert \tilde{\mathcal{O}}_{0} ) \equiv \vert \mathcal{O}(0))~, \quad \quad  \vert \mathcal{O}_{1}) := b_{1}^{-1}\mathcal{L}\vert \tilde{\mathcal{O}}_{0} )~,
\end{equation}
where $b_{1}:= (\tilde{\mathcal{O}}_{0} \mathcal{L}\vert\mathcal{L} \tilde{\mathcal{O}}_{0})^{1/2}$. One then constructs the next states $n>1$ by an iterative algorithm
\begin{equation}
    \label{eq:KrylovBasisOn}
    \vert \mathcal{O}_{n}) :=  b_{n}^{-1}\vert A_{n})~,\quad\quad b_{n}:= (A_{n} \vert A_{n})^{1/2}~,
\end{equation}
where
\begin{equation}
    \label{eq:KrylovBasisAn}
    \vert A_{n}) := \mathcal{L} \vert \mathcal{O}_{n-1})-b_{n-1}\vert \mathcal{O}_{n-2})~.
\end{equation}
In this way, the elements of the Krylov basis are orthogonal to each other and properly normalized $(\mathcal{O}_{m} \vert \mathcal{O}_{n})=\delta_{mn}$. The $\lbrace b_{n}\rbrace$ are called \emph{Lanczos coefficients} and encode the information about the growth of the operator $\mathcal{O}(0)$.

In the Krylov basis, the time-evolution of $\vert\mathcal{O}(0) )$~\eqref{eq:HeisenbergSolSeries} takes the form
\begin{equation}
    \label{eq:HeisenbergSolSeriesV2}
     \vert\mathcal{O}(t) )= \sum_{n=0}^{\infty}i^{n}\varphi_{n}(t)\vert\mathcal{O}_{n})~,
\end{equation}
where the $\varphi_{n}(t):=i^{-n}(\mathcal{O}_{n}\vert \mathcal{O}(t))$ are ``probability amplitudes'' whose square is conserved in time
\begin{equation}
    \label{eq:PhiSquare}
      \sum_{n=0}^{\infty}\vert\varphi_{n}(t)\vert^{2}=1~.
\end{equation}
These amplitudes are determined by iteratively solving a discretized ``Schr\"{o}dinger equation'' 
\begin{equation}
    \label{eq:PhiSchroed}
      \frac{\textrm{d}\,\varphi_{n}(t)}{\textrm{d}t}=b_{n}\varphi_{n-1}(t)-b_{n+1}\varphi_{n+1}(t)~,
\end{equation}
with initial condition $\varphi_{n}(0)=\delta_{n,0}$ and where $\varphi_{-1}(t)\equiv 0 \equiv b_{0}$ by convention. This equation is interpreted as the hopping of a quantum-mechanical particle on a $1$-dimensional chain. In the Krylov basis, the Krylov complexity of the operator $\mathcal{O}$ is defined as
\begin{equation}
    \label{eq:KrylovCDef}
    K_{\mathcal{O}}(t):=(\mathcal{O}(t)\vert n\vert\mathcal{O}(t))=\sum_{n=0}^{\infty}n \vert \varphi_{n}(t)\vert^{2}~.
\end{equation}
 In the language of~\cite{Parker:2018yvk}, the Krylov complexity belongs to a class of~  ``\emph{quelconque}-complexities'' defined for local operators $H$ and $\mathcal{O}$ on lattice systems, and its growth rate bounds the growth rate other notions of operator growth, such as OTOCs and operator size. 

A central object in the Lanczos algorithm is the \emph{thermal Wightman $2$-point function}, also known as the auto-correlation function
\begin{equation}
    \label{eq:AutoCorr}
    \begin{split}
   C(t)&:= \varphi_{0}(t) \equiv ( \mathcal{O}(t)\vert \mathcal{O}(0))  \equiv \\
   & \equiv\langle e^{i\,(t-i\,\beta /2)H}\mathcal{O}^{\dagger}(0)e^{-i\,(t-i\,\beta /2)H}\mathcal{O}(0) \rangle_{\beta} =\\
   & = \langle \mathcal{O}^{\dagger}(t-i\beta/2)\mathcal{O}(0) \rangle_{\beta} :=\Pi^{W}(t)~. 
   \end{split}
\end{equation}
$\Pi^{W}(t)$ can also be thought of as a two-sided correlation function in the thermofield double state $\vert\textrm{TFD} \rangle$\footnote{This identification is possible due to the choice of inner product~\eqref{eq:WightInnProd}.}. From~\eqref{eq:PhiSchroed} it is clear that the K-complexity~\eqref{eq:KrylovCDef} is obtained from the derivatives of the auto-correlation function. As a consequence, the physical information contained in the auto-correlation function, and therefore in the K-complexity, is equivalently encoded in the \emph{moments} $\lbrace\mu_{2n}\rbrace$ which are the Taylor expansion coefficients of  $\Pi^{W}(t)$ around $t=0$ 
\begin{equation}
    \label{eq:MomentsAutoC}
    \Pi^{W}(t):=\sum_{n=0}^{\infty} \mu_{2n}\frac{(it)^{2n}}{(2n)!}\quad , \quad \mu_{2n}:=(\mathcal{O}(0)\vert \mathcal{L}^{2n}\vert \mathcal{O}(0))=\frac{1}{i^{2n}}\frac{\textrm{d}^{2n}\Pi^{W}(t)}{\textrm{d}t^{2n}}\Big\vert_{t=0}~. 
\end{equation}
These can also be derived from the \emph{Wightman power spectrum} $f^{W}(\omega)$
\begin{equation}
    \label{eq:MomentsSpectral}
    \mu_{2n}=\frac{1}{2\pi}\int_{-\infty}^{\infty}\textrm{d}\omega\,\omega^{2n}f^{W}(\omega)~,
\end{equation}
which is related to the thermal Wightman $2$-point function via a Fourier transformation
\begin{equation}
    \label{eq:SpectralAutoC}
    f^{W}(\omega)=\int_{-\infty}^{\infty}\textrm{d}t\,e^{i\omega t}\Pi^{W}(t)~.
\end{equation}
In this sense, the information about the growth of $\mathcal{O}$ is equivalently encoded in the auto-correlation function, the Lanczos coefficients, the moments, and the power spectrum. In particular there is a non-linear relation between the moments $\mu_{2n}$ and the Lanczos coefficients $b_{n}$~\cite{RecursionBook}
\begin{equation}
    \label{eq:Mu2nToBn}
    b_{1}^{2n}\cdots b_{n}^{2}=\det\left(\mu_{i+j}\right)_{0\leq i,j\leq n}~,
\end{equation}
where $\mu_{i+j}$ is a Hankel matrix constructed from the moments. This expression can alternatively be represented via a recursion relation
\begin{subequations}
\label{eq:Mu2nToBnRecursion}
\begin{equation}
\label{eq:Mu2nToBnRecursion1}
    b_{n}=\sqrt{M^{(n)}_{2n}}~,
\end{equation}
\begin{equation}
\label{eq:Mu2nToBnRecursion2}
    M^{(j)}_{2l}=\frac{M^{(j-1)}_{2l}}{b_{j-1}^{2}}-\frac{M^{(j-2)}_{2l-2}}{b_{j-2}^{2}}~\quad \textrm{with} \quad l=j,\ldots,n~,
\end{equation}
\begin{equation}
\label{eq:Mu2nToBnRecursion3}
    M^{(0)}_{2l}=\mu_{2l}~\quad,\quad b_{-1}\equiv b_{0}:=1\quad,\quad  M^{(-1)}_{2l}=0~.
\end{equation}
\end{subequations}

The usefulness of the Lanczos algorithm lies in the fact that it is applicable to dynamical systems with either finite- or infinite-sized Hilbert spaces. In the former case, the dimension of the Krylov space is also finite and constrained by the dimension of the Hilbert space of the original quantum system~\cite{Rabinovici:2021qqt}. In this case, one performs a finite number of iterations to find the Krylov basis and probability amplitudes. If the Hilbert space is infinite, the Krylov basis is in principle infinite, unless one terminates the algorithm at an arbitrary value of $n$. Similarly, the Lanczos algorithm is applicable to any quantum system with unitary time evolution. 

\subsection{Krylov Complexity and Quantum Chaos}
\label{subsec:LanczosChaos}

In~\cite{Elsayed:2014chaos}, it was proposed that an exponential decay of the power spectrum 
\begin{equation}
    \label{eq:SpectralDecay}
    f^{W}(\omega)\sim e^{-\omega /\omega_{0}}~,
\end{equation}
for $\omega\rightarrow \infty$ could be interpreted as a signature of chaos in \emph{classical} spin systems, where the decay rate $1/\omega_{0}>0$ is associated with the pole of $\Pi^{W}(t)$ along the imaginary axis.  
A necessary condition for this behavior is the analyticity of $\Pi^{W}(t)$ around $t=0$. 

In~\cite{Parker:2018yvk}, authors conjectured that the Lanczos coefficients of a generic chaotic quantum many-body system with local interactions should grow as fast as possible, namely
\begin{equation}
    \label{eq:UnivOperGrowthHyp}
    b_{n}\,\sim\,\alpha n + \gamma = \left(\frac{\pi\omega_{0}}{2}\right)n+\gamma \quad , \quad n \rightarrow \infty~.
\end{equation}
where $\gamma$ includes sub-leading terms in $n$. This is referred to as the \emph{universal operator growth hypothesis}. The linear growth of Lanczos coefficients~\eqref{eq:UnivOperGrowthHyp} implies an exponential decay of the power spectrum~\eqref{eq:SpectralDecay} and can therefore be seen as a more general condition for chaotic behavior. The authors of \cite{Parker:2018yvk} furthermore found a family of $b_n$ for which the Krylov complexity $K_{\mathcal{O}}(t)$ could be exactly obtained. The solutions they found are given by
\begin{align}
\label{ess}
C(t)=\frac{1}{(\cosh(\alpha t))^{\eta}}, \;\;\; b_n=\alpha\sqrt{n(n-1+\eta)}, \;\;\; K_{\mathcal{O}}(t)=\eta \sinh^2 (\alpha t),
\end{align}
where $b_n$ has a linear growth behavior for $n\to\infty$, and $K_{\mathcal{O}}(t)$ grows exponentially at late times. From this exact example, the authors argued the following assertion: whenever the Lanczos coefficients have a smooth linear behavior (\ref{eq:UnivOperGrowthHyp}), the Krylov complexity is expected to grow exponentially
\begin{equation}
    \label{eq:KrylovExp}
    K_{\mathcal{O}}(t)\propto e^{\lambda_{K}t}~,
\end{equation}
where $\lambda_{K}=2\alpha=\pi\omega_{0}$\footnote{This has been verified to hold for many cases, such as those with artificially generated sequences of Lanczos coefficients $b_n = \alpha n + \gamma$ with various kinds of terms $\gamma \sim O(1)$. See~\cite{Parker:2018yvk} for more details.}.

As we will discuss in detail in Sec.~\ref{subsec:KrylovSpectral}, in QFTs at finite temperature $1/\beta$, the thermal $2$-point function $\Pi^{W}(t)$ has poles at $it=\pm\beta/2$ and as a consequence, the power spectrum has a leading exponential decay according to~\eqref{eq:SpectralDecay} with $\omega_{0}=2/\beta$. Assuming that the Lanczos coefficients behave smoothly, the Krylov complexity behaves exponentially with $\lambda_{K}=\pi\omega_{0}=2\pi/\beta$ and the conjectured bound on Lyapunov exponents~\eqref{eq:BoundLyapKrylov} reduces to the MSS bound~\eqref{eq:MSS}. Since this analysis does not rely on any details of the integrability/chaotic behavior of the QFT, the exponential growth of the Krylov complexity~\eqref{eq:KrylovExp} \emph{cannot} be interpreted as a feature of chaotic behavior in QFTs at finite temperature \cite{Dymarsky:2021bjq}.

In~\cite{Avdoshkin:2019trj}, the authors proposed a \emph{quantum} chaos bound for quantum many-body systems at finite temperature. Assuming the Lanczos coefficients have smooth linear behavior for large $n$~\eqref{eq:UnivOperGrowthHyp} they showed that the following bound should hold
\begin{equation}
    \label{eq:BoundAlphaTemp}
     \alpha \leq \frac{\pi}{\beta+\beta^{\ast}(\beta)}~,
\end{equation}
where $\alpha$ is the growth rate of the Lanczos coefficients~\eqref{eq:UnivOperGrowthHyp} and where $\beta^{\ast}(\beta)$ depends on the pole of $\Pi^{W}(t)$. Furthermore, assuming the bound $\lambda_{L}\leq 2\alpha$ is valid at finite temperature $\beta^{-1}$, the authors conjectured the following bound on the Lyapunov exponent
\begin{equation}
    \label{eq:BoundKrylovTemp}
     \lambda_{L}\leq \frac{2\pi}{\beta+\beta^{\ast}(\beta)}~.
\end{equation}
In usual QFTs, $1/\beta^{\ast}(\beta)$ will be of the order of the UV cutoff $\Lambda\rightarrow \infty$, in which case the bound~\eqref{eq:BoundKrylovTemp} reduces to the original MSS bound. However, in lattice theories $\beta^{\ast}(\beta)$ can be non-zero.

A proof of~\eqref{eq:BoundKrylovTemp} was obtained in~\cite{Gu:2021xaj}, where the authors studied the behavior of OTOCs for many-body systems, such as the SYK model, in the large-$N$ regime. They showed that in this case, the OTOC can be written as an integral of a product of two $3$-point functions; consistent with a summation of ladder diagrams. From certain analytic properties of the said functions, which depend on positive semi-definiteness for the fastest-growing scrambling modes, they proved the following bound
\begin{equation}
    \label{eq:BoundGu}
    \frac{1}{\omega^{\ast}} \leq \frac{\pi}{\lambda_{L}}-\frac{\beta}{2}~,
\end{equation}
where $\omega^{\ast}=2\omega_{0}/(2-\beta\omega_0)$ and where $1/\omega_{0}$ is the decay rate of the power spectrum, as in~\eqref{eq:SpectralDecay}. In usual QFTs\footnote{By usual QFTs here we mean that the thermal $2$-point function $\Pi^{W}(t)$ has a pole at $it=\beta/2$ as explained in the  paragraph after (\ref{eq:KrylovExp}), which leads $\omega_0=2/\beta$ and $1/\omega^{\ast}\rightarrow 0$.}, where $\omega_0=2/\beta$ and $1/\omega^{\ast}\rightarrow 0$, this expression similarly reduces to the MSS bound. Substituting the expression for $\omega^{\ast}(\omega_{0})$ in~\eqref{eq:BoundGu} we obtain
\begin{equation}
    \label{eq:BoundGu2}
    \lambda_{L}\leq \pi\omega_{0}~,
\end{equation}
and thus, this expression can be interpreted as an inequality between the Lyapunov exponent $\lambda_{L}$ and the decay rate $1/\omega_{0}$ of the Wightman power spectrum~\eqref{eq:SpectralDecay}. Furthermore, if the Lanczos coefficients behave smoothly as a function of $n$, then $\omega_{0}$, $\alpha$ and $\lambda_{K}$ can be shown to be related from the assertion (\ref{eq:KrylovExp}) via
\begin{equation}
    \label{eq:Omega0LambdaKAlpha}
    \lambda_{K} =2\alpha=\pi\omega_{0}~,
\end{equation}
as discussed in~\cite{Dymarsky:2021bjq}. As a consequence, in such systems at finite temperature, the Lyapunov exponent $\lambda_{L}$ and growth rate of the Krylov complexity $\lambda_{K}$ satisfy 
\begin{equation}
    \label{eq:BoundGu3}
    \lambda_{L}\leq \lambda_{K}=\pi\omega_{0}~.
\end{equation}
It is then plausible to conjecture that the following inequality should hold for any quantum system at finite temperature $1/\beta$~\cite{Parker:2018yvk, Avdoshkin:2019trj,Avdoshkin:2022xuw}
\begin{equation}
    \label{eq:ImprovedChaosBound}
    \lambda_{L}\leq \lambda_{K}\leq \frac{2\pi}{\beta}~.
\end{equation}
However, if the Lanczos coefficients are not smooth functions of $n$, then the bound~\eqref{eq:BoundGu} does not lead to a direct inequality between $\lambda_{L}$ and $\lambda_{K}$. Nevertheless, the fact that $\lambda_{L}$ is generally expected to be bounded by the decay rate of the power spectrum~\eqref{eq:BoundGu} at finite temperature could also be argued from the properties of $2$- and $4$-point correlation functions as follows: To derive the MSS bound~\eqref{eq:MSS}, the analytic property of thermal $4$-point functions (OTOCs) is used (see e.g.~\cite{Maldacena:2015waa}). Since $4$-point functions can be constructed from integrals of $2$-point functions with interaction vertices, the bound ({\ref{eq:BoundGu}}) of $\lambda_L$ by the analytic property of $\Pi^{W}(t)$ is somehow expected. The derivation of ({\ref{eq:BoundGu}}) is therefore similar to the derivation of the MSS bound from the Eigenstate Thermalization Hypothesis (ETH)~\cite{Murthy:2019fgs} in that they use the Ansatz of OTOCs with a fastest-growing scrambling mode for $\lambda_{L}$ and the analytic property of $\Pi^{W}(t)$.

\subsection{Krylov Complexity and Spectral Statistics}
\label{subsec:KrylovSpectral}

The structure of the thermal two-point function $\Pi^{W}(t)$, and consequently of the power spectrum $f^{W}(\omega)$, in an energy eigenbasis $\lbrace \vert E \rangle \rbrace$ provides details about how the Lanczos coefficients and the Krylov complexity capture information from different sectors in the energy spectrum, as well as UV or IR cutoffs. To see this, it is useful to think of $f^{W}(\omega)$ as a transition amplitude between states with different energies. As we will also discuss below, from the physical point of view, the fact that in QFTs the power spectrum~\eqref{eq:SpectralDecay} has a leading exponential decay for large $\omega$ is a consequence of the singular behavior of $\mathcal{O}(t)\mathcal{O}(0)$ at $t=0$ in local QFTs~\cite{Wilson:1969zs, Wilson:1972ee}. For a single particle state in the canonical (Gibbs) ensemble, we can write $f^{W}(\omega)$ in the case where the energy spectrum is discrete in the following way
\begin{equation}
    \label{eq:PowerSpectrumEnergyDisc}
    f^{W}(\omega)=  \frac{2\pi}{\mathcal{Z}_{\beta}}\sum_{i,j}e^{-\frac{\beta}{2} \left(E_{i}+E_{j}\right)}\vert \mathcal{O}_{ij}\vert^{2}\delta(E_{i}-E_{j}+\omega)~,
\end{equation}
since in an energy eigenbasis $\lbrace \vert E_{i}\rangle \rbrace$ and for the normalized thermal state
\begin{equation}\label{eq:ThermalState}
    \rho_{\beta}=\frac{e^{-\beta H}}{\mathcal{Z}_{\beta}}=\frac{1}{\mathcal{Z}_{\beta}}\sum_{i}e^{-\beta E_{i}}\vert E_{i}\rangle \langle E_{i} \vert~,
\end{equation}
the (Wightman) thermal two-point function~\eqref{eq:AutoCorr} is given by
\begin{equation}
    \label{eq:WightTherm2pPFn}
    \Pi^{W}(t)= \frac{1}{\mathcal{Z}_{\beta}}\sum_{i,j}e^{-\frac{\beta}{2} \left(E_{i}+E_{j}\right)}\,e^{i\,t\left(E_{i}-E_{j}\right)}\vert \mathcal{O}_{ij}\vert^{2}~,
\end{equation}
where $\left(\mathcal{O}(t)\right)_{ij}=\langle E_{i}\vert \mathcal{O}(t)\vert E_{j}\rangle=e^{i(E_{i}-E_{j})t}\mathcal{O}_{ij}$, with $\mathcal{O}_{ij}\equiv\left(\mathcal{O}(0)\right)_{ij}=\mathcal{O}(E_{i},E_{j})$ are the matrix elements in the energy eigenbasis and $\vert \mathcal{O}_{ij}\vert^{2}:=\mathcal{O}^{\dagger}_{ij}\mathcal{O}_{ji}$. In this case the partition function is simply given by $\mathcal{Z}_{\beta}=$Tr$(e^{-\beta H})=\sum_{i}e^{-\beta E_{i}}$.
Note that the power spectrum~\eqref{eq:PowerSpectrumEnergyDisc} is in reality a distribution consisting of a series of delta distributions, each centered at $\omega=E_{j}-E_{i}:=\Delta E_{ji}\equiv-\Delta E_{ij}$. The explicit form of $f^{W}(\omega)$ depends on the dynamics arising from $H$ and the operator $\mathcal{O}(0)$ encoded in the square of the matrix elements $\vert \mathcal{O}_{ij}\vert^{2}$. Indeed, rigorously obtaining analytic constraints on~\eqref{eq:PowerSpectrumEnergyDisc} is challenging without making assumptions about the structure of the matrix elements. Nevertheless, it is possible to obtain general upper bounds on the moments $\mu_{2n}$~\eqref{eq:MomentsSpectral}. Assuming that $0\leq\omega_{I}\leq \omega \leq \omega_{F}$ we can write
\begin{equation}
    \label{eq:IntegralSpectrumEnergyDiscPos}
    \begin{split}
  & \frac{1}{2\pi}\int_{\omega_{I}}^{\omega_{F}}\textrm{d}\omega \,\omega^{2n}f^{W}(\omega)= \\
  &= \frac{1}{\mathcal{Z}_{\beta}}\sum_{i,j}\int_{\omega_{I}}^{\omega_{F}}\textrm{d}\omega \,\omega^{2n}\,e^{-\frac{\beta}{2}(E_{i}+E_{j})}\vert \mathcal{O}_{ij}\vert^{2}\delta(E_{i}-E_{j}+\omega)= \\
   & = \frac{1}{\mathcal{Z}_{\beta}}\sum_{E_{j}\geq E_{i}+\omega_{I}}^{E_{j}\leq E_{i}+\omega_{F}}(\Delta E_{ji})^{2n}\,e^{-\frac{\beta}{2} \Delta E_{ji}}\,e^{-\beta E_{i}}\vert \mathcal{O}_{ij}\vert^{2}\Big\vert_{E_{i}+\omega_{I}\leq E_{j}\leq E_{i}+\omega_{F}} \leq \\
   &\leq \frac{\omega_{F}^{2n}\,e^{-\frac{\beta}{2} \omega_{I}}}{\mathcal{Z}_{\beta}}\sum_{E_{j}\geq E_{i}+\omega_{I}}^{E_{j}\leq E_{i}+\omega_{F}}\,e^{-\beta E_{i}}\vert \mathcal{O}_{ij}\vert^{2}\Big\vert_{E_{i}+\omega_{I}\leq E_{j}\leq E_{i}+\omega_{F}} \leq \\
   &\leq\frac{\omega_{F}^{2n}\,e^{-\frac{\beta}{2} \omega_{I}}}{\mathcal{Z}_{\beta}}\sum_{i,j}\,e^{-\beta E_{i}}\vert \mathcal{O}_{ij}\vert^{2}=\omega_{F}^{2n}\,e^{-\frac{\beta}{2} \omega_{I}}\textrm{Tr}\left(\mathcal{O}^{\dagger}(0)\rho_{\beta}\mathcal{O}(0)\right)~,
    \end{split}
\end{equation}
where we used $\omega_{I}\leq \Delta E_{ji} \leq \omega_{F}$ and $\beta>0$. Similarly, for $0\geq-\omega_{I}\geq \omega \geq- \omega_{F}$ we have
\begin{equation}
    \label{eq:IntegralSpectrumEnergyDiscNeg}
    \frac{1}{2\pi}\int_{-\omega_{F}}^{-\omega_{I}}\textrm{d}\omega \,\omega^{2n}f^{W}(\omega)\leq \omega_{F}^{2n}\,e^{-\frac{\beta}{2} \omega_{I}}\textrm{Tr}\left(\mathcal{O}(0)\rho_{\beta}\mathcal{O}^{\dagger}(0)\right)~,
\end{equation}
since in this case $\omega_{I}\leq -\Delta E_{ji} \leq \omega_{F}$. Thus, the moments $\mu_{2n}$ are bounded according to
\begin{equation}
\label{eq:MomSpectrumEnergyDisc}
\begin{split}
    &\mu_{2n}(\omega_{I},\omega_{F})= \frac{1}{2\pi}\int_{\omega_{I}\leq \vert \omega\vert \leq\omega_{F}}\textrm{d}\omega \,\omega^{2n}f^{W}(\omega) \leq \\
    & \leq \omega_{F}^{2n}\,e^{-\frac{\beta}{2} \omega_{I}}\left(\textrm{Tr}\left(\mathcal{O}(0)\rho_{\beta}\mathcal{O}^{\dagger}(0)\right)+\textrm{Tr}\left(\mathcal{O}^{\dagger}(0)\rho_{\beta}\mathcal{O}(0)\right)\right)
    \end{split}
\end{equation}
In particular, if the initial operator $\mathcal{O}(0)$ is Hermitian then $\mu_{0}(\omega_{I})\leq 2e^{-\beta \omega_{I}/2}\times$ $\textrm{Tr}\left(\mathcal{O}(0)\rho_{\beta}\mathcal{O}(0)\right)$. In usual QFTs, this thermal correlation function would diverge due to the singular behavior of $\mathcal{O}(t)\mathcal{O}(0)$ at $t=0$. However, in quantum many-body systems with finite-dimensional Hilbert spaces, one could expect this to yield a finite value\footnote{Even though the dimension of the Hilbert space and matrix representation of $\mathcal{O}$ is infinite, the $2$-point function can be finite when only a finite number of matrix components of $\mathcal{O}$ are non-zero due to the $r$-local property of $\mathcal{O}$ and $H$.}.

Here, $\omega_{I}$ and $\omega_{F}$ can be thought of as cutoffs of the energy transfer associated with $\mathcal{O}$. In a free scalar theory, $\omega_{I}$ and $\omega_{F}$ are the minimum and maximum energies of a particle, respectively. In a lattice system, $\omega_{F}$ would be associated with the existence of a minimum length scale: the lattice spacing $a$. For instance, $\omega_{F}$ of a free massless scalar on a periodic lattice may be given by the maximum value of the dispersion relation $\omega=\frac{2}{a}\sin[pa/2]$, where $p$ is discrete momentum. If the energy spectrum is bounded, the power spectrum is also bounded as $\omega_F\le E_{\textrm{UV}}-E_{\textrm{IR}}\sim O(S/a)$, where $S$ is the number of degrees of freedom of the system, e.g. the number of lattice sites. The order of $\omega_F$ is $O(1/a)$ in a free lattice theory and $O(S/a)$ in a chaotic lattice system~\cite{Barbon:2019wsy}. In the continuum limit $a\to0$, both $1/a$ and $S/a$ diverge, and we cannot practically distinguish them. This is an explanation of why the linear growth behavior of $b_n$ (\ref{eq:UnivOperGrowthHyp}) and the exponential growth behavior of $K_\mathcal{O}(t)$ (\ref{eq:KrylovExp}) cannot be interpreted as features of chaotic behavior in QFTs. Nevertheless, Eq.~\eqref{eq:MomSpectrumEnergyDisc} shows that we generically expect the moments to be sensitive to the energy transfer cutoffs and as a consequence, also the Lanczos coefficients and the Krylov complexity. For instance, suppose that $f^{W}(\omega)$ decays exponentially at large $\vert\omega\vert$. Then, for small $n$, the integration in the region where $\vert\omega\vert$ is large does not contribute much to $\mu_{2n}$ (\ref{eq:MomSpectrumEnergyDisc}). As $n$ increases, the contribution from $\vert\omega\vert\sim\omega_F$ also comes into $\mu_{2n}$ due to the factor of $\omega^{2n}$. This means that the effect of IR cutoff $\omega_I$ to the Lanczos coefficients $b_n$ is visible even when $n$ is small, while the effect of UV cutoff $\omega_F$ is not visible unless $n$ is sufficiently large.

Heuristically, if the power spectrum is unbounded $\omega_{F}\rightarrow \infty$, the particle with power spectrum $f^{W}(\omega)$~\eqref{eq:PowerSpectrumEnergyDisc} can access states with arbitrarily high energies. If $\Pi^{W}(t)$ is analytic, its integral representation $\Pi^{W}(t)=\frac{1}{2\pi}\int_{-\infty}^{\infty}\textrm{d}\omega\,e^{-i\omega t}f^W(\omega)$ should yield a finite value. This implies that for large $\omega$, $f^{W}(\omega)$ should decay faster than the growth rate of $e^{-i\omega t}$ along $t=i \tau$. This condition connects the decay rate $1/\omega_0$ and the pole of $\Pi^{W}(t)$ along the imaginary axis as explained around (\ref{eq:SpectralDecay}). An argument based on ETH~\cite{Murthy:2019fgs} can also be used to provide a bound on the matrix elements $\vert \mathcal{O}_{ij}\vert^{2}$.

This analysis can be made more precise if we consider a continuous energy spectrum. This would be the case, for example, if the energy depends on a continuous variable as is the case in QFTs, which are the main focus of this work, or if we are interested in studying random-matrix models in the context of Jackiw--Teitelboim (JT) gravity~\cite{Saad:2018bqo,Saad:2019lba,Saad:2019pqd,Mertens:2022irh}, black hole microstates in AdS/CFT~\cite{Cotler:2020ugk,Cotler:2021cqa,Cotler:2022rud} or Schwarzian quantum mechanics~\cite{Mertens:2017mtv}. For a continuous energy spectrum with a given Hamiltonian $H$, the discrete sum $\sum_i$ is replaced by an integral $\int $d$E\, \rho_H(E)$, where $\rho_H(E)$ is the density of states for $H$ at a given energy $E$ (see e.g. \cite{Saad:2019pqd}). The continuous version of \eqref{eq:WightTherm2pPFn}, which is not normalized by the partition function, is given by
\begin{align}
    \Pi_H^{W}(t)= \int\textrm{d}E\int\textrm{d}E'\rho_H(E)\rho_H(E')e^{-\frac{\beta}{2} \left(E+E'\right)}\,e^{i\,t\left(E-E'\right)}\vert \mathcal{O}(E,E')\vert^{2}~,\label{eq:EnsemblePi}
\end{align}
where $\mathcal{O}(E,E')\equiv \langle E \vert \mathcal{O}(0)\vert E' \rangle$ and $\vert \mathcal{O}(E,E')\vert^{2}:=\mathcal{O}^{\dagger}(E,E')\mathcal{O}(E',E)$\footnote{Strictly speaking, Eq.~\eqref{eq:EnsemblePi} is valid if the matrix elements depend only the absolute value of energy. For a free scalar theory, we should consider the sum over particle number.}. The integration is performed over the absolute value of the energy. Here we are implicitly considering the case where we have a system described by an ensemble of two identical Hamiltonians $H_{1}=H_{2}=H$. 

Let $P(H)$ be the probability density for a single Hamiltonian $H$, and suppose that $P(H)$ factorizes into a measure for the density of states and a measure for the matrix elements like in RMT~\cite{Saad:2019pqd}. Then the pair correlation function can be defined by 
\begin{equation}
\label{eq:DensityPair}
   \rho(E,E'):=\int \textrm{d}H \,P(H)\, \rho_H(E)\rho_H(E')~,
\end{equation}
and we can write the un-normalized \emph{ensemble averaged} thermal two-point function as
\begin{align}
\begin{split}
\overline{\Pi^{W}}(t)&=\int \textrm{d}H \,P(H)\,\Pi_H^{W}(t)=\\
&=\int\textrm{d}E\,\int\textrm{d}E'\,\rho(E,E')e^{-\frac{\beta}{2} \left(E+E'\right)}\,e^{i\,t\left(E-E'\right)}\vert \mathcal{O}(E,E')\vert^{2}~.\label{eq:UnNormEnsAvPiW}
\end{split}
\end{align}
Here, $\vert \mathcal{O}(E,E')\vert^{2}$ is the average of the squared matrix element. For a continuous energy spectrum, the partition function is given by
\begin{align}
\label{eq:ThermalZDensityRho}
 \mathcal{Z}_{H}(\beta):=\textrm{Tr}\left(e^{-\beta H}\right)=\int \textrm{d}E\, \rho_H(E)e^{-\beta E}\langle E\vert E\rangle~.
\end{align}
The normalization of $\langle E\vert E'\rangle$ can be determined from
\begin{align}
\int \textrm{d}E \,\rho_H(E) \,\vert E\rangle \langle E\vert E'\rangle=\vert E'\rangle~.
\end{align}
This condition leads to 
\begin{align}
\label{eq:normEE}
\langle E\vert E'\rangle=\frac{\delta(E-E')}{\rho_H(E)}~.
\end{align} 
With this normalization, we can verify that
\begin{align}
&\int\textrm{d}E\int\textrm{d}E'\rho_H(E)\rho_H(E')e^{-\frac{\beta}{2} \left(E+E'\right)}\,e^{i\,t\left(E-E'\right)}\langle E\vert E'\rangle\langle E'\vert E\rangle\\
=&\int \textrm{d}E\, \rho_H(E)e^{-\beta E}\langle E\vert E\rangle=\mathcal{Z}_{H}(\beta)~,
\end{align}
where $\mathcal{Z}_{H}(\beta)$ is defined by (\ref{eq:ThermalZDensityRho}).
We define the ensemble-averaged partition function in the following way
\begin{align}
\label{eq:EnsPartFunc}
\begin{split}
\overline{\mathcal{Z}}(\beta)&:=\int \textrm{d}H\,P(H)[\mathcal{Z}_{H}(\beta)]^2=\\
&=\int\textrm{d}E\int\textrm{d}E'\,\rho(E,E')\,e^{-\beta \left(E+E'\right)}\langle E\vert E\rangle\langle E'\vert E'\rangle~,
\end{split}
\end{align}
where we used the fact that the ensemble of theories consists of two copies of the same theory arising from the Hamiltonian $H$. We note that $\langle E\vert E\rangle\langle E'\vert E'\rangle$ in (\ref{eq:EnsPartFunc}) is the averaged one, and its normalization is different from (\ref{eq:normEE})\footnote{In RMT, the normalization can be determined from (2.18) of~\cite{Saad:2019pqd} by choosing $\mathcal{O}$ as the identity operator.}. In~\eqref{eq:DensityPair} and~\eqref{eq:EnsPartFunc} the integration over $H$ can be thought of as an integration over the couplings appearing in the Hamiltonian.
The Fourier transform of~\eqref{eq:UnNormEnsAvPiW} yields the un-normalized ensemble-averaged power spectrum
\begin{equation}
    \label{eq:PowerSpectrumEnergyCont}
    \begin{split}
    \overline{f^{W}}(\omega) = &2\pi\int_{E_{\textrm{IR}}}^{E_{\textrm{UV}}}\textrm{d}E\int_{E_{\textrm{IR}}}^{E_\textrm{UV}}\textrm{d}E'\,\rho(E,E')\,e^{-\frac{\beta}{2} \left(E+E'\right)}\vert \mathcal{O}(E,E')\vert^{2}\,\delta(E-E'+\omega)~.
    \end{split}
\end{equation}
where we introduced integration limits $E_{\textrm{IR}}$ and $E_{\textrm{UV}}$, corresponding to the IR and UV bounds in the energy spectrum respectively. While $E_{\textrm{UV}}$ is typically unbounded, $E_{\textrm{IR}}$ can be thought of as related to the ground state energy $E_{0}$. 
Evaluating the integral over $E'$ in~\eqref{eq:PowerSpectrumEnergyCont} yields
\begin{equation}
    \label{eq:PowerSpectrumEnergyContV2}
    \overline{f^{W}}(\omega) = 2\pi\,e^{-\frac{\beta \omega}{2}}\int_{E_{\textrm{IR}}}^{E_{\textrm{UV}}-\omega}\textrm{d}E\,\rho(E,\omega)\,e^{-\beta E}\vert \mathcal{O}(E,\omega)\vert^{2}\big\vert_{E_{\textrm{IR}}\leq E+\omega \leq E_{\textrm{UV}}}~,
\end{equation}
where $\rho(E,\omega)\equiv\rho(E,E+\omega)$ and $\mathcal{O}(E,\omega)\equiv\mathcal{O}(E,E+\omega):=\langle E \vert \mathcal{O}(0) \vert  E+\omega\rangle$ and where we assumed $0\leq \omega_{I}\leq \omega\leq \omega_{F}$. For $-\omega_{F}\leq\omega\leq -\omega_{I}\leq0$, (\ref{eq:PowerSpectrumEnergyContV2}) is given by
\begin{equation}
    \overline{f^{W}}(\omega) = 2\pi\,e^{-\frac{\beta \omega}{2}}\int_{E_{\textrm{IR}}-\omega}^{E_{\textrm{UV}}}\textrm{d}E\,\rho(E,\omega)\,e^{-\beta E}\vert \mathcal{O}(E,\omega)\vert^{2}\big\vert_{E_{\textrm{IR}}\leq E+\omega \leq E_{\textrm{UV}}}~.
\end{equation}
By using a change of variables $E'=E+\omega$, we obtain
\begin{equation}
    \overline{f^{W}}(\omega) = 2\pi\,e^{\frac{\beta \omega}{2}}\int_{E_{\textrm{IR}}}^{E_{\textrm{UV}}+\omega}\textrm{d}E'\,\rho(E',-\omega)\,e^{-\beta E'}\vert \mathcal{O}(E',-\omega)\vert^{2}\big\vert_{E_{\textrm{IR}}\leq E-\omega \leq E_{\textrm{UV}}}~,\label{eq:fWNegativeOm}
\end{equation}
where we use $\rho(E,E')=\rho(E',E)$ and $\vert \mathcal{O}(E,E')\vert^{2}=\vert \mathcal{O}(E',E)\vert^{2}$. This expression can also be obtained from (\ref{eq:PowerSpectrumEnergyCont}) by integrating over $E$ instead.
Though it is tempting to state that the Boltzmann prefactors in Eqs.\eqref{eq:PowerSpectrumEnergyContV2} and~\eqref{eq:fWNegativeOm} are sufficient to guarantee the exponential decay of the power spectrum at $\vert \omega \vert \rightarrow \infty$, the situation is more subtle due to the presence of the density pair correlation function $\rho(E,E')$~\eqref{eq:DensityPair}. 

In this case, the ensemble-averaged moments can be written as:
\begin{align}
    \label{eq:MomentsSpectralCont}
    \begin{split}
   & \overline{\mu}_{2n}(\omega_{I},\omega_{F},E_{\textrm{IR}},E_{\textrm{UV}})=\frac{1}{2\pi}\int_{\omega_{I}\leq \vert \omega \vert \leq  \omega_{F}}\textrm{d}\omega\,\omega^{2n}\,\overline{f^{W}}(\omega)=\\
    &\int_{\omega_{I}\leq \vert \omega \vert \leq  \omega_{F}}\textrm{d}\omega\,\int_{E_{\textrm{IR}}}^{E_{\textrm{UV}}}\textrm{d}E\int_{E_{\textrm{IR}}}^{E_\textrm{UV}}\textrm{d}E'\,\rho(E,E')\,e^{-\frac{\beta}{2} \left(E+E'\right)}\vert \mathcal{O}(E,E')\vert^{2}\,\delta(E-E'+\omega)\,\omega^{2n}~.
    \end{split}
\end{align}
If the expressions for the density pair correlation function and the matrix elements are known, such an integral could be evaluated using the saddle-point approximation. This, however, is beyond the scope of the present discussion. The relevant point is the following: as argued in the above paragraphs, it is important to distinguish the bounds in the power spectrum $\lbrace\omega_{I},\omega_{F}\rbrace$ (energy transfer cutoffs) from the bounds in the energy spectrum $\lbrace E_{\textrm{IR}},E_{\textrm{UV}}\rbrace$. While the latter did not appear explicitly in the discrete case, they are implicitly present in the double sum in~\eqref{eq:IntegralSpectrumEnergyDiscPos}. Indeed, assuming that the $\omega$ and $E$ integrals can be exchanged and assuming that $0<\omega_{I}<\omega_{F}$,
\begin{equation}
    \label{eq:MOmentsSpectralContV2}
    \begin{split}
   & \overline{\mu}_{2n}(\omega_{I},\omega_{F},E_{\textrm{IR}},E_{\textrm{UV}})=\\ &\int_{E_{\textrm{IR}}}^{E_{\textrm{UV}}}\textrm{d}E\int_{E_{\textrm{IR}}}^{E_\textrm{UV}}\textrm{d}E'\,\rho(E,E')\,e^{-\frac{\beta}{2} \left(E+E'\right)}\vert \mathcal{O}(E,E')\vert^{2}\,(-E+E')^{2n}\times\\
   &\times \left(\Theta\left(E-E'+\omega_{F},-E+E'-\omega_{I}\right)+\Theta\left(-E+E'+\omega_{F},E-E'-\omega_{I}\right)\right)~,
   \end{split}
\end{equation}
where $\Theta$ is the Heaviside step function\footnote{The Heaviside step function is defined as $\Theta(x)=1$ for $x>0$ and $\Theta(x)=0$ for $x<0$. $\Theta(x,y)\equiv\Theta(x)\Theta(y)$ is a generalization of the Heaviside step function for two variables.}. As we will see in later sections, the behavior of the Lanczos coefficients as well as the Krylov complexity will be influenced by the presence of such cutoffs due to the relation~\eqref{eq:Mu2nToBn}.

We emphasize again that the moments $\mu_{2n}$ are sensitive to cutoffs of the power spectrum as (\ref{eq:MomSpectrumEnergyDisc}) and (\ref{eq:MomentsSpectralCont}). Thus, the Lanczos coefficients $b_n$ and the Krylov complexity $K_\mathcal{O}(t)$ are also expected to be sensitive to them. In this work, we focus on a momentum cutoff in the continuous power spectrum of QFTs on a non-compact space without the average of $H$ as explained below. We also note that there is another approach that introduces a UV cutoff by means of a microcanonical version of the Krylov complexity \cite{Kar:2021nbm}. In that case, fixing $E$ for the microcanonical ensemble yields a bounded power spectrum.

At the same time and as mentioned in Sec.~\ref{sec:introduction}, although Krylov complexity has been studied for CFTs and other systems with a high-degree of symmetry, it has not yet been studied for more general QFTs. To go beyond symmetry-dependent scenarios, in the following sections we study how the Lanczos coefficients and the Krylov complexity behave when we break the conformal symmetry by considering a massive scalar field. The mass can be thought of as an IR cutoff $\omega_{I}\sim m$, and it is expected to affect the K-complexity at early times. We will also consider the effect of introducing a UV cutoff $\Lambda$ in two ways: one by bounding the power spectrum from above with a hard UV cutoff $\omega_{F}\sim \Lambda$ and another by smoothly modifying the behavior of $f^{W}(\omega)$ for $\vert \omega \vert \rightarrow \infty$ with an exponential correction. As explained in Sec.~\ref{sec:introduction}, the introduction of such cutoffs is motivated by an argument made by Dymarsky and his collaborators based on a discretized lattice model~\cite{DymarskyTalk, Avdoshkin:2022xuw}. For finite $\omega_F$, we introduce a momentum cutoff in free QFTs, which is one of the commonly used UV regularization methods in QFTs. Afterward, we will study the behavior of the Lanczos coefficients for interacting QFTs using perturbation theory. In this case, we will start with a 4-dimensional massless scalar field theory and deform it with a relevant or a marginally irrelevant operator ($g \, \phi^3$ or $g \, \phi^4$ respectively). Since these two types of deformation affect UV physics in different ways, we expect them to leave a different imprint on the Lanczos coefficients.

\section{Free Massive Scalar in $d$-dimensions}
\label{sec:FreeMassScalar}

Consider a real-valued massive scalar field $\phi$ in $d$ spacetime dimensions with Euclidean Lagrangian of the form
\begin{equation}
\label{eq:FreeScalarLag}
    \mathcal{L}^{\textrm{free}}_{E} = \frac{1}{2} (\partial \phi)^2+\frac{1}{2} m^2 \phi^2~,
\end{equation}
where $m$ is the bare mass. We will apply the Lanczos algorithm at finite temperature $\beta^{-1}$ to the field operator $\phi$ in order to find the Lanczos coefficients $b_{n}$ and K-complexity $K_{\mathcal{O}}$. We follow the notation and conventions of~\cite{Laine:2016hma}.

\subsection{The Wightman Power Spectrum $f^W(\omega)$}
\label{subsec:MassScalarMoments}

Our goal in this section is to express the Wightman power spectrum $f^W(\omega)$ in terms of the \emph{spectral function} $\rho(\omega,\mathbf{k})$ for free massive scalar theories with Lagrangian~\eqref{eq:FreeScalarLag} at finite temperature. For thermal correlators in momentum space, such as advanced and retarded correlators, it is useful to express these correlators as a function of $\rho(\omega,\mathbf{k})$ (see e.g. \cite{Stanford:2015owe}).

For a real scalar field $\phi(t,\mathbf{x})$ in $\mathbb{R}^{1,d-1}$, consider the following two-point functions with two different time orderings
\begin{align}
\Pi^{>}(t,\mathbf{x}):=&\langle\phi(t,\mathbf{x})\phi(0,\mathbf{0})\rangle_\beta~,\\
\Pi^{<}(t,\mathbf{x}):=&\langle\phi(0,\mathbf{0})\phi(t,\mathbf{x})\rangle_\beta~,
\end{align}
where $\langle\cdot\rangle_\beta$ is the thermal expectation value~\eqref{eq:WightInnProd}. Consider their Fourier transforms as well as the spectral function $\rho(\omega,\mathbf{k})$ given by
\begin{align}
\Pi^{>}(\omega,\mathbf{k}):=&\int \textrm{d}t\int \textrm{d}^{d-1}\mathbf{x}\, e^{i\omega t-i\mathbf{k}\cdot\mathbf{x}}\,\Pi^{>}(t,\mathbf{x})~,\\
\Pi^{<}(\omega,\mathbf{k}):=&\int \textrm{d}t\int \textrm{d}^{d-1}\mathbf{x}\, e^{i\omega t-i\mathbf{k}\cdot\mathbf{x}}\,\Pi^{<}(t,\mathbf{x})~,\\
\rho(\omega,\mathbf{k}):=&\frac{1}{2}[\Pi^{>}(\omega,\mathbf{k})-\Pi^{<}(\omega,\mathbf{k})]~.
\end{align}
Due to the Kubo--Martin--Schwinger (KMS) conditions, in thermal equilibrium, the above two-point functions satisfy~\cite{Laine:2016hma}
\begin{align}
\Pi^{<}(t,\mathbf{x})=&\Pi^{>}(t-i\beta,\mathbf{x})~, \;\;\; \Pi^{<}(\omega,\mathbf{k})=e^{-\beta\omega}\Pi^{>}(\omega,\mathbf{k})~,\\
\rho(\omega,\mathbf{k})=&\frac{1}{2}(1-e^{-\beta\omega})\Pi^{>}(\omega,\mathbf{k})~.\label{rosf}
\end{align}

In order to construct the K-complexity of the field operator $\phi$, we need to compute the auto-correlation function $C(t):=\Pi^{W}(t,\mathbf{0})$~\eqref{eq:AutoCorr} defined by 
\begin{equation}
\begin{split}
&\Pi^{W}(t,\mathbf{x}):=\langle\phi(t-i\beta/2,\mathbf{x})\phi(0,\mathbf{0})\rangle_\beta~, \\ 
&\Pi^{W}(\omega,\mathbf{k}):=\int \textrm{d}t\int \textrm{d}^{d-1}\mathbf{x}\, e^{i\omega t-i\mathbf{k}\cdot\mathbf{x}}\,\Pi^{W}(t,\mathbf{x})~,
\end{split}
\end{equation}
where we chose the normalization $C(0)=1$. By using $\Pi^{W}(\omega,\mathbf{k})=e^{-\beta\omega/2}\Pi^{>}(\omega,\mathbf{k})$ and (\ref{rosf}), we obtain
\begin{align}
\Pi^{W}(\omega,\mathbf{k})=\frac{1}{\sinh[\beta\omega/2]}\rho(\omega,\mathbf{k}).\label{wicf}
\end{align}
The Wightman power spectrum $f^W(\omega)$ for $C(t)=\Pi^{W}(t,\mathbf{0})$ is given by
\begin{align}
f^W(\omega):=\int \textrm{d}t \,C(t) e^{i\omega t}=\int \textrm{d}t\, \Pi^{W}(t,\mathbf{0}) e^{i\omega t}=\int \frac{\textrm{d}^{d-1}\mathbf{k}}{(2\pi)^{d-1}}\,\Pi^{W}(\omega,\mathbf{k}).
\end{align}
Using (\ref{wicf}), we thus obtain a formula for the Wightman power spectrum $f^W(\omega)$ in terms of the spectral function $\rho(\omega,\mathbf{k})$
\begin{align}
f^W(\omega)=\frac{1}{\sinh[\beta\omega/2]}\int \frac{\textrm{d}^{d-1}\mathbf{k}}{(2\pi)^{d-1}}\rho(\omega,\mathbf{k}).\label{fw}
\end{align}
The spectral function $\rho(\omega,\mathbf{k})$ of $d$-dimensional free massive scalar theories is given by~\cite{Laine:2016hma}
\begin{align}
\rho(\omega, \mathbf{k})=\frac{N}{\epsilon_k}[\delta(\omega-\epsilon_k)-\delta(\omega+\epsilon_k)],
\end{align}
where $\epsilon_k:=\sqrt{|\mathbf{k}|^2+m^2}$, and where $N$ is a normalization factor. One can compute the integral $\int \textrm{d}^{d-1}\mathbf{k} \,\rho(\omega, \mathbf{k})$ as follows
\begin{align}
\int \textrm{d}^{d-1}\mathbf{k}\, \rho(\omega, \mathbf{k})&=N\Omega_{d-2}\int_0^\infty \textrm{d}k\,\frac{k^{d-2}}{\epsilon_k}[\delta(\omega-\epsilon_k)-\delta(\omega+\epsilon_k)]\notag\\
&=N\Omega_{d-2}\int_m^\infty \textrm{d}\epsilon_k\,k^{d-3}[\delta(\omega-\epsilon_k)-\delta(\omega+\epsilon_k)]\notag\\
&=\begin{cases}\label{f2}
N\Omega_{d-2}(\omega^2-m^2)^{(d-3)/2} & \omega\ge m\\
0 &|\omega|< m\\
-N\Omega_{d-2}(\omega^2-m^2)^{(d-3)/2} & \omega\le-m
\end{cases}~,
\end{align}
where $\Omega_{d-2}$ is the surface area of the $(d-2)$-dimensional unit sphere. By using (\ref{fw}) and (\ref{f2}), we obtain $f^W(\omega)$ of $d$-dimensional free massive scalar theories at finite temperature\footnote{This expression has subtleties at $d=2,3$. In this work we consider $d\ge 4$.}
\begin{align}
f^W(\omega)=\begin{cases}
N(m,\beta,d)\,(\omega^2-m^2)^{(d-3)/2}/|\sinh (\frac{\beta \omega}{2})| & |\omega|\ge m\\
0 &|\omega|< m\label{fw2}
\end{cases},
\end{align}
where the normalization factor $N(m,\beta,d)$, which includes a contribution from $\Omega_{d-2}$, is determined by the condition
\begin{align}
\int\frac{\textrm{d}\omega}{2\pi}f^W(\omega)=1~.\label{eq:NormfW}
\end{align}
As a consistency check, let us compare (\ref{fw2}) and the CFT results in~\cite{Dymarsky:2021bjq}. The asymptotic behavior of $f^W(\omega)$ for  $|\omega|\gg m, 1/\beta$ is given by
\begin{align}
f^W(\omega)\sim 2N(m,\beta,d)|\omega|^{d-3} e^{-\frac{\beta |\omega|}{2}}~,\label{fwms}
\end{align}
which agrees with (34) in \cite{Dymarsky:2021bjq} for scaling dimensional $\Delta=d/2-1$. In the massless limit $m\rightarrow0$ with $d=4$, the Wightman power spectrum $f^W(\omega)$ is given by
\begin{align}
f^W(\omega)=\frac{\beta^2\omega}{\pi \sinh (\frac{\beta \omega}{2})}~,\label{wpsm0d4}
\end{align}
which agrees with the CFT result with $\Delta=1$ \cite{Dymarsky:2021bjq}
\begin{align}
f^W(\omega)=\int^{\infty}_{-\infty}\textrm{d}t\frac{1}{\cosh (\pi t/\beta)^2}e^{i\omega t}=\frac{\beta^2\omega}{\pi \sinh (\frac{\beta \omega}{2})}~.\label{fw4dcft}
\end{align}

Having obtained the power spectrum, we can compute the moments $\mu_{2n}$ following~\eqref{eq:MomentsSpectral}. While it is possible to obtain numerical results for arbitrary values of $\beta m$ and for different spacetime dimension $d$, we will focus on specific examples in the following sections which will allow for analytic treatment. In particular, we will focus on the large $\beta m$ regime and odd $d$ dimensions, since in this case the power spectrum and the auto-correlation function have a simpler expression. 

As mentioned in the previous section, we are interested in understanding how the Lanczos coefficients and Krylov complexity behave as a function of IR and UV cutoffs. The bare mass $m$ of the scalar field~\eqref{eq:FreeScalarLag} can be seen as an IR cutoff associated with the divergence of the zero mode $k=0$ in the massless limit. At the same time, we can introduce UV cutoffs in momentum space by modifying~\eqref{f2}. One way to do this is to include a hard UV cutoff $\Lambda$ as an integration limit in the momentum integrals, like $\int_m^\infty \textrm{d}\epsilon_k\to \int_m^\Lambda \textrm{d}\epsilon_k$. This will be the focus of Sec.~\ref{subsec:MassScalarHeavyLimitHardUVCutoff}. Another approach is to introduce a smooth UV cutoff by modifying the power spectrum by hand. This in turn will be the focus of Sec.~\ref{subsec:MassScalarHeavyLimitSmoothUVCutoff}. However, we will first analyze the case where there is no UV cutoff in Sec.~\ref{subsec:MassScalarHeavyLimitNoUVCutoff}. For numerical computations, one can choose a value of one length scale, and we set $\beta=1$ in our numerical computations.

\subsection{Krylov Complexity with Unbounded Power Spectrum}
\label{subsec:MassScalarHeavyLimitNoUVCutoff}

In the case where $\beta m \gg 1$ and for $d>4$, the normalized power spectrum~\eqref{fw2}   takes the form
\begin{equation}
    \label{eq:PowSpLargeBetaMass}
f^{W}(\omega)\approx N(m,\beta,d)\,e^{-\beta \vert \omega \vert/2}\left(\omega^{2}-m^{2}\right)^{(d-3)/2}\Theta(\vert \omega \vert - m )~,
\end{equation}
where
\begin{equation}
    \label{eq:NormPowSpLargeBetaMass}
N(m,\beta,d)=\frac{\pi^{3/2}\beta^{(d-2)/2}}{2^{d-2}\,m^{(d-2)/2}K_{\frac{d-2}{2}}\left(\frac{m\beta}{2}\right)\Gamma\left(\frac{d-1}{2}\right)}~.
\end{equation}
Here $K_{n}(z)$ is the modified Bessel function of the second kind and $\Gamma(z)$ is the usual gamma function.

For concreteness, let us focus on $d=5$ for the moment. An analogous analysis can be performed for any $d>4$. The moments $\mu_{2n}$ can be computed directly from~\eqref{eq:PowSpLargeBetaMass} via~\eqref{eq:MomentsSpectral} and are given by
\begin{equation}
\label{eq:Moments5dLargeBetaMass}
    \mu_{2n} =\frac{2^{-2} e^{\frac{m \beta}{2}} }{2+m \beta} \left( \frac{2}{\beta}\right)^{2n} \left[-m^2 \beta^2 \, \tilde{\Gamma} \left(2n+1,\frac{m \beta}{2}\right)+4 \tilde{\Gamma} \left( 2n+3,\frac{m \beta}{2}\right) \right]~,
\end{equation}
where $\tilde{\Gamma}(n,z)$ is the incomplete Gamma function.

\begin{figure}
    \centering
    \includegraphics[width=7cm]{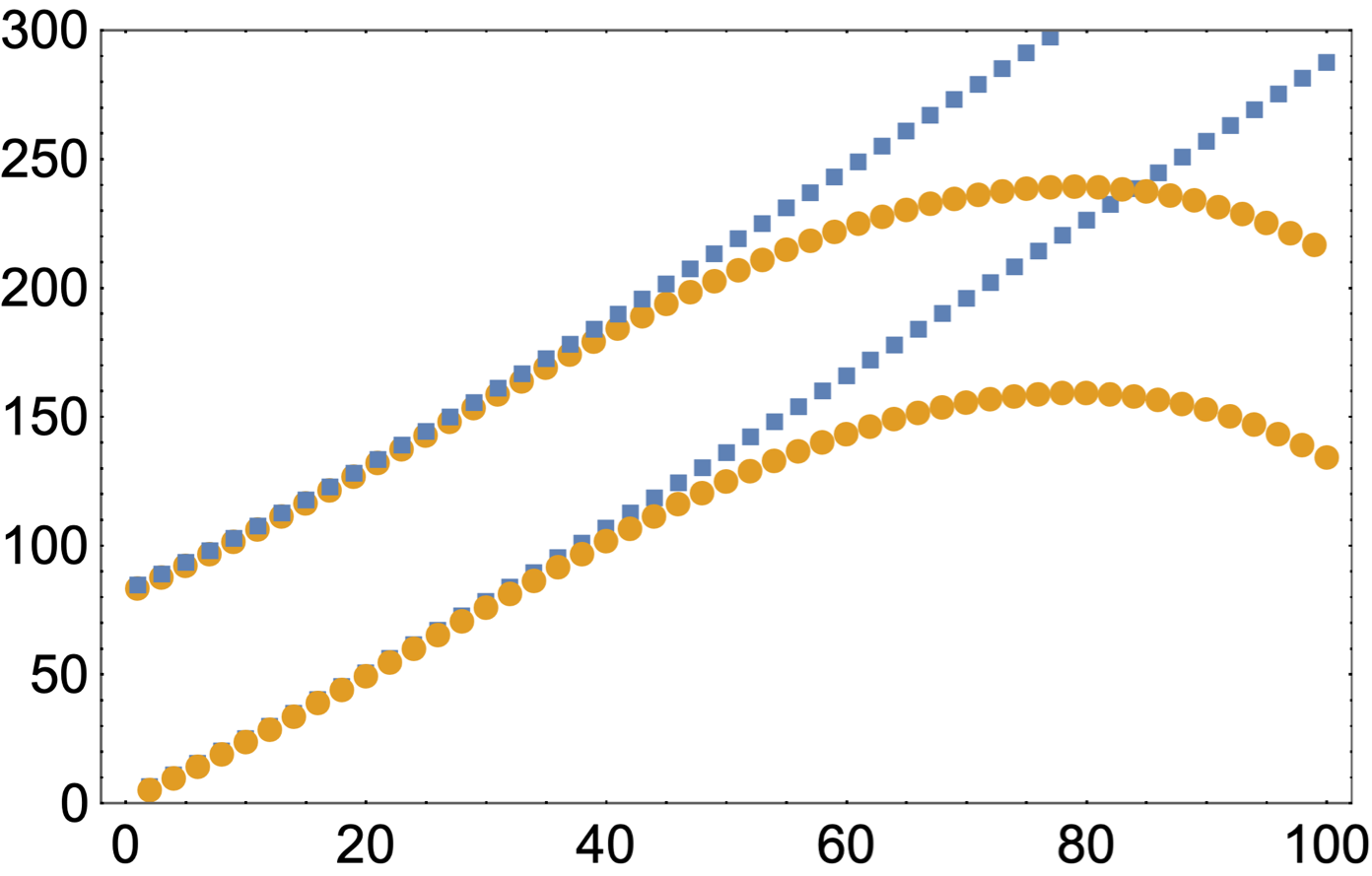}
    \put(5,0){$n$}
    \put(-200,135){$b_n$}
    \caption{Lanczos coefficients $b_{n}$ for $m \beta=80$ in $d=5$. The squares represent the results obtained numerically from~\eqref{eq:Moments5dLargeBetaMass}, while the circles represent the approximate results~Eq.~\eqref{eq-bn-largemassv2}, which are valid for $m \beta \gg n$. In both cases, the top family of $b_{n}$ corresponds to odd $n$, while the lower one corresponds to even $n$.}
    
    \label{fig:bn5dLargeBetaMassNoCutoff}
\end{figure}

Using the non-linear recursion relation~\eqref{eq:Mu2nToBnRecursion}, we can obtain the Lanczos coefficients $b_{n}$ from the moments~\eqref{eq:Moments5dLargeBetaMass}. These coefficients exhibit \emph{staggering}, which means that they can be separated into two, approximately smooth, families of $b_{n}$; one for even $n$ and one for odd $n$.  Fig.~\ref{fig:bn5dLargeBetaMassNoCutoff} shows the Lanczos coefficients obtained from~\eqref{eq:Moments5dLargeBetaMass} with arbitrarily large numerical precision as well as the approximate coefficients obtained in Appendix \ref{app:analyticbn} for $m\beta \gg n$.

Assuming that $\frac{n}{\beta m}$ is small, we show in Appendix \ref{app:analyticbn} that the Lanczos coefficients can be written as follows
\begin{equation} \label{eq-bn-largemassv2}
  \beta^{2}b_{n}^{2}= m^{2}\beta^{2}
\begin{cases}
1+4\frac{1+n}{m\beta}+8\frac{(n+1)^2}{m^{2}\beta^{2}}+12\frac{(n+1)^3}{m^{3}\beta^{3}}+\cdots\,, \,\,\text{for}\,\,n\,\,\text{odd}~,\\
4\frac{n(n+2)}{m^{2}\beta^{2}}+8\frac{n(n+1)(n+2)}{m^{3}\beta^{3}}+\cdots\,,\,\,\text{for}\,\, n\,\,\text{even}~,
\end{cases}  
\end{equation}
which shows that the staggering increases with $m \beta$ while the $b_{n}$ remain smooth for odd and even $n$\footnote{This formula is valid up to $n=35$, so higher powers of $n$ are not neglected. The results are grouped in terms of powers of $1/(\beta m)$.}. Eq.~\eqref{eq-bn-largemassv2} furthermore shows that the separation between the even and odd coefficients $\Delta b_{n}:=\vert b^{\textrm{odd}}_{n}-b^{\textrm{even}}_{n}\vert$ is of the order of the mass $O(m)$. 
Similar behavior of the Lanczos coefficients is also observed for different $d>4$.

\begin{figure}
      \centering
     \begin{subfigure}[b]{0.45\textwidth}
         \centering
         \includegraphics[width=\textwidth]{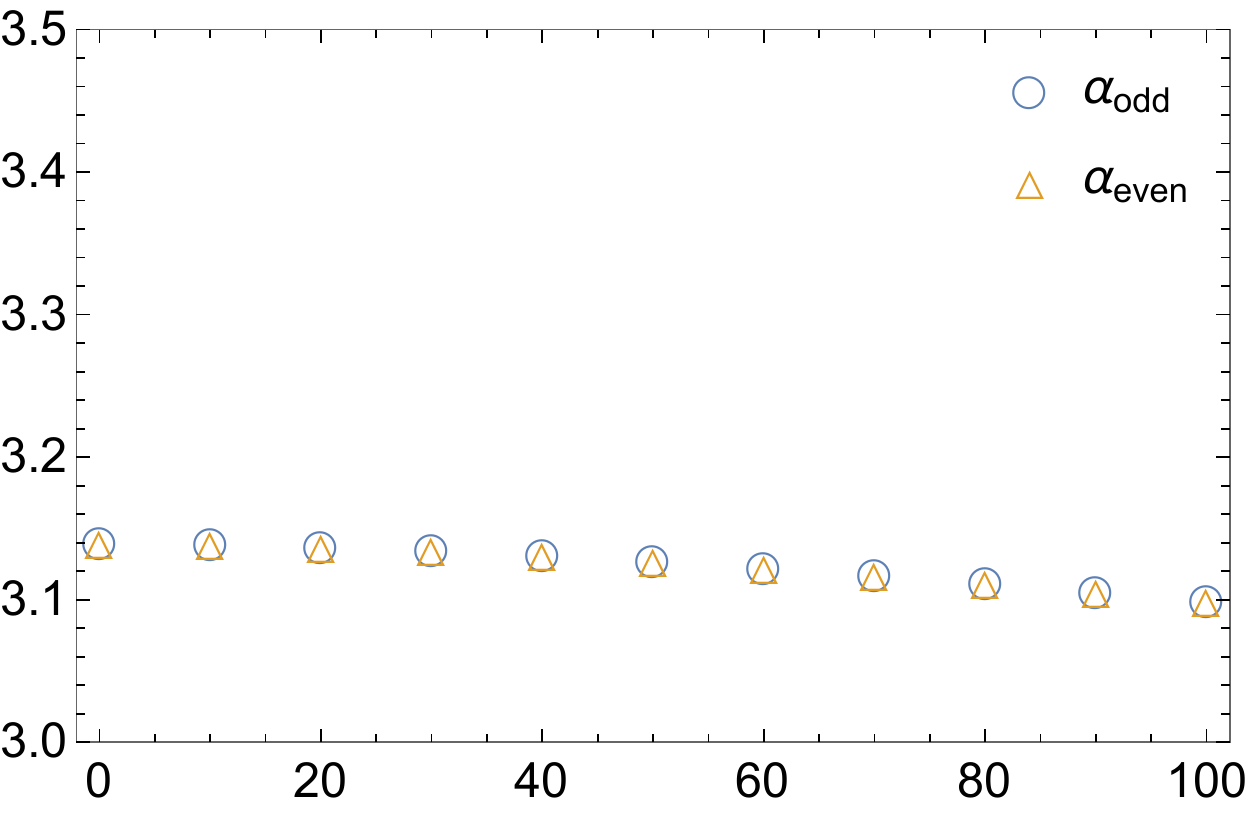}
          \put(0,0){$\beta m $}
    \put(-195,130){$\beta \alpha$}
         \caption{Mass-dependence of $\alpha_{\textrm{odd}}$ and $\alpha_{\textrm{even}}$}\label{fig:massdepconstantsbn(a)}
     \end{subfigure}
     \hfill
     \begin{subfigure}[b]{0.45\textwidth}
         \centering
         \includegraphics[width=\textwidth]{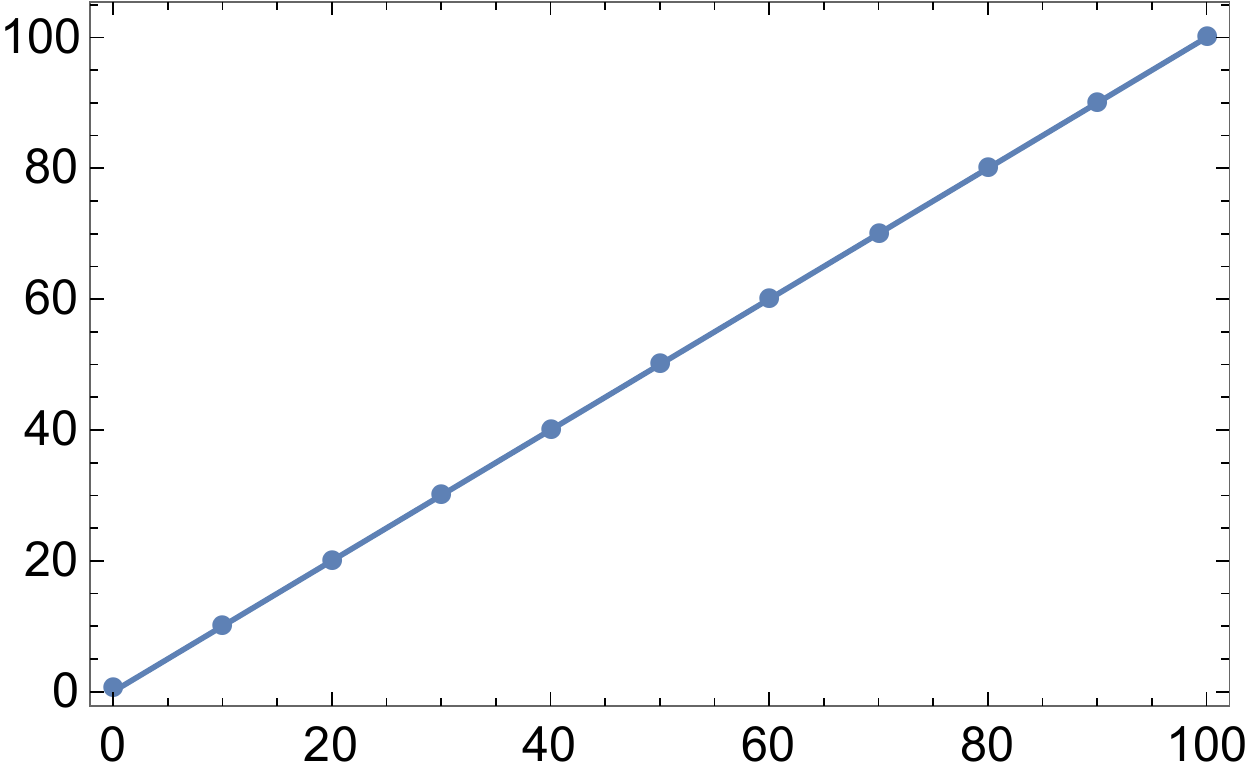}
           \put(0,0){$\beta m $}
    \put(-220,130){$\beta(\gamma_{\textrm{odd}}-\gamma_{\textrm{even}})$}
         \caption{Mass-dependence of $\gamma_{\textrm{odd}}-\gamma_{\textrm{even}}$}\label{fig:massdepconstantsbn(b)}
     \end{subfigure}
             \caption{Mass-dependence of the constants in the linear fitting (\ref{linearfittingbnd5}) for $d=5$ and $\beta=1$. We also plot a straight line $\gamma_{\textrm{odd}}-\gamma_{\textrm{even}}=m$ in the right figure to compare the mass-dependence.}
        \label{fig:massdepconstantsbn}
\end{figure}

Let us perform a linear fitting of $b_n$ for odd and even $n$ as follows
\begin{subequations}
\label{linearfittingbnd5}
\begin{equation}
b_{n}\sim\alpha_{\textrm{odd}}\,n+\gamma_{\textrm{odd}} \;\;\; (\text{odd} \; n)~,\label{linearfittingbnd5odd}
\end{equation}
\begin{equation}
b_{n}\sim\alpha_{\textrm{even}}\,n+\gamma_{\textrm{even}} \;\;\; (\text{even} \; n)~,\label{linearfittingbnd5even}
\end{equation}
\end{subequations}
where $\alpha_{\textrm{odd}}$, $\alpha_{\textrm{even}}$, $\gamma_{\textrm{odd}}$, and $\gamma_{\textrm{even}}$ are constants that do not depend on $n$.
We do the linear fitting for $n\in[151,200]$ and analyze the mass-dependence of the constants. Fig.~\ref{fig:massdepconstantsbn} shows the mass-dependence of the constants for $d=5$ and $\beta=1$. From Fig.~\ref{fig:massdepconstantsbn(a)}, one can see that $\alpha_{\textrm{odd}}\sim\alpha_{\textrm{even}}\sim\pi/\beta$ at $\beta m=0$ and they decrease slightly as $\beta m$ increases, which is consistent with the bound (\ref{eq:BoundAlphaTemp}). Moreover, Fig.~\ref{fig:massdepconstantsbn(b)} shows that the separation of $b_n$ between odd and even $n$ is $\Delta b_{n}\sim m$ even though $n$ is large.

We note that the fitting value in Fig.~\ref{fig:massdepconstantsbn} is for $n\in[151,200]$. If we do the linear fitting in a region of much larger $n$, the fitting value of $\alpha$ may change. To confirm this speculation, we also do the linear fitting of $\alpha$ for $(\beta=1, d=5, m=100)$ in a region of much larger $n$. The result is
\begin{align}
\alpha_{\textrm{odd}}\sim\alpha_{\textrm{even}}\sim&3.101\;\;\; \left(n\in[151,200]\right),\\
\alpha_{\textrm{odd}}\sim\alpha_{\textrm{even}}\sim&3.133\;\;\; \left(n\in[401,500]\right),\\
\alpha_{\textrm{odd}}\sim\alpha_{\textrm{even}}\sim&3.139\;\;\; \left(n\in[901,1000]\right).
\end{align}
The fitting value of $\alpha$ seems to approach $\pi/\beta=\pi$ as we consider a region of larger $n$, which is consistent with the analysis in \cite{Avdoshkin:2022xuw}.\footnote{We thank Anatoly Dymarsky for his comment on this point.} Our numerical result in Fig.~\ref{fig:massdepconstantsbn} suggests that mass has the effect of decreasing the fitting value of $\alpha$ in some finite $n$ regions. However, when we consider the fitting in a region of larger $n$, the mass effect on $\alpha$ seems to be negligible.

The phenomenon of staggering was also observed previously in~\cite{Dymarsky:2021bjq} where authors studied the Krylov complexity of generalized free fields with conformal dimension $\Delta$ in $d$ spacetime dimensions. For $d>4$, the Lanczos coefficients of said operators exhibited a similar splitting into two families of coefficients. However, the growth of the staggering in this case was attributed to the large scaling dimension $\Delta \gg 1$. It should also be remarked that the staggering of Lanczos coefficients in quantum many-body models was also discussed in~\cite{RecursionBook,ViswanathPhysRevB1994}. Here, it was shown that for power spectra of the form
\begin{equation}
    \label{eq:PowerSpectrumStaggering}
    f(\omega)=N(\omega_{0},\delta,\lambda)\left\vert\frac{\omega}{\omega_{0}}\right\vert^{\lambda}e^{-\left\vert\frac{\omega}{\omega_{0}}\right\vert^{\frac{2}{\delta}}}~,
\end{equation}
the Lanczos coefficients would generically exhibit staggering for $\delta=1$ as the parameter $\lambda$ is changed since in this case the moments are of the form
\begin{equation}
    \label{eq:MomentsStaggering}
    \mu_{2n}=\omega_{0}^{2n}\frac{\Gamma\left(\frac{\delta}{2}(1+\lambda+2n)\right)}{\Gamma\left(\frac{\delta}{2}(1+\lambda)\right)}~.
\end{equation}
Comparing~\eqref{eq:PowerSpectrumStaggering} with~\eqref{eq:PowSpLargeBetaMass} we see that the origin of the staggering is due both to the existence of the pole $\omega_{0}\propto 1/\beta$ as well as the dimension of the spacetime $\lambda \propto d-3$. This analysis is likewise applicable to the asymptotic case~\eqref{fwms}, which is relevant in higher-dimensional CFTs. 

In order to compute the Krylov complexity~\eqref{eq:KrylovCDef} from the power spectrum~\eqref{eq:PowSpLargeBetaMass} we first need to compute the auto-correlation function~\eqref{eq:AutoCorr}. This can be done in a straightforward way for $m\beta \gg 1$ and for odd $d>4$ by considering the Fourier transform of Eq.~\eqref{eq:PowSpLargeBetaMass}. In this case, the general form of $C(t)$ is the following.
\begin{equation}
    \label{eq:AutoCPhi0Lma}
   C^{(d)}(t)\equiv \varphi^{(d)}_{0}(t)=c^{(d)}_{1}(t)\left(c^{(d)}_{2}(t)\sin(m\,t)+c^{(d)}_{3}(t)\cos(m\,t)\right)~,
\end{equation}
and where the exact form of the functions $\lbrace c^{(d)}_{i}(t)\rbrace$ for $d=5,7,9$ can be found in Appendix~\ref{app:AutocOddd}. It is also possible to find an expression of $C(t)$ in even $d$, however as mentioned previously, its general expression is more involved.

\begin{figure}
     \centering
     \begin{subfigure}[b]{0.4\textwidth}
         \centering
         \includegraphics[width=\textwidth]{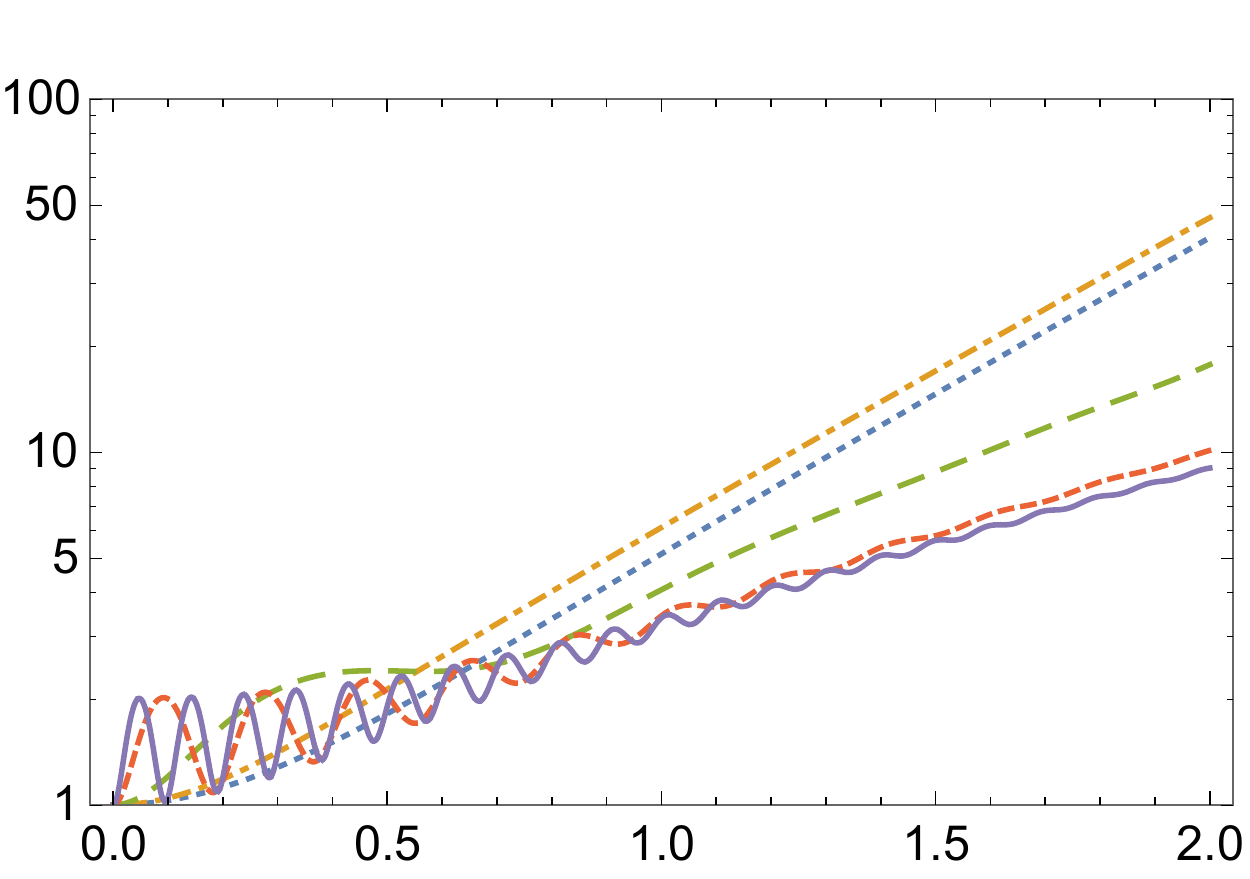}
 \put(0,0){$\frac{\pi t}{\beta}$}
    \put(-180,115){$1+K_{\mathcal{O}}(t)$}
         \caption{$d=5$}
     \end{subfigure}
     \hfill
     \begin{subfigure}[b]{0.4\textwidth}
         \centering
         \includegraphics[width=\textwidth]{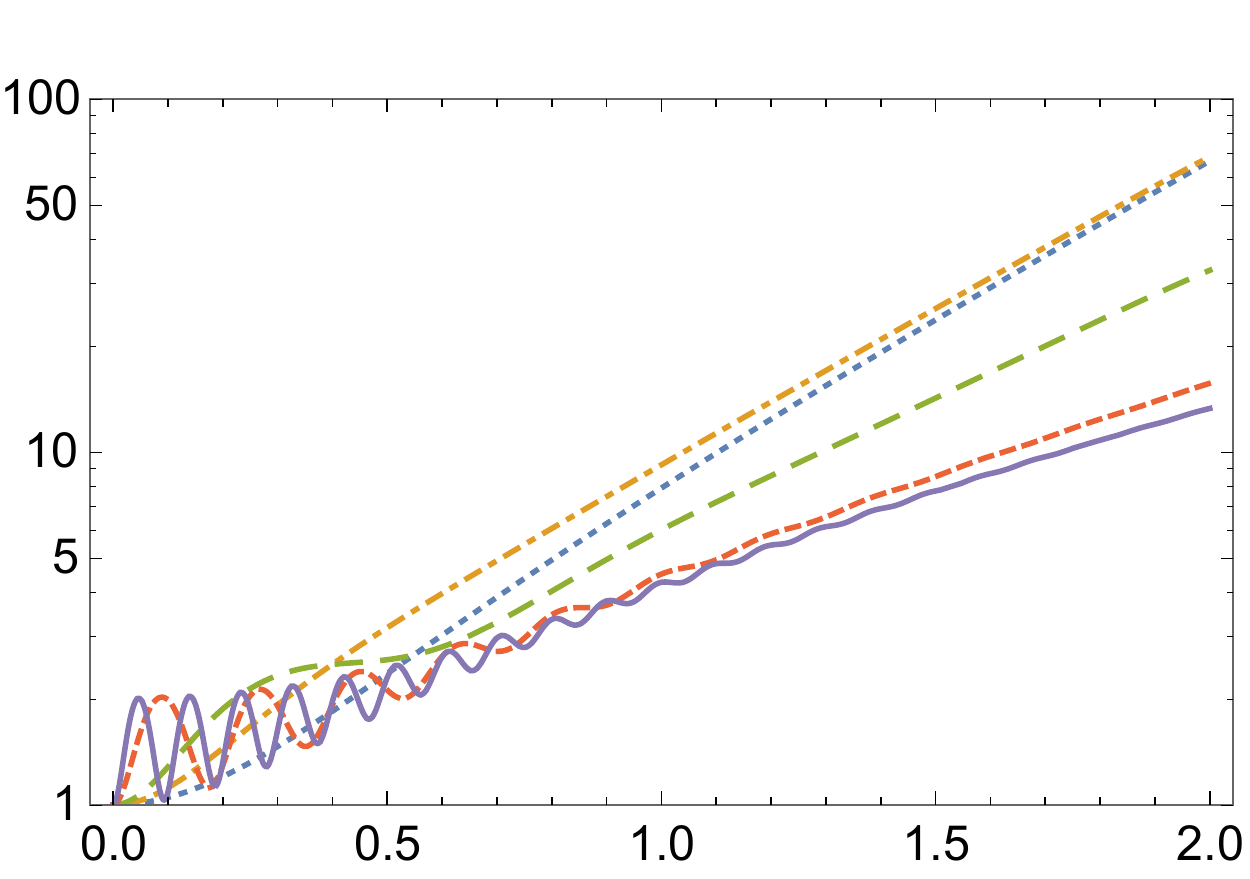}
          \put(0,0){$\frac{\pi t}{\beta}$}
    \put(-180,115){$1+K_{\mathcal{O}}(t)$}
         \caption{$d=7$}
     \end{subfigure}
     \hfill
     \begin{subfigure}[b]{0.4\textwidth}
         \centering
         \includegraphics[width=\textwidth]{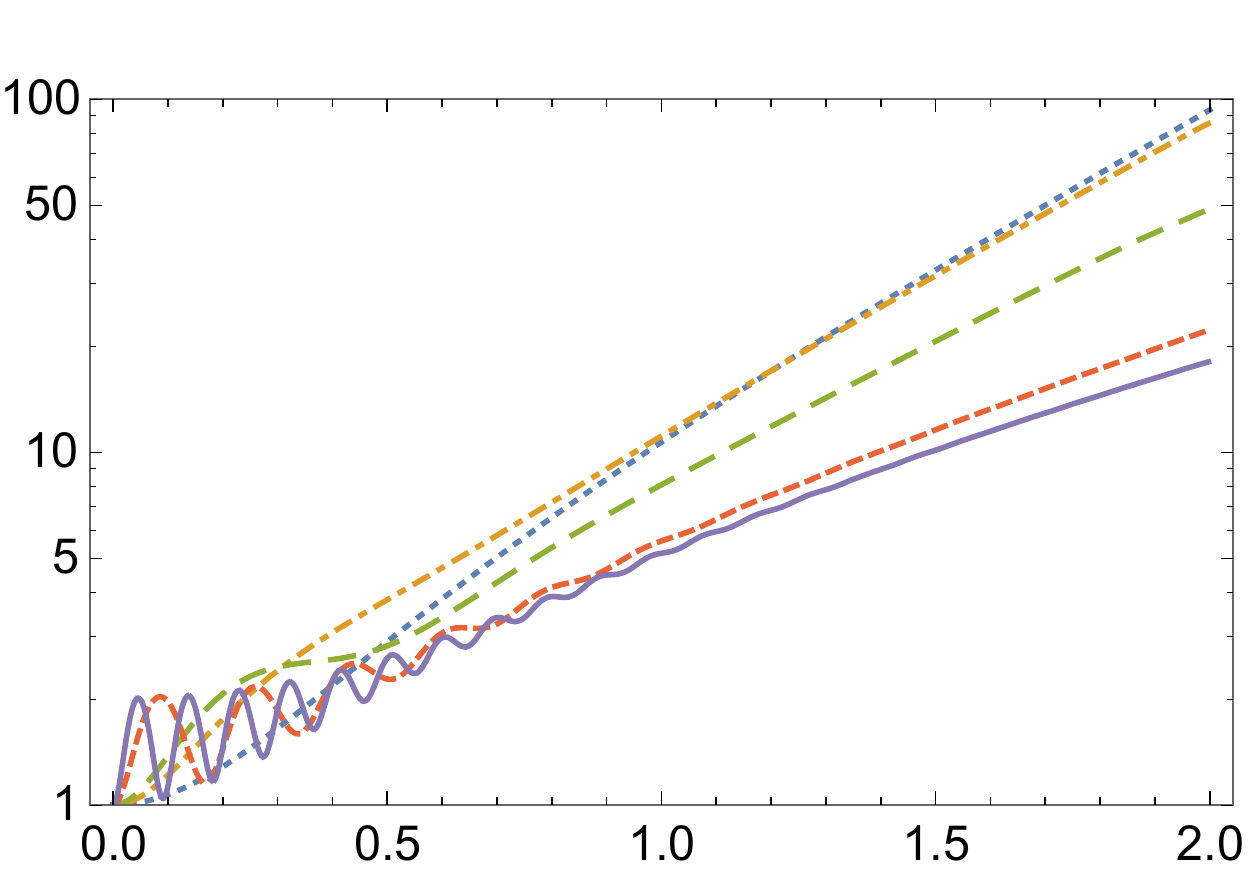}
          \put(0,0){$\frac{\pi t}{\beta}$}
    \put(-180,115){$1+K_{\mathcal{O}}(t)$}
         \caption{$d=9$}
     \end{subfigure}
       \hfill
     \begin{subfigure}[b]{0.3\textwidth}
         \centering
         \includegraphics[width=\textwidth]{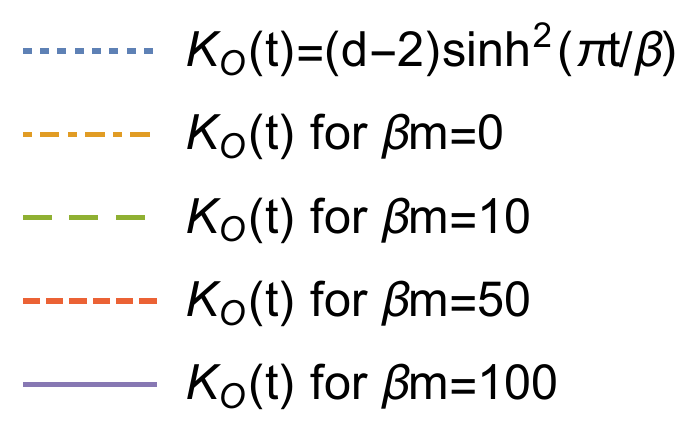}
       \hfill
     \end{subfigure}
        \caption{Krylov complexity of free scalar theories with $\beta=1$. The vertical axis is in a logarithmic scale. Linear growth in the log plot implies an exponential growth of the Krylov complexity.}
        \label{fig:KrylovComplexity}
\end{figure}

The closed-form expression~\eqref{eq:AutoCPhi0Lma} allows us to compute the probability amplitudes $\varphi_{n}^{(d)}(t)$ by solving~\eqref{eq:PhiSchroed}, where the corresponding Lanczos coefficients $b_{n}$ are computed numerically from the recursion relation~\eqref{eq:Mu2nToBnRecursion}. Figure~\ref{fig:KrylovComplexity} shows the Krylov complexity of free scalar theories with $d=5,7,9$ and for $\beta=1$. For the log plots, the vertical axis is set to $1+K_{\mathcal{O}}(t)$ for convenience. 
Following the discussion in this section, we compute the Krylov complexity with non-zero $\beta m$ in the large mass regime. On the other hand, in order to compute the Krylov complexity with $\beta m=0$, which was first examined in~\cite{Dymarsky:2021bjq}, we use the following auto-correlation function 
\begin{align}
\varphi_0(t)=\frac{\zeta(d-2,1/2+it/\beta)+\zeta(d-2,1/2-it/\beta)}{2\zeta(d-2,1/2)},
\end{align}
where $\zeta(s,a)$ is the Hurwitz zeta function. For completeness, we also plot the Krylov complexity $K_{\mathcal{O}}(t)=(d-2)\sinh^2(\pi t/\beta)$~\eqref{kotcft} of conformal $2$-point functions in CFTs on a hyperbolic space computed from Euclidean CFTs on $S^{1}\times H^{d-1}$ and compare their exponential growth behavior. A detailed analysis of the construction of K-complexity in this geometry is given in Appendix~\ref{app:KcompHyperbolic}. Since we are not able to find a closed-form expression of $\varphi_{n}$, we can only numerically compute an approximate expression for the Krylov complexity using $\sum_{n=0}^{n_{\text{max}}}n \vert \varphi_{n}(t)\vert^{2}$ for some finite $n_{\text{max}}$. For instance, we chose $n_{\text{max}}=200$ for $d=9$. We are confident that this is a very good approximation of the Krylov complexity for the range $\pi t/\beta\in [0,2]$ since we confirmed that the sum of the squares of the probability amplitudes~\eqref{eq:PhiSquare} in this case is almost equal to $1$ as $\sum_{n=0}^{n_{\text{max}}}\vert\varphi_{n}(t)\vert^{2}\sim1$ for $\pi t/\beta\in [0,2]$.

From Figure~\ref{fig:KrylovComplexity} we can distinguish several properties of the K-complexity of a free massive scalar field $\phi$ in odd-dimensions. These are as follows:
\begin{itemize}
    \item For non-zero $\beta m$, the K-complexity generically oscillates. These oscillations can be traced back to the trigonometric functions appearing in the auto-correlation function~\eqref{eq:AutoCPhi0Lma}. These oscillations are furthermore inherited to the subsequent probability amplitudes $\varphi_{n}(t)$ through the discretized Schr\"{o}dinger equation~\eqref{eq:PhiSchroed}. The larger $m\beta$, the shorter the period of oscillation $\pi/m$ (see e.g.~\eqref{eq:AutoCPhi0Lma}).
    \item As $t$ increases, the amplitude of oscillation becomes smaller. This property can be explained by the following reason: At early times, only the $\varphi_n(t)$ with small $n$, such as $\varphi_1(t)$, contribute to $K_{\mathcal{O}}(t)$. In the late-time region, $\varphi_n(t)$ with various $n$ contributes to $K_{\mathcal{O}}(t)$, and the oscillations cancel out. 
    \item In the log plots, the slope of $K_{\mathcal{O}}(t)$ for $\beta m=0$ at late times with respect to $t$ is very similar to the slope of $K_{\mathcal{O}}(t)=(d-2)\sinh^2(\pi t/\beta)$. This means that the exponential growth rate of $K_{\mathcal{O}}(t)$ at late times for $\beta m=0$ is $\lambda_K\sim2\pi/\beta$, which is consistent with the results obtained by~\cite{Dymarsky:2021bjq}.
    \item For non-zero $\beta m$, the slope of $K_{\mathcal{O}}(t)$ in the range $1.5\le \pi t/\beta \le 2.0$, which we will denote by $\tilde{\lambda}_{K}$, seems to be different from $2\pi/\beta$~\footnote{Here we use $\tilde{\lambda}_{K}$ to refer to the slope obtained for the finite range in $t$, in order to distinguish it from the asymptotic slope $\lambda_{K}$ defined at $t\rightarrow \infty$.}. This may be attributed to the fact that $b_n$ cannot be regarded as a sufficiently smooth function of $n$ due to the large staggering from non-zero $\beta m$ (see Eq.~\eqref{eq-bn-largemassv2}). We note that our numerical results for non-zero $\beta m$ \emph{do not} prove that the exponential growth rate at $t\to\infty$ \emph{is not} $2\pi/\beta$. There remains a possibility that $\lambda_K$, which is determined at $t\to\infty$, is $2\pi/\beta$.
\end{itemize}

\begin{figure}
    \centering
    \includegraphics[width=7cm]{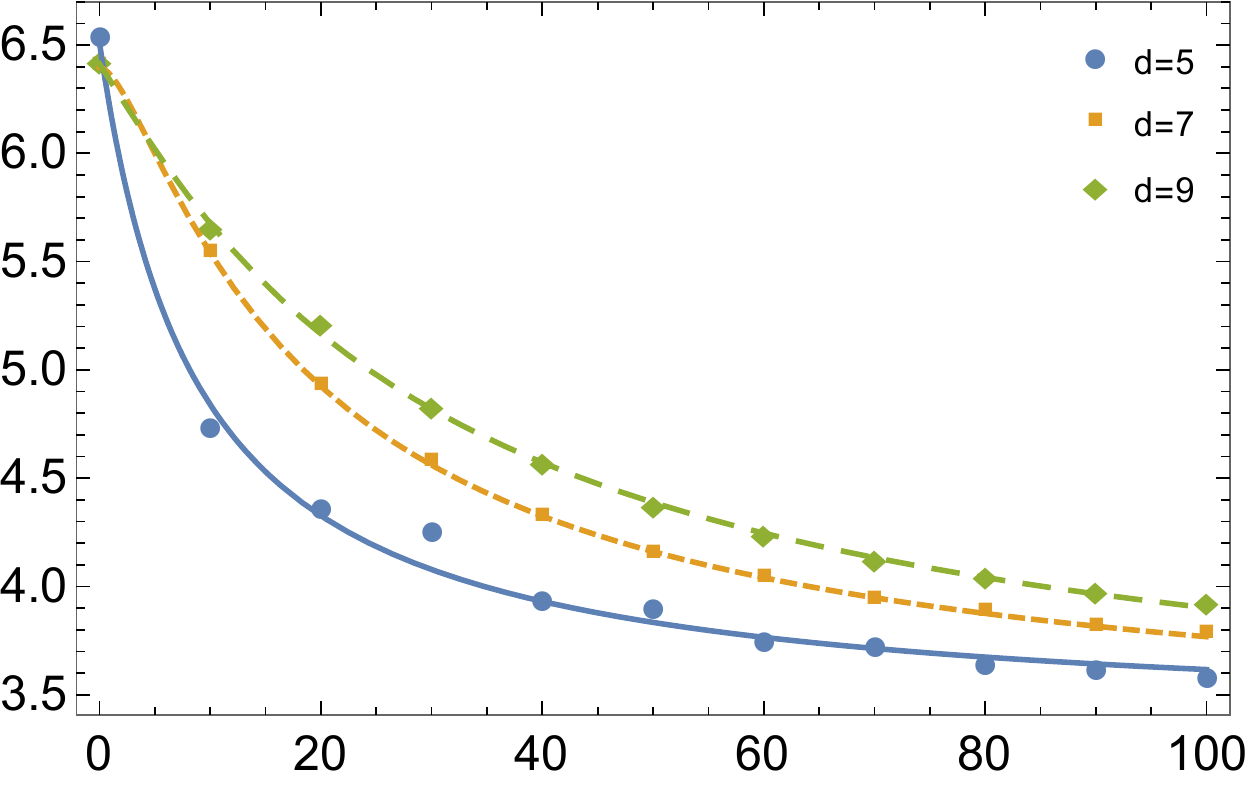}
       \put(0,0){$\beta m $}
    \put(-195,130){$\beta \tilde{\lambda}_{K}$}
    \caption{Mass-dependence of $\tilde{\lambda}_{K}$ for $\beta=1$. The linear fitting to determine $\tilde{\lambda}_{K}$ is done in the range $1.5\le \pi t/\beta \le 2.0$. We also plot the fitting curves of $\tilde{\lambda}^{(d)}_{K}$ (\ref{eq:LambdaKMass}).}
    \label{fig:MassDependenceLambdaK}
\end{figure}

In Figure~\ref{fig:MassDependenceLambdaK}, we plot the mass-dependence of $\tilde{\lambda}_{K}$ with respect to $m\beta$ for $\beta=1$. To determine $\tilde{\lambda}_{K}$, we do a linear fitting in the range $1.5\le \pi t/\beta \le 2.0$. Figure \ref{fig:MassDependenceLambdaK} shows that $\tilde{\lambda}_{K}$ with non-zero $\beta m$ differs from $2\pi/\beta$ and decreases as $m\beta$ increases, which is consistent with the conjectured bound $\lambda_K\le 2\pi/\beta$. Note that the slope with $\beta m=0$ is larger than $2\pi/\beta$ because the linear fitting is done in a finite $t$ region. In fact, the slope of $\text{Log}[K_{\mathcal{O}}(t)]$ with $K_{\mathcal{O}}(t)=(d-2)\sinh^2(\pi t/\beta)$ is given by $\frac{2\pi}{\beta}\coth(\pi t/\beta)$ which yields $2\pi/\beta$ in the limit $t\rightarrow \infty$ but which is larger than $2\pi/\beta$ when $t$ is finite and positive.

A numerical fit of slope of $\text{Log}[K_{\mathcal{O}}(t)]\approx \tilde{\lambda}_{K}\,t$ as a function of the mass $m$ for $\beta=1$, as displayed in Figure~\ref{fig:MassDependenceLambdaK}, is given by 
\begin{equation}
    \label{eq:LambdaKMass}
    \beta\tilde{\lambda}^{(d)}_{K}=k^{(d)}_{1}+\frac{k_{2}^{(d)}}{k^{(d)}_{3}+\beta m}+\frac{k_{4}^{(d)}}{\left(k^{(d)}_{3}+\beta m\right)^{2}}~,
\end{equation}
where $\lbrace k_{i}^{(d)} \rbrace$ are dimensionless constants that depend on $d$. The specific values of $\lbrace k_{i}^{(d)} \rbrace$ for Figure~\ref{fig:MassDependenceLambdaK} are given by
\begin{align}
k^{(5)}_{1}=3.36, \;\;\; k^{(5)}_{2}=0.0000110, \;\;\; k^{(5)}_{3}=8.85, \;\;\; k^{(5)}_{4}=27.9, \\
k^{(7)}_{1}=3.26, \;\;\; k^{(7)}_{2}=-268, \;\;\; k^{(7)}_{3}=9.36, \;\;\; k^{(7)}_{4}=58.0, \\
k^{(9)}_{1}=3.16, \;\;\; k^{(9)}_{2}=-607, \;\;\; k^{(9)}_{3}=18.5, \;\;\; k^{(9)}_{4}=93.0.
\end{align}
Suppose that the large $n$ behavior of the Lanczos coefficients is given by Eq.~\eqref{linearfittingbnd5}.
As discussed in the paragraph below~\eqref{eq-bn-largemassv2}, the staggering of the Lanczos coefficients is proportional to the mass
\begin{equation}
    \label{eq:StaggeringLanczosMassiveFreeQFT}
   \Delta b_{n}:=\vert b_{n}^{\textrm{odd}}-b_{n}^{\textrm{even}}\vert\equiv \vert\gamma_{\textrm{odd}}-\gamma_{\textrm{even}}\vert\propto m~,
\end{equation}
where we used the fact that $\vert\alpha_{\textrm{odd}}-\alpha_{\textrm{even}}\vert\approx\vert\alpha-\alpha\vert= 0$. Thus, we can write the slope of Log$[K_{\mathcal{O}}(t)]$ as
\begin{equation}
\begin{split}
    \label{eq:LambdaKMassV2}
\beta\tilde{\lambda}^{(d)}_{K}&=\beta(\alpha_{\textrm{odd}}+\alpha_{\textrm{even}})+k_{2}^{(d)}\left(\frac{1}{k^{(d)}_{3}+\beta\vert\gamma_{\textrm{odd}}-\gamma_{\textrm{even}}\vert}-\frac{1}{k^{(d)}_{3}}\right)+\\
    &+k_{4}^{(d)}\left(\frac{1}{\left(k^{(d)}_{3}+\beta\vert\gamma_{\textrm{odd}}-\gamma_{\textrm{even}}\vert\right)^{2}}-\frac{1}{(k^{(d)}_{3})^{2}}\right)~,
    \end{split}
\end{equation}
where we assumed that we recover the CFT behavior $\lim_{m\rightarrow 0}\tilde{\lambda}_{K}=\alpha_{\textrm{odd}}+\alpha_{\textrm{even}}= 2\alpha$ in the massless limit. This empirical formula for the growth rate of the K-complexity shows how this quantity is sensitive to the smoothness (or lack thereof) of the Lanczos coefficients. At the same time, this formula shows that the growth rate of the K-complexity is sensitive to an IR cutoff through the subleading behavior of the Lanczos coefficients. Our numerical computation with the non-zero mass implies one of the following two possibilities. (1) $\lambda_K$ at $t\to\infty$ is different from $\lambda_K=2\pi/\beta$ due to the non-zero mass. (2) $\lambda_K$ at $t\to\infty$ is not changed from $\lambda_K=2\pi/\beta$, but a time scale when the approximation $\tilde{\lambda}_K\sim2\pi/\beta$ is valid becomes later due to the non-zero mass. To understand in what way the K-complexity would be sensitive to a UV cutoff, in the next section we will introduce a hard cutoff $\Lambda$ in the momentum integrals.

\begin{figure}
     \centering
     \begin{subfigure}[b]{0.4\textwidth}
         \centering
         \includegraphics[width=\textwidth]{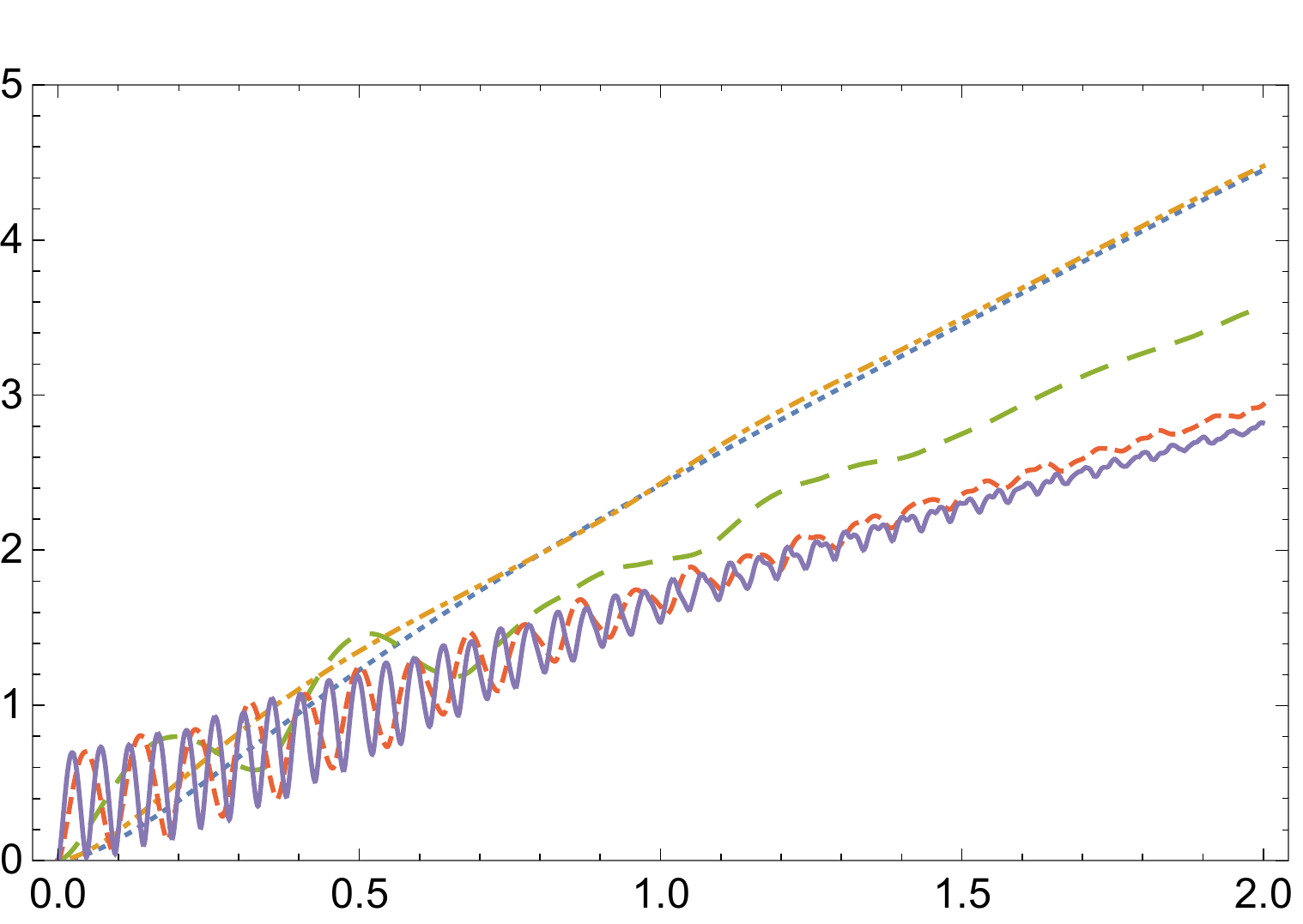}
 \put(0,5){$\frac{\pi t}{\beta}$}
    \put(-180,120){$S_K(t)$}
     \end{subfigure}
          \hfill
     \begin{subfigure}[b]{0.4\textwidth}
         \centering
         \includegraphics[width=\textwidth]{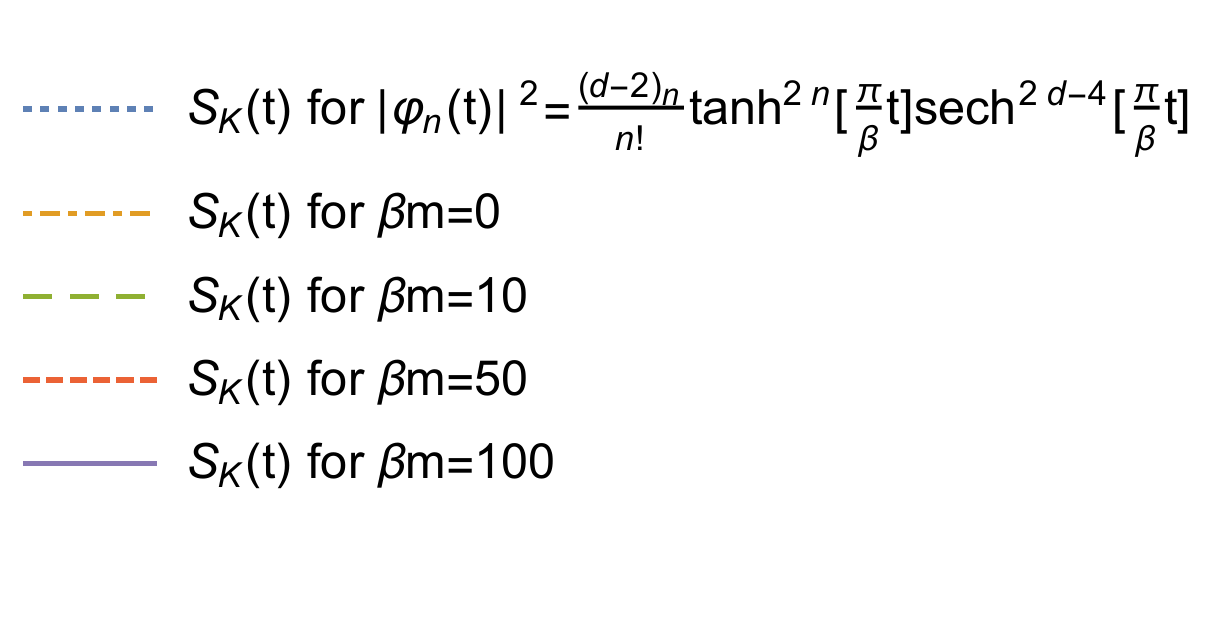}
       \hfill
     \end{subfigure}
        \caption{Krylov entropy of free scalar theories with $\beta=1$ and $d=5$.}
        \label{fig:KrylovEntropy}
\end{figure}

It is also helpful to compute Krylov entropy $S_K(t)$, defined by \cite{Barbon:2019wsy}
\begin{align}
S_K(t):=-\sum_{n=0}^\infty\vert\varphi_{n}(t)\vert^{2}\log\vert\varphi_{n}(t)\vert^{2},\label{Kentropy}
\end{align}
for confirming the decrease of slopes because of non-zero mass. In Figure~\ref{fig:KrylovEntropy}, we plot the Krylov entropy of free scalar theories with $\beta=1$ and $d=5$. For comparison, we also plot the Krylov entropy for (\ref{wtfh}) with $2\Delta=d-2$ whose $\vert\varphi_{n}(t)\vert^{2}$ is given by \cite{Parker:2018yvk}
\begin{align}
\vert\varphi_{n}(t)\vert^{2}=\frac{(d-2)_n}{n!}\tanh^{2n}\left(\frac{\pi}{\beta}t\right)\sech^{2d-4}\left(\frac{\pi}{\beta}t\right),
\end{align}
where $(d-2)_n:=(d-2)(d-1)\cdots(d+n-3)$ is the Pochhammer symbol. For non-zero $\beta m$, one can observe oscillations in $S_K(t)$ although the period is not simple because of the product of $\vert\varphi_{n}(t)\vert^{2}$ and $\log\vert\varphi_{n}(t)\vert^{2}$. Excluding the oscillations, $S_K(t)$ seems to grow linearly with respect to time, which is the expected behavior with the linear growth of $b_n$ \cite{Barbon:2019wsy}. However, because of the staggering, their slopes for non-zero $\beta m$ are smaller than $2\pi/\beta$, which is similar to the decrease of $\tilde{\lambda}_K$. If $b_n$ can be approximated by a continuous function of $n$ that increases asymptotically linearly, the Krylov entropy and the Krylov complexity satisfy $S_K(t)\sim \tilde{\eta}\log K_\mathcal{O}(t)$ at late times with $\tilde{\eta}\sim1$~\cite{Fan:2022xaa}. The staggering effect breaks the continuous property of $b_n$, and $\tilde{\eta}$ for non-zero $\beta m$ seems to shift from 1.

\subsection{Krylov Complexity for Bounded Power Spectrum with Hard UV Cutoff}
\label{subsec:MassScalarHeavyLimitHardUVCutoff}

\begin{figure}
    \centering
    \includegraphics[width=7cm]{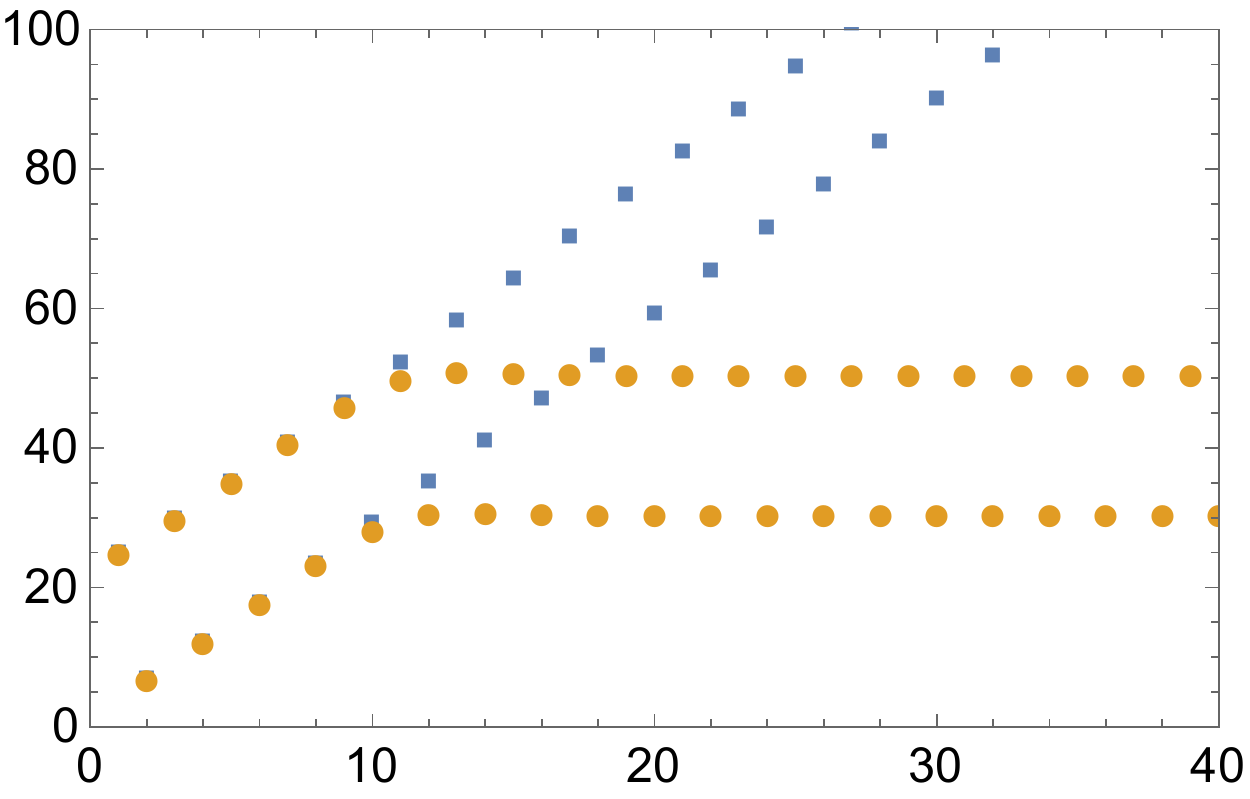}
    \put(5,0){$n$}
    \put(-200,135){$b_n$}
    \caption{Comparison of Lanczos coefficients $b_{n}$ in $d=5$ for a massive scalar field $m\beta\gg1$ for finite UV cutoff (circles) and infinite UV cutoff (squares). Here we set $\Lambda=80$, $m=20$ and $\beta=1$.}
    \label{fig:lanczoscutoff}
\end{figure}

Following~\eqref{f2}, we introduce a UV cutoff in the $\epsilon_{k}$ integral for $d=5$ as follows
\begin{equation}
    \label{eq:PowerSpectruHardUVCutoff}
    f^{W}(\omega)=\frac{N(m,\beta,\Lambda)}{ \sinh[\beta\,\omega/2]}\int_{m}^{\Lambda}\textrm{d}\epsilon_{k}\,k^{2}\,[\delta(\omega-\epsilon_{k})-\delta(\omega+\epsilon_{k})]~,
\end{equation}
where $\Lambda$ is a UV cutoff. In the limit $\Lambda > \beta m >> 1$, the power spectrum is given by
\begin{align}
f^W(\omega)=
N(m,\beta,\Lambda)\,(\omega^2-m^2)\,e^{-\frac{\beta \vert\omega\vert}{2}}\Theta\left(\vert \omega \vert - m \,,\, \Lambda- \vert \omega \vert\right) \label{fw3}~,
\end{align}
where the normalization constant in this case is given by
\begin{equation}
\label{eq:NormfWHArdCutoff}
    N(m,\beta,\Lambda)= \frac{e^{\frac{1}{2} \beta  (\Lambda +m)}\pi \, \beta ^3}{\left(2 e^{\frac{\beta  m}{2}} \left(\beta ^2 m^2-\beta  \Lambda  (\beta  \Lambda +4)-8\right)+ 8 e^{\frac{\beta  \Lambda }{2}} (\beta  m+2)\right)}~,
\end{equation}
which in the limit $\Lambda\rightarrow \infty$ reduces to~\eqref{eq:NormPowSpLargeBetaMass} for $d=5$. The moments $\mu_{2n}$ can be obtained from~\eqref{eq:MomentsSpectral} and are given as follows
\begin{equation}
\label{eq:Moments5dHardCutoff}
\begin{split}
    \mu_{2n} &=\frac{2^{2(n-1)}e^{\frac{1}{2}\beta(m+\Lambda)}}{\beta^{2n}e^{\frac{\beta\Lambda}{2}}(m\beta+2)+e^{\frac{m\beta}{2}}\left(m^{2}\beta^{2}-\beta\Lambda(\beta\Lambda+4)-8\right)}\Bigg(4\tilde{\Gamma}\left(3+2n,\frac{m\beta}{2}\right)-\\
    &-4\tilde{\Gamma}\left(3+2n,\frac{\beta\Lambda}{2}\right)+m^{2}\beta^{2}\left(\tilde{\Gamma}\left(1+2n,\frac{\beta\Lambda}{2}\right)-\tilde{\Gamma}\left(1+2n,\frac{m\beta}{2}\right)\right)\Bigg)~.
\end{split}
\end{equation}
Similarly to the normalization~\eqref{eq:NormfWHArdCutoff}, the moments~\eqref{eq:Moments5dHardCutoff} reduce to~\eqref{eq:Moments5dLargeBetaMass} in the limit $\Lambda\rightarrow\infty$. The corresponding Lanczos coefficients, computed numerically from the recursion relation~\eqref{eq:Mu2nToBnRecursion} for finite $\Lambda$ and for $\beta=1$, are shown in Fig.~\ref{fig:lanczoscutoff}.
From Figure~\ref{fig:lanczoscutoff} it can be seen that for finite UV cutoff $\Lambda$ both families of Lanczos coefficients saturate for a finite value $n_{\textrm{sat}}$. Some general observations of the behavior of the Lanczos coefficients are the following:
\begin{itemize}
    \item The saturation value of the Lanczos coefficients is given by $b_{\pm}^{\textrm{sat}}=\frac{\Lambda \pm m}{2}$, where $b_{+}^{\textrm{sat}}$ corresponds to odd $n$ and $b_{-}^{\textrm{sat}}$ for even $n$. In the massless case, both families saturate to the same value $b^{\textrm{sat}}=\Lambda/2$.
    \item The value of $n$ at which the Lanczos coefficients saturate, $n_{\textrm{sat}}$, scales linearly with $\Lambda$, but it also depends on $m$ in a subleading way. In particular, $n_{\textrm{sat}}$ decreases as $m$ is taken closer to $\Lambda$. At the same time, $n_{\textrm{sat}} \sim \mathcal{O}(1)$ for $m \sim \Lambda$~.
\end{itemize}

An explanation of why the Lanczos coefficients saturate can be obtained by analyzing the power spectrum. With the UV cutoff $\Lambda$, the power spectrum $f^W(\omega)$ (\ref{fw3}) is non-zero only if $\Lambda\ge|\omega|\ge m$. The saturation of $b_n$ for $f^W(\omega)$ with such bounded support can be explained by an exact example in \cite{RecursionBook}. Suppose that $b_n$ for odd $n$ has constant value $b_n=b_{\textrm{odd}}$ and that $b_n$ for even $n$ has constant value $b_n=b_{\textrm{even}}$.
Assuming $b_{\textrm{odd}}\ge b_{\textrm{even}}>0$, the closed form of $f^W(\omega)$ is given by \cite{RecursionBook}
\begin{align}
\label{eq:SatfW}
\begin{split}
f^W(\omega)&=
\frac{1}{ b_{\text{even}}^2}\sqrt{2(b_{\text{odd}}^2+b_{\text{even}}^2)-\omega^2-(b_{\text{odd}}^2-b_{\text{even}}^2)^2/\omega^2}\times\\
&\Theta(b_{\text{odd}}+b_{\text{even}}-|\omega|,|\omega|- (b_{\text{odd}}-b_{\text{even}}))~,
\end{split}
\end{align}
which implies a relation between the saturation of $b_n$ and the bounded support of $f^W(\omega)$.

Another explanation of this phenomenon can be obtained from the perspective of Dyck paths\footnote{A similar analysis was done in \cite{Barbon:2019wsy}, but here we also derive the coefficient $1/2$ of $b^{\textrm{sat}}=\Lambda/2$ and the saturation value with the staggering.}. First, suppose that all $b_n$ have the same value $b_n=b^{\textrm{sat}}$. By using a formula for $\mu_{2n}$ in terms of Dyck paths (see Eq. (A7) of~\cite{Parker:2018yvk}), we obtain
\begin{align}
\mu_{2n}=&(b^{\textrm{sat}})^{2n}C_n~,\\
\frac{\mu_{2n+2}}{\mu_{2n}}=&(b^{\textrm{sat}})^{2}\frac{C_{n+1}}{C_n}~,\label{rmu}
\end{align}
where $C_n:=\frac{(2n)!}{(n+1)!n!}$ is the Catalan number. Next, suppose that $b_n$ saturates to the maximum value $b^{\textrm{sat}}$ when $n\ge n_{0}$ for some $n_{0}>0$. In this case, if $n$ is large enough then~\eqref{rmu} may be approximated by
\begin{align}
\lim_{n\to\infty}\frac{\mu_{2n+2}}{\mu_{2n}}\sim(b^{\textrm{sat}})^{2}\lim_{n\to\infty}\frac{C_{n+1}}{C_n}=4(b^{\textrm{sat}})^{2}~.\label{4b2}
\end{align}
With a UV cutoff $\Lambda$, the moments $\mu_{2n}$  (\ref{eq:MomentsSpectral}) are given by
\begin{align}
    \mu_{2n}=\frac{1}{2\pi}\int_{-\Lambda}^{\Lambda}\textrm{d}\omega\,\omega^{2n}f^{W}(\omega)~.
\end{align}
Due to the contribution from $\omega^{2n}$, if $n$ is sufficiently large then the peak of the integrand $\omega^{2n}f^{W}(\omega)$ may exist at $|\omega|=\Lambda$. Thus, in the integral, the contribution from $|\omega|=\Lambda$ may be dominant, and we can estimate
\begin{align}
\lim_{n\to\infty}\frac{\mu_{2n+2}}{\mu_{2n}}\sim\frac{\omega^{2n+2}}{\omega^{2n}}\Big|_{|\omega|=\Lambda}=\Lambda^2~.\label{l2}
\end{align}
Combining (\ref{4b2}) and (\ref{l2}), we conclude that
\begin{align}
b^{\textrm{sat}}\sim\Lambda/2~,\label{bsatl/2}
\end{align}
which provides a naive explanation of why $b_n$ saturates to $\Lambda/2$ in the massless case. 
Finally, suppose that $b_n$ saturates to $b^{\textrm{sat}}_{\text{odd}}$ for odd $n$ and to $b^{\textrm{sat}}_{\text{even}}$ for even $n$. From the viewpoint of Dyck paths, $4(b^{\textrm{sat}})^{2}$ in~\eqref{4b2} may be modified to 
\begin{align}
4(b^{\textrm{sat}})^{2}\to (b^{\textrm{sat}}_{\text{odd}})^2+(b^{\textrm{sat}}_{\text{even}})^2+b^{\textrm{sat}}_{\text{odd}}b^{\textrm{sat}}_{\text{even}}+b^{\textrm{sat}}_{\text{even}}b^{\textrm{sat}}_{\text{odd}}=(b^{\textrm{sat}}_{\text{odd}}+b^{\textrm{sat}}_{\text{even}})^2~,
\end{align}
and (\ref{bsatl/2}) may be modified to
\begin{align}
\frac{b^{\textrm{sat}}_{\text{odd}}+b^{\textrm{sat}}_{\text{even}}}{2}\sim\frac{\Lambda}{2},
\end{align}
which is consistent with our observation in the massive case. However, in this estimation we cannot determine the value of $(b^{\textrm{sat}}_{\text{odd}}-b^{\textrm{sat}}_{\text{even}})/2$.

\begin{figure}
      \centering
    \includegraphics[width=7cm]{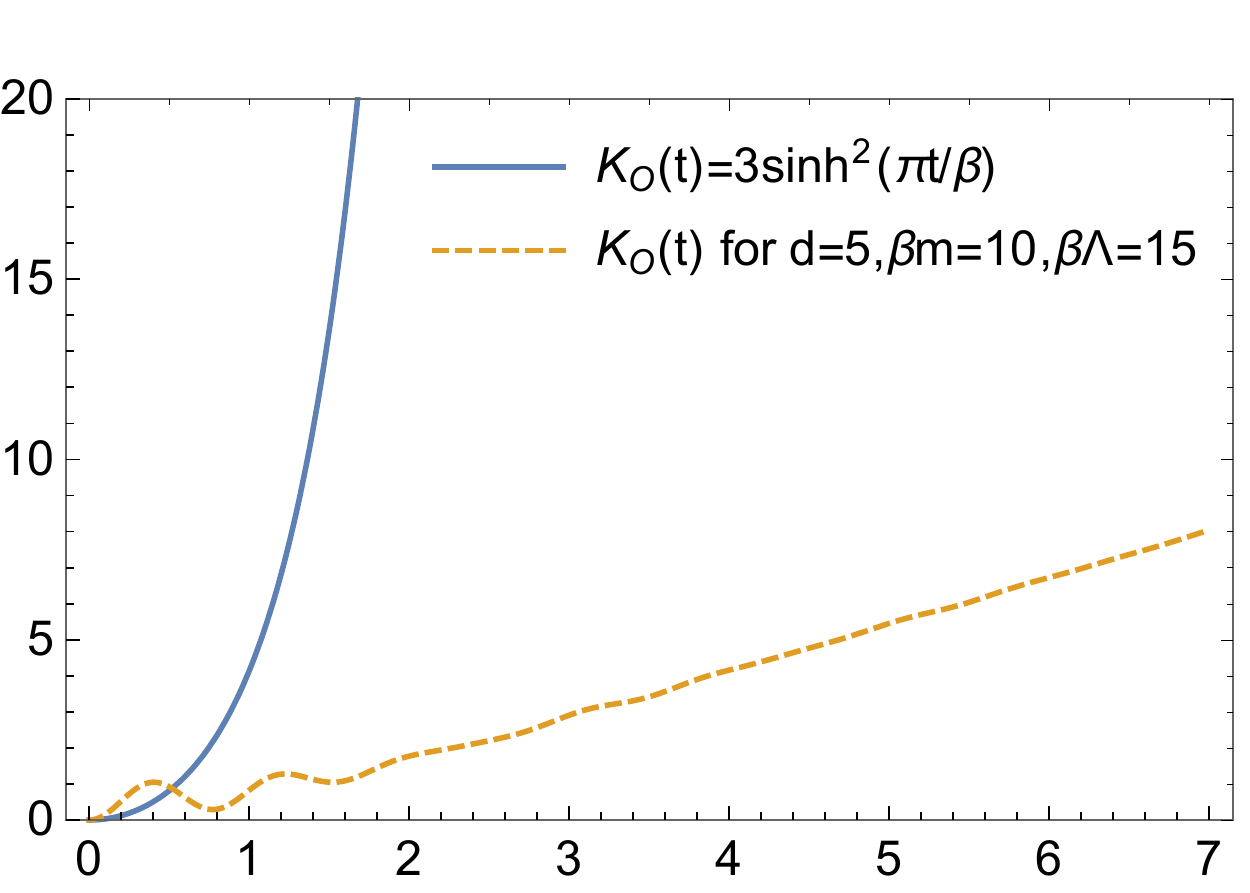}
    \put(0,0){$\pi t/\beta$}
    \put(-205,135){$K_{\mathcal{O}}(t)$}
    \caption{Linear growth of $K_{\mathcal{O}}(t)$ $(\beta=1)$ with $d=5, \beta m=10, \beta \Lambda=15$.}
    \label{fig:KCUVcutoff}
\end{figure}

\begin{figure}
      \centering
    \includegraphics[width=7cm]{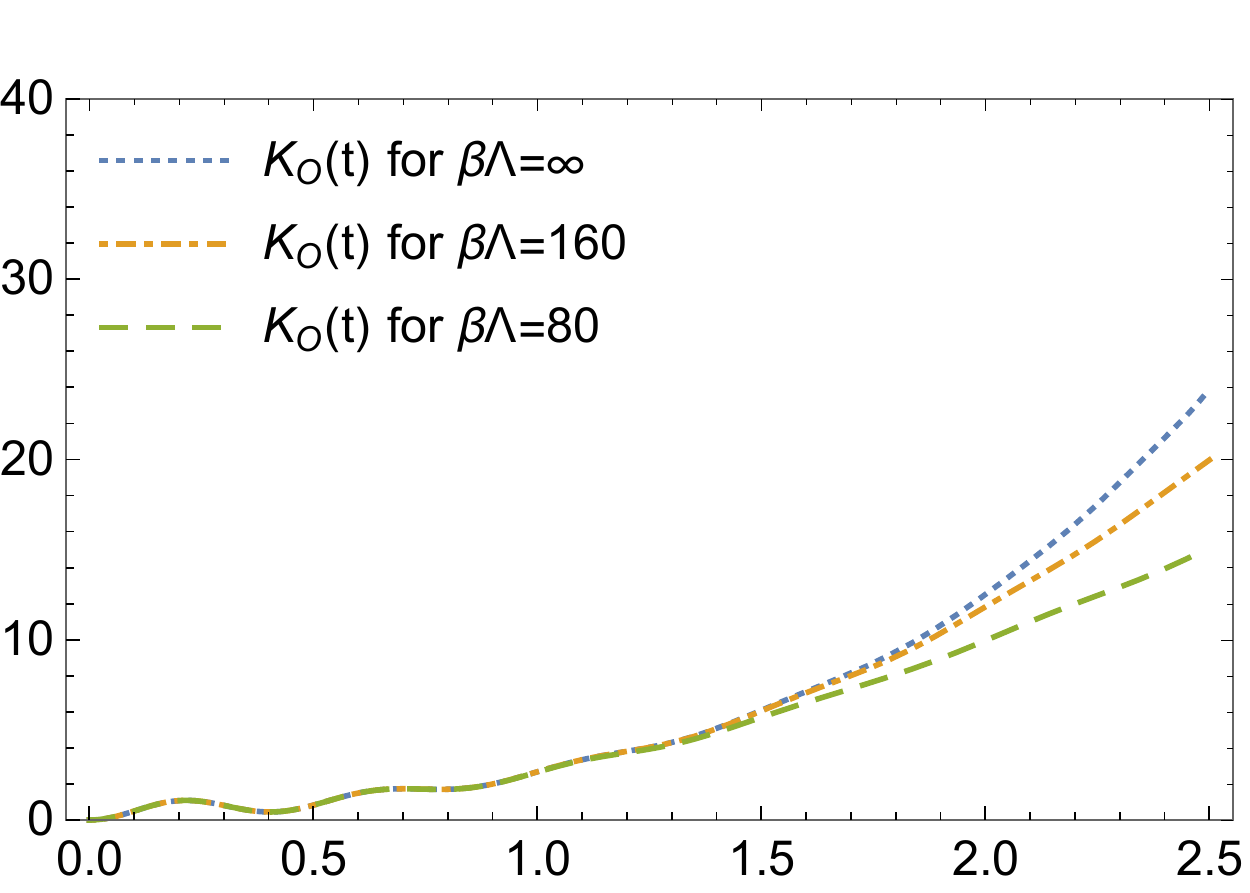}
    \put(0,0){$\pi t/\beta$}
    \put(-205,135){$K_{\mathcal{O}}(t)$}
    \caption{Growth behaviors of Krylov complexity $K_{\mathcal{O}}(t)$ $(\beta=1, d=5, \beta m=20)$ with different UV cutoff $\Lambda$.}
    \label{fig:KCUVcutoff2}
\end{figure}

Whenever the Lanczos coefficient $b_n$ saturates, the Krylov complexity $K_{\mathcal{O}}(t)$ is expected to increase linearly rather than exponentially~\cite{Parker:2018yvk,Barbon:2019wsy}. With the large mass approximation $2\sinh (\frac{\beta m}{2})\sim\exp(\frac{\beta m}{2})$, the auto-correlation function $\varphi_0(t)$ in the presence of a UV cutoff $\Lambda$ for $d=5$ can be computed from~\eqref{fw3} and is given by
\begin{align}
\varphi^{d=5}_0(t)=&\frac{\beta ^3}{\left(\beta ^2+4 t^2\right)^3 \left(e^{\frac{\beta  m}{2}} \left(-\beta 
   \Lambda  (\beta  \Lambda +4)+\beta ^2 m^2-8\right)+4 e^{\frac{\beta  \Lambda }{2}}
   (\beta  m+2)\right)}\notag\\
   \times& \Big( 2 te^{\frac{\beta  m}{2}} \sin (\Lambda  t) (\beta ^2
   \left(\beta  \Lambda  (\beta  \Lambda +8)-\beta ^2 m^2+24\right)+16 t^4
   \left(\Lambda ^2-m^2\right)\notag\\
   &+8 t^2 \left(\beta  \Lambda  (\beta  \Lambda +4)-\beta ^2
   m^2-4\right))\notag\\
   &+e^{\frac{\beta  m}{2}}\cos (\Lambda  t) (\beta ^3 \left(-\beta 
   \Lambda  (\beta  \Lambda +4)+\beta ^2 m^2-8\right)+16 t^4 \left(\Lambda  (4-\beta 
   \Lambda )+\beta  m^2\right)\notag\\
   &+8 \beta  t^2 \left(\beta ^2 (m-\Lambda ) (\Lambda
   +m)+12\right))\notag\\
   &-16 t e^{\frac{\beta  \Lambda }{2}} \left(\beta ^2 (\beta 
   m+3)+4 t^2 (\beta  m-1)\right) \sin (m t)\notag\\
   &+4 e^{\frac{\beta  \Lambda }{2}} \left(2
   \beta ^3+m \left(\beta ^4-16 t^4\right)-24 \beta  t^2\right) \cos (m
   t)\Big)~,\label{phi0lmauvc}
   \end{align}
which reduces to~\eqref{eq:AutoCPhi0Lma} with $d=5$ in the limit $\Lambda\to\infty$. Starting from~\eqref{phi0lmauvc}, one can iteratively solve~\eqref{eq:PhiSchroed} in order to find the subsequent probability amplitudes $\varphi^{d=5}_{n}$. Figure~\ref{fig:KCUVcutoff}, which is not a log plot, shows the time evolution of $K_{\mathcal{O}}(t)$ $(\beta=1)$ with $d=5, \beta m=10$ and $ \beta \Lambda=15$. Compared to the exponential growth of $K_{\mathcal{O}}(t)=3\sinh^2(\pi t/\beta)$, we can see the liner growth of $K_{\mathcal{O}}(t)$ with finite $\beta \Lambda$. We plot $K_{\mathcal{O}}(t)$ $(\beta=1, d=5, \beta m=20)$ for different values of UV cutoff $\Lambda$ in Figure~\ref{fig:KCUVcutoff2}. At early times, each plot is identical since $b_n$ at small $n$ is the same as seen in Figure~\ref{fig:lanczoscutoff}. At late times, growth behaviors of $K_{\mathcal{O}}(t)$ with finite $\Lambda$ are different from exponential growth for infinite $\Lambda$ due to the saturation of $b_n$.

\begin{figure}
      \centering
    \includegraphics[width=7cm]{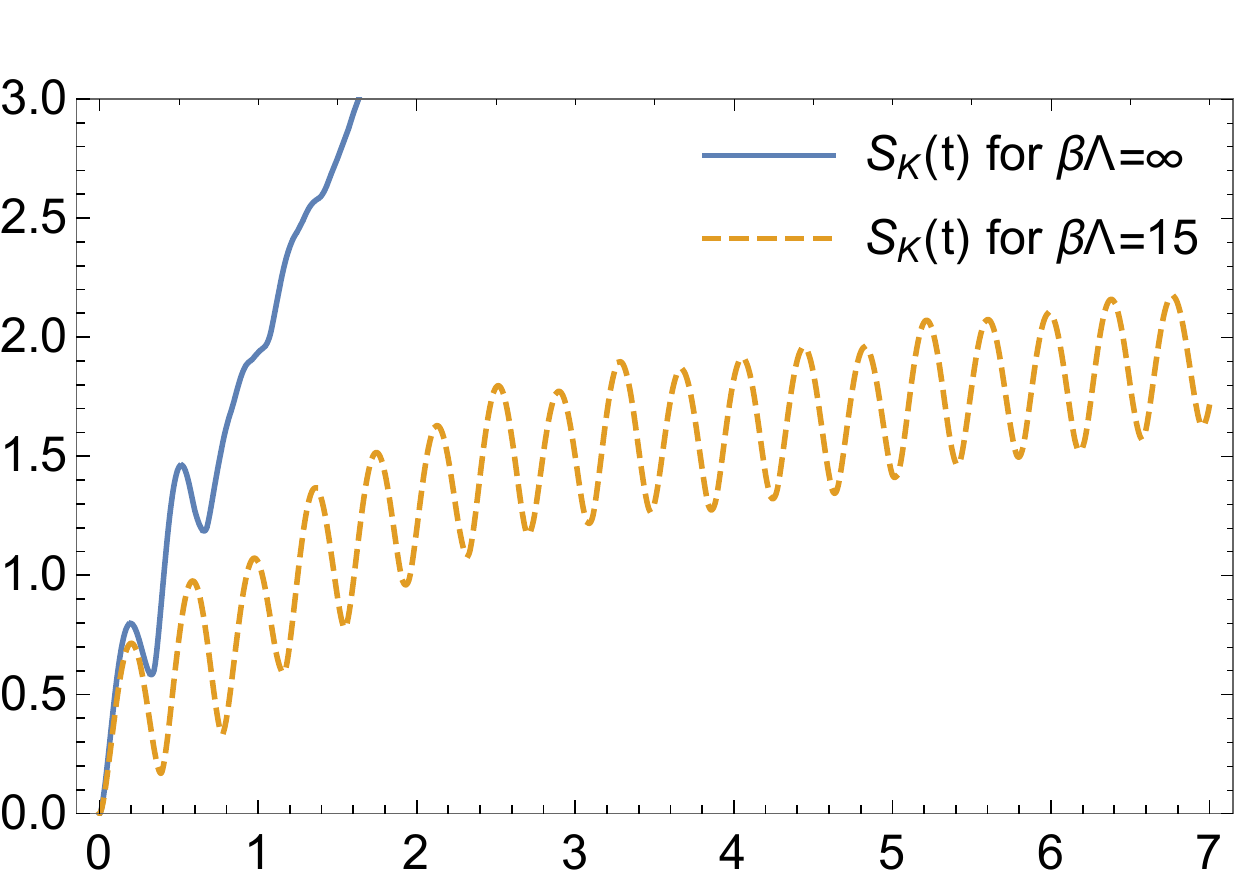}
    \put(0,0){$\pi t/\beta$}
    \put(-205,135){$S_{K}(t)$}
    \caption{Krylov entropy $S_{K}(t)$ $(\beta=1, d=5, \beta m=10)$ with infinite and finite $\beta\Lambda$.}
    \label{fig:KEUVcutoff}
\end{figure}

We can also calculate the Krylov entropy (\ref{Kentropy}) for $\varphi^{d=5}_{n}$. When $b_n$ saturates, the expected behavior of $S_K(t)$ is logarithmic growth \cite{Barbon:2019wsy}. Figure~\ref{fig:KEUVcutoff} shows the time evolution of $S_{K}(t)$ $(\beta=1)$ with $d=5, \beta m=10$ and $ \beta \Lambda=15$. In this figure, the Krylov entropy with finite $\beta \Lambda$ seems to grow logarithmically with oscillations due to non-zero $\beta m$.

It is interesting to compare the behavior of the Lanczos coefficients (Fig.~\ref{fig:lanczoscutoff}) and the Krylov complexity (Fig.~\ref{fig:KCUVcutoff}) with the situation in free and chaotic quantum many-body systems with a finite number of degrees of freedom $S$. In~\cite{Parker:2018yvk,Barbon:2019wsy,Rabinovici:2020ryf,Rabinovici:2022beu} it was shown that in fast-scrambling systems with finite $S$, the K-complexity $K_{\mathcal{O}}$ grows exponentially until the scrambling time $t_{s}\sim \log(S)$, at which point $K_{\mathcal{O}}$ reaches a value of order $O(S)$. It then switches to a linear growth until a time of order $\exp(O(S))$ when it settles around a plateau. In the case of the Lanczos coefficients, these have a linear growth until $n$ becomes of the order $O(S)$, at which point they saturate in $n$ according to $b_{n}\sim \Delta E \,S$, where $\Delta E$ is called the spectral bandwidth, which is of the order of $O(1/a)$ for a lattice spacing of size $a$. 

In the language of Section~\ref{subsec:LanczosChaos}, this means that for $n\sim O(S/a\alpha)$, $b_{n}$ saturate to a value of the order of $\omega_{F}\sim O(S/a)$. When the number of degrees of freedom $S$ is infinite, the K-complexity does not appear to transition from an exponential growth to a linear growth at finite $t$ because the scrambling time $t_{s}$ is not well-defined in this case, like $t_{s}\rightarrow \infty$. Nevertheless, introducing a cutoff for finite $\omega_{F}$ in the continuum theory, such as the $N$-scaling limit in \cite{Kar:2021nbm}, leads to a plateau similar to that in finite-sized chaotic quantum many-body systems, although the plateau in the $N$-scaling limit is infinite. However, the situation is more closely related to the behavior of free lattice theories with finite lattice spacing $a$, for which the Lanczos coefficients also saturate to a value $b_{n}\sim\omega_{F}\sim O(1/a)$, leading to a linear growth of the K-complexity. Introducing a hard UV cutoff in the continuous momentum integrals~\eqref{f2} does not alter the fact that the continuum theory on a non-compact space has an infinite number of degrees of freedom $S\rightarrow \infty$. As a consequence, we are tempted to associate the continuous free scalar theory~\eqref{eq:FreeScalarLag} with finite momentum hard-cutoff $\Lambda$ instead with the large $N$ limit of a free lattice theory with finite lattice spacing $a\sim 1/\Lambda$, where $N$ is the number of lattice sites. In fact, the saturation behavior of the Lanczos coefficients in our computations with the momentum hard-cutoff is very similar to the saturation behavior in a discretized free lattice model~\cite{DymarskyTalk, Avdoshkin:2022xuw}. Thus, the saturation of the Lanczos coefficients displayed in Fig.~\ref{fig:lanczoscutoff} and the linear growth of the K-complexity in Fig.~\ref{fig:KCUVcutoff} can be understood as a reflection of the underlying coarse-grained free lattice theory with finite lattice spacing describing the same IR physics.

Let us comment on a correction caused by finite $N$. Consider free bosons on a $1d$ periodic lattice on $S^1$ of length $aN$. In the large-$N$ limit $N\to\infty$, the Lanczos coefficients of this lattice model may be similar to the ones of a free scalar QFT on non-compact space $R^1$ with a finite UV cutoff because $aN$ diverges\footnote{We thank Anatoly Dymarsky for his comment on the large-$N$ limit.}. When $aN$ is of the same order as other scales such as $\beta$, the correction due to the compactness of $S^1$ comes into play. In particular, our plots in Fig.~\ref{fig:lanczoscutoff} are qualitatively similar to plots for $N/\beta=1$ in Fig.~6 of~\cite{Avdoshkin:2022xuw} but differ significantly  from plots with $N/\beta=0.5$\footnote{For our plots in Fig.~\ref{fig:lanczoscutoff}, $\beta=1$ is fixed and the value of $\Lambda$ is changed. In Fig.~6 of~\cite{Avdoshkin:2022xuw}, the lattice spacing $a=2$ is fixed and the values of other parameters are changed.}.

\subsection{Lanczos Coefficients with a Smooth UV Cutoff}
\label{subsec:MassScalarHeavyLimitSmoothUVCutoff}

Another way of introducing a UV cutoff consists in modifying the power spectrum by hand. Given that the exponential decay of the power spectrum $f^{W}(\omega)\sim e^{-\beta\omega/2}$ for $\omega \rightarrow \infty$ is expected for a large class of QFTs at finite temperature $1/\beta$, the only way to significantly modify the growth behavior of the Lanczos coefficients and therefore of the K-complexity is to modify the features of such an exponential decay. As mentioned previously, Dymarsky and his collaborators have pointed out that the behavior of K-complexity after the UV cutoff scale can be used as a probe of quantum chaos~\cite{DymarskyTalk, Avdoshkin:2022xuw}. If we introduce a hard cutoff, the Lanczos coefficients saturate, and we cannot obtain the growth behavior $b_n\sim n^\delta$ for integrable lattice models, where $0<\delta<1$. In this section, we briefly analyze how to reproduce such behavior. We focus on the massless case $m\rightarrow 0$ in $d=4$. The motivation for focusing on this particular case is that we do not expect there to be any staggering of the Lanczos coefficients that may arise from the mass or the spacetime dimension. Our starting point is thus the Wightman power spectrum~\eqref{wpsm0d4}. We introduce a smooth cutoff by hand as follows
\begin{align}
f^W(\omega)=N(\beta,\Lambda,\delta)\frac{\omega}{\sinh (\frac{\beta \omega}{2})}\exp (-|\omega/\Lambda|^{1/\delta})~,\label{wpssc}
\end{align}
where $N(\beta,\Lambda,\delta)$ is a normalization factor determined by~\eqref{eq:NormfW}.
The original Wightman power spectrum~\eqref{wpsm0d4} corresponds to $\delta\to\infty$, and the hard cutoff limit corresponds to $\delta\to0$.

\begin{figure}
     \centering
    \includegraphics[width=7cm]{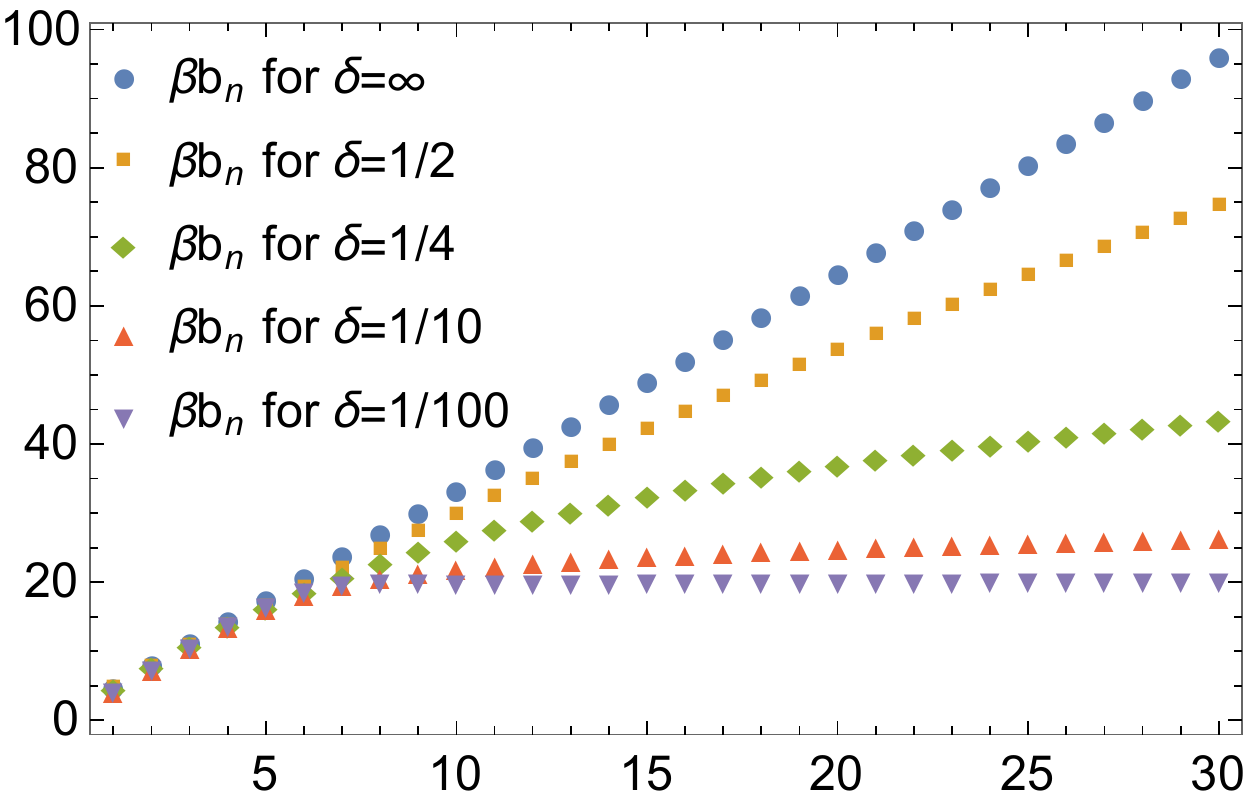}
      \put(5,0){$n$}
    \put(-205,135){$\beta b_n$}
    \caption{Lanczos coefficients $b_n$ computed from (\ref{wpssc}) with $\beta=1$ and $\beta\Lambda=40$. We plot $b_n$ for various values of $\delta$, which is a parameter of the smooth cutoff.}
    \label{fig:LCSC}
\end{figure}

Figure~\ref{fig:LCSC} shows the Lanczos coefficients $b_n$ computed from (\ref{wpssc}) with $\beta=1$ and $\beta\Lambda=40$. One can see that the asymptotic behavior of the Lanczos coefficients $b_n$ changes as we vary the value of $\delta$. Figure~\ref{fig:LCSC} thus gives an explicit numerical check of the connection between the growth rate of $b_n$ and the exponential decay of $f^W(\omega)$~\cite{RecursionBook, Parker:2018yvk}. 

While we are unaware of any physical situation where one should expect this exponential correction to the power spectrum, it is worth pointing out that Eq.~\eqref{wpssc} provides an expectation on how to deform QFTs to encode information about the UV structure of lattice models, e.g. via their lattice spacing $a\sim 1/\Lambda$.

\section{Interacting Massive Scalar in $4$-dimensions}
\label{sec:InterMassScalar}

In this section, we consider a real-valued scalar field $\phi$ with Euclidean Lagrangian of the form
\begin{equation} \label{eq:Lag}
    \mathcal{L}_E = \frac{1}{2} (\partial \phi)^2+\frac{1}{2} m^2 \phi^2+\frac{g}{\ell!}\phi^{\ell}~.
\end{equation}
We would like to understand how the interaction term $\frac{g}{\ell!}\phi^{\ell}$ affects the Lanczos coefficients associated to the auto-correlation $C(t)= \langle \phi(t-i \beta/2,{\bf 0}) \phi(0,{\bf 0}) \rangle_{\beta}$. For simplicity, we set $d=4$ and consider the cases in which the interaction term is a relevant ($\ell=3$) and a marginally irrelevant ($\ell=4$) deformation of the free theory.

\subsection{Marginally irrelevant deformation ($\ell=4$)}
\label{subsec:MargIrrelDefo}

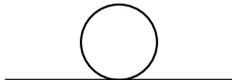
\begin{figure}[h!]
    \centering
    \begin{tikzpicture}[scale=1.]

\draw [thick]  (0,0) -- (3,0);
\draw [thick]  (1.5,0.5) circle (.5cm);
\end{tikzpicture}
    \caption{Diagram relevant to the computation of the one-loop self-energy $\Pi_E$ in the theory with quartic interactions.}
    \label{fig:selfenergy1}
\end{figure}

In this case, the Euclidean one-loop self-energy comes from the diagram of Fig.~\ref{fig:selfenergy1}, which gives
\begin{equation} \label{eq-Pi1}
    \Pi_E =  \frac{g\, T}{2} \sum_{p_0} \int \frac{\textrm{d}^3 {\bf p}}{(2\pi)^3} \frac{1}{p_0^2+{\bf p}^2+m^2}~,
\end{equation}
where the Matsubara frequencies $p_0 = 2 \pi T\, n$ are summed over all integers $n$. 

\begin{figure}[h!]
    \centering
    \includegraphics[width=6cm]{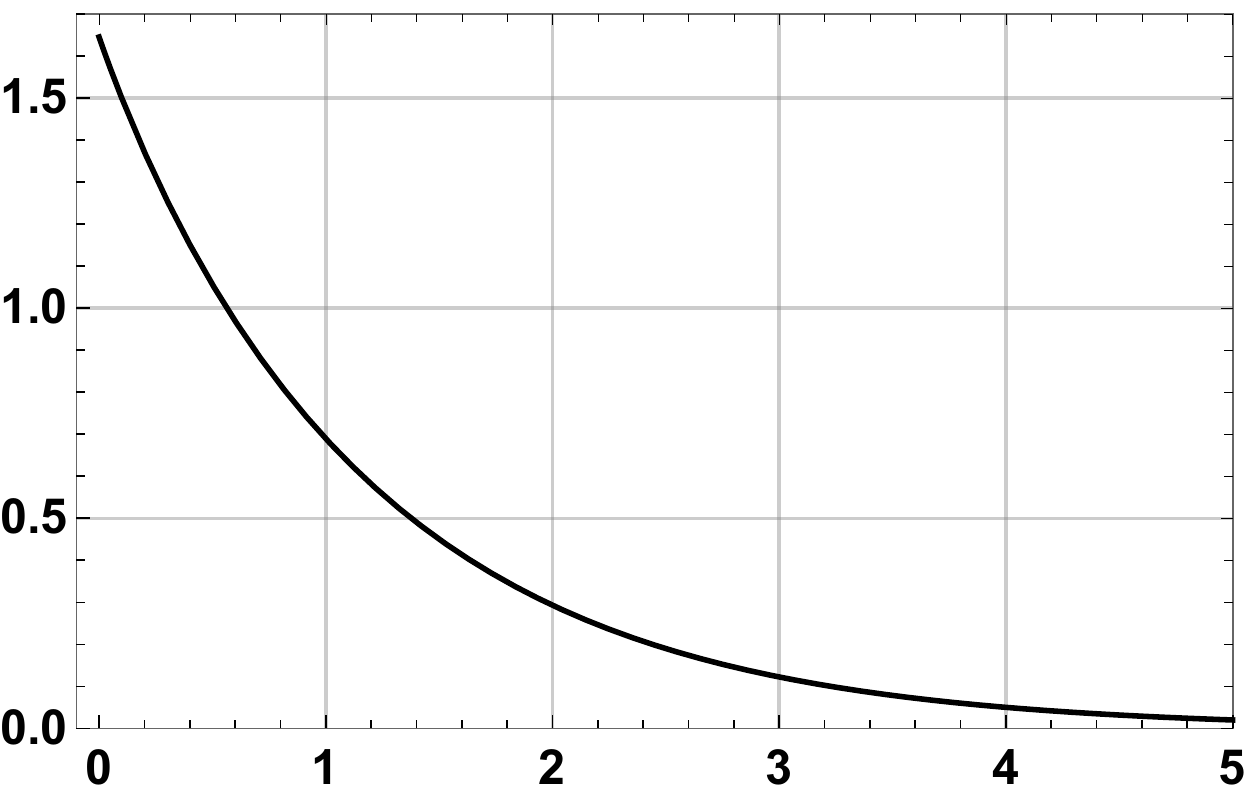}
    \put(0,0){$\beta m $}
    \put(-190,25){\rotatebox{90}{$4 \pi^2 \beta^2 \Pi^{(T)}_E/g$}}
     \caption{Thermal mass (in units of $g/(4 \pi^2 \beta^2)$) as a function of $\beta m$. In the massless case, the integral in (\ref{eq-Pi2}) gives $\pi^2/6 \approx 1.645$. The thermal mass decreases exponentially as we increase the value of $\beta m$.}
    \label{fig-mth}
\end{figure}

The sum in (\ref{eq-Pi1}) can be evaluated using the contour integral method\footnote{See for instance \cite{Kapusta:1989tk, Laine:2016hma}.}, and the self-energy can be written as
\begin{equation}
      \Pi_E = \Pi^{(0)}_E+\Pi^{(T)}_E~,
\end{equation}
where $\Pi^{(0)}_E$ is the vacuum zero temperature contribution, while $\Pi^{(T)}_E$ is a finite temperature part. The vacuum contribution is actually divergent, but it can be regularized by a mass counter-term $\delta m$. After regularization, one obtains $\Pi^{(0)}_E=0$. The temperature-dependent contribution can be written as
\begin{align} \label{eq-Pi2}
   \Pi^{(T)}_E&=  \frac{g}{2}  \int \frac{\textrm{d}^3 {\bf p}}{(2\pi)^3} \frac{1}{ \sqrt{{\bf p}^2+m^2}}\frac{1}{e^{ \beta \sqrt{{\bf p}^2+m^2}}-1} \nonumber\\
   &= \frac{g}{4\pi^2 \beta^2}  \int_{0}^{\infty} \textrm{d}u \frac{u^2}{\sqrt{u^2 + (\beta m)^2}}\frac{1}{e^{\sqrt{u^2 + (\beta m)^2}}-1}~,
\end{align}
where $u =\beta |{\bf p}|$, with $\beta=1/T$. At one-loop, the self-energy only depends on the dimensionless parameter $\beta m$. For $\beta m =0$, the integral in (\ref{eq-Pi2}) gives $\pi^2/6$, and the self-energy becomes
\begin{equation} \label{eq-mth}
    \Pi^{(T)}_E = \frac{g}{24 \beta^2}~.
\end{equation}
More generally, when $\beta m \neq 0$, the integral (\ref{eq-Pi2}) can be computed numerically and its value decreases very quickly as we increase the value of $\beta m$. See Fig.~\ref{fig-mth}. 

By summing all one-particle irreducible diagrams, the resulting propagator can be obtained from the free one as follows
\begin{equation}
    \frac{1}{-\omega^2 + {\bf k}^2+m^2} \rightarrow \frac{1}{-\omega^2 + {\bf k}^2+m^2+\Pi(\omega,{\bf k})}~,
\end{equation}
which shows that $ \Pi^{(T)}_E$ can be interpreted as a (squared) thermal mass $m_\text{th}^2$.
Therefore, the effects of interactions in the $\ell=4$ case can be taken into account by shifting the scalar field mass as $m^2 \rightarrow m^2+m_\text{th}^2$, where $m_\text{th}^2=\Pi^{(T)}$. From the point of view of the Lanczos coefficients and K-complexity, this case is almost indistinguishable from the free theory case, except in the case where $m=0$, in which the thermal mass (\ref{eq-mth}) produces a staggering effect that is absent when there are no interactions in $d=4$. In particular, by replacing  $m$ by an \emph{effective mass} $\sqrt{m^2+m_\text{th}^2}$ in Eq.~\eqref{eq:LambdaKMass} one can see that $\tilde{\lambda}_K$ decreases under the presence of interactions. 

\subsection{Relevant deformation ($\ell=3$)}
\label{subsec:RelDefo}

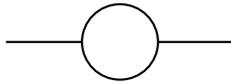
\begin{figure}[h!]
    \centering
    \begin{tikzpicture}[scale=1.]

\draw [thick]  (0,0) -- (1,0);
\draw [thick]  (2,0) -- (3,0);
\draw [thick]  (1.5,0) circle (.5cm);

\end{tikzpicture}
    \caption{Diagram relevant to the computation of the one-loop self-energy $\Pi_E$ in the theory with cubic interactions.}
    \label{fig:selfenergy2}
\end{figure}

In this case, the Euclidean one-loop self-energy can be computed by considering the diagram shown in Fig.~\ref{fig:selfenergy2}, which gives
\begin{equation} \label{eq-Pi3}
    \Pi_E(p_0,{\bf p})=  \frac{g^2 T}{2} \sum_{q_0} \int \frac{\textrm{d}^3 {\bf q}}{(2\pi)^3} \frac{1}{q_0^2+{\bf q}^2+m^2}\frac{1}{(p_0+q_0)^2+({\bf p+q})^2+m^2}~,
\end{equation}
where here once again the Matsubara frequencies $q_0 =  \, 2 \pi T\, n$ are summed over all integers $n$. 

For simplicity, let us consider the $m=0$ case.
Following~\cite{Arnold:1994ps}, we start by writing the propagators appearing in (\ref{eq-Pi3}) in configuration space
\begin{equation}
    \int \frac{\textrm{d}^3 {\bf q}}{(2\pi)^3} \frac{e^{i {\bf q \cdot r}}}{-q_0^2+{\bf q}^2} = \frac{e^{-|p_0|r}}{4 \pi r}~,
\end{equation}
where $r=|{\bf r}|$. We then rewrite (\ref{eq-Pi3}) as follows
\begin{equation}
    \Pi_E(p_0,{\bf p})=  \frac{g^2 T}{2} \sum_{q_0} \int \textrm{d}^3{\bf r} \, \frac{e^{i {\bf p \cdot r}}}{r^2} e^{-|p_0|r} e^{-|p_0+q_0|r}\,.
\end{equation}
Performing the sum in $q_0$, we find
\begin{equation}
    \Pi_E(p_0,{\bf p})=  \frac{g^2 T}{2(4 \pi)^2}  \int \textrm{d}^3{\bf r} \, \frac{e^{i {\bf p \cdot r}}}{r^2} e^{-|p_0|r} \left( \coth(2 \pi T \, r) +\frac{|p_0|}{2\pi T} \right) \,.
\end{equation}
The zero-temperature vacuum contribution
\begin{equation}
    \Pi^{(0)}_E = \frac{g^2 }{2(4 \pi)^2} \int \textrm{d}^3{\bf r} \, \frac{e^{i {\bf p \cdot r}}}{r^2} e^{-|p_0|r} \left( \frac{1}{2 \pi r} +\frac{|p_0|}{2\pi} \right)~, 
\end{equation}
diverges, but it can be regularized with a mass counter-term. After regularization, one obtains $\Pi^{(0)}=0$. The finite-temperature contribution reads
\begin{align} \label{eq-Pi4}
    \Pi^{(T)}_E(p_0,{\bf p})&= \Pi_E - \Pi^{(0)}_E =  \frac{g^2 T}{2(4 \pi)^2}  \int \textrm{d}^3{\bf r} \, \frac{e^{i {\bf p \cdot r}}}{r^2} e^{-|p_0|r} \left( \coth(2 \pi T \, r) -\frac{1}{2\pi T r} \right)\,,  \nonumber\\
    &=\frac{g^2 T}{16 \pi}  \int_{0}^{\infty} \textrm{d} r\, r^2 \int_{0}^{\pi} \textrm{d}\theta\, \sin \theta \, \frac{e^{i {\bf |p|} r \cos \theta}}{r^2} e^{-|p_0|r} \left( \coth(2 \pi T \, r) -\frac{1}{2\pi T r} \right)\,, \nonumber \\
    &=\frac{g^2 T}{8 \pi}  \int_{0}^{\infty} \textrm{d} r\,  \frac{ \sin (|{\bf p}|r)}{|{\bf p}|r} e^{-|p_0|r} \left( \coth(2 \pi T \, r) -\frac{1}{2\pi T r} \right)\,.
\end{align}
The above expression can be evaluated numerically once one specifies $(p_0,{\bf p})$. The corresponding Lorentzian correlators can be computed by analytic continuation.

The first quantum correction to the spectral function is given by \cite{Srednicki:1019751}
\begin{equation}
    \rho_1 (\omega, {\bf p})= \frac{ \text{sgn}(\omega) \, \text{Im}\, \Pi^{(T)}(\omega, {\bf p})}{(-\omega^2+{\bf p}^2+\text{Re}\, \Pi^{(T)}( \omega,\mathbf{p}))^2+(\text{Im}\, \Pi^{(T)}(\omega,\mathbf{p}))^2}~.
\end{equation}
where $\Pi^{(T)}(\omega, {\bf p})=\Pi^{(T)}_E( p_0 \rightarrow -i \omega, {\bf p})$.

The one-loop corrected spectral power is then computed as
\begin{equation} \label{eq-oneloopfw}
    f^W(\omega) =  \frac{N(\beta, g)}{\sinh[\beta\omega/2]} \left( \frac{\Omega_2}{16\pi^2}\omega + \int \frac{\textrm{d}^{3}\bf k}{(2\pi)^{3}}\rho_1(\omega,\mathbf{k}) \right) ~,
\end{equation}
where the first term in the parentheses represents the free theory contribution while the second term represents the one-loop correction, which depends on $g^2$.

\begin{figure}
     \centering
     \begin{subfigure}[b]{0.45\textwidth}
         \centering
         \includegraphics[width=\textwidth]{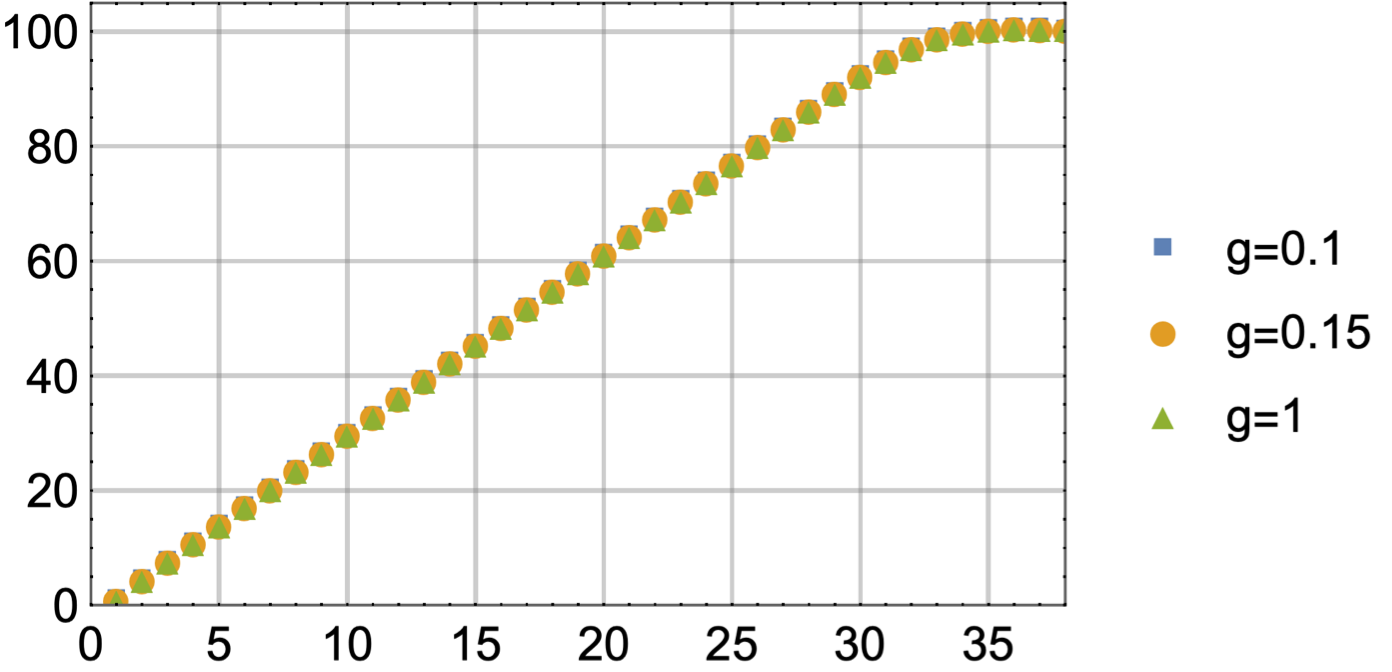}
         \caption{}
     \end{subfigure}
     \hfill
     \begin{subfigure}[b]{0.48\textwidth}
         \centering
         \includegraphics[width=\textwidth]{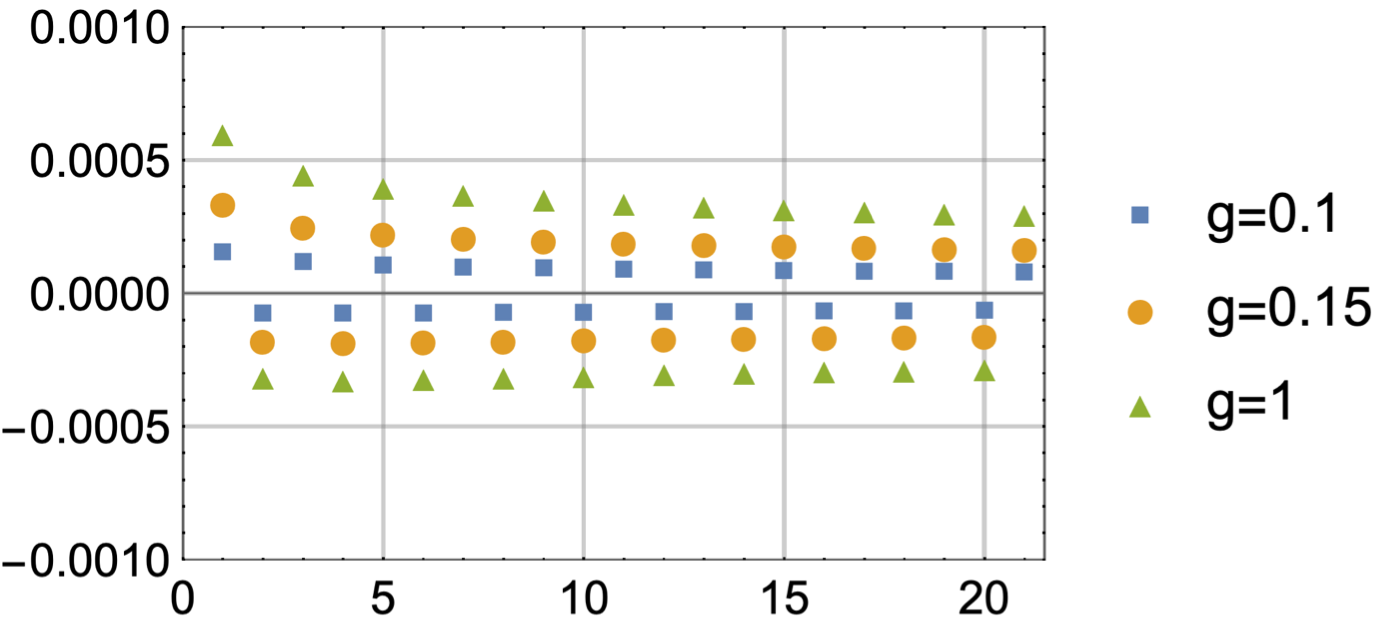}
         \caption{}
     \end{subfigure}
     \put(-280,15){\small $n$}
     \put(-50,15){\small $n$}
     \put(-205,120){\small $\delta b_n$}
     \put(-430,120){\small $b_n$}
     \hfill
           \caption{(a) Lanczos coefficients computed from the one-loop corrected spectral power. (b) Staggering, measure as $\delta b_n =b_n(g)-b_n(0)$, as a function of $n$ for several values of the coupling constant $g$. In both panels, we set $\beta=1$ and $\Lambda=200$. }
        \label{fig:lfskc}
\end{figure}

From \eqref{eq-oneloopfw} we numerically compute the Lanczos coefficients $b_n$ for several values of the coupling constant $g$. The results are shown in Fig.~\ref{fig:lfskc} (a). Since our calculation is perturbative in $g$, the curves of $b_n$ as a function of $n$ are not very different from the one we obtain for the free theory in $d=4$. But the effects of the interactions can be seen in the curves of  $b_n(g)-b_n(0)$ as a function of $n$, which shows a staggering effect that decreases as $n$ increases. See Fig.~\ref{fig:lfskc} (b). 
A staggering effect that decreases as $n$ increases is known to be associated with systems in which the power spectrum has bounded support and no mass gap~\cite{RecursionBook}. Considering a free massless field with $d\ne4$, we checked that this feature is present even when we remove the condition of bounded support - i.e. when we take $\omega_F$ to infinity. This feature is also shown in Fig.~3 of \cite{Dymarsky:2021bjq}.

\begin{figure}
    \centering
    \includegraphics[width=10cm]{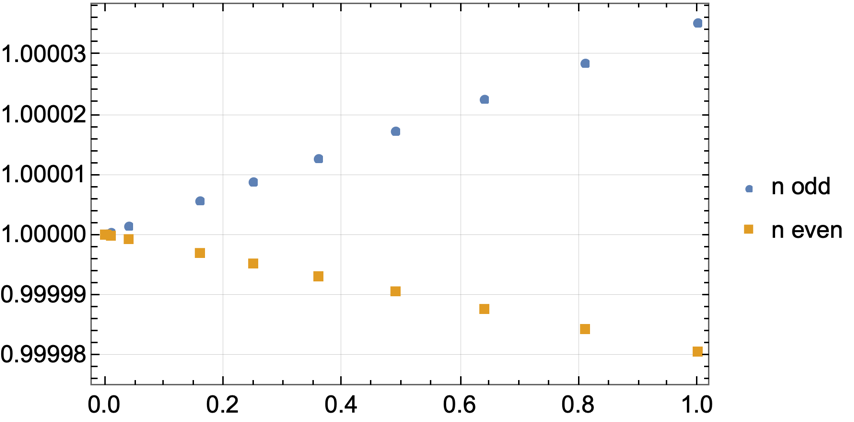}
    \put(-35,5){\small $g^2$}
    \put(-285,150){\small $\alpha(g)/\alpha(0)$}
    \caption{Log-plot of the normalized slope $\alpha(g)/\alpha(0)$ as a function of the squared coupling $g^2$ for Lanczos coefficients $b_n$ with odd and even values of $n$. }
    \label{fig:slopevscoupling}
\end{figure}

To understand how the interactions affect the slope of the curves of $b_n$ as a function of $n$, we fit a line of the form $b_n = \alpha(g) \, n + \gamma(g)$ to the data, separating odd and even Lanczos coefficients\footnote{We separate the data $\{n,b_n \}$ into two sets, $\{ n_\text{odd},b_{n_\text{odd}} \}$ and $\{n_\text{even},b_{n_\text{even}}\}$, and fit a line to each one. }. Fig.~\ref{fig:slopevscoupling} shows the slope $\alpha(g)$ for odd and even Lanczos coefficients as a function of $g^2$. Note that the slope of the odd coefficients increases under the presence of interactions, while the slope of the even coefficients decreases. 

\section{Discussion and Conclusions}
\label{sec:DiscussionConclusion}

In this paper, we discussed the behavior of the Lanczos coefficients $b_n$ and the Krylov complexity $K_\mathcal{O}(t)$ for free and interacting scalar QFTs at finite temperature for several spacetime dimensions. For free scalar QFTs, we studied the effects of introducing IR and UV cutoffs in the power spectrum induced by a mass term in the Lagrangian~\eqref{eq:FreeScalarLag} and a cutoff $\Lambda$ in the momentum integral of the power spectrum~\eqref{eq:PowerSpectruHardUVCutoff}. The bare mass $m$ causes a staggering of the Lanczos coefficients $b_n$, separating them into two smooth families, one for odd $n$ and one for even $n$. This effect is seen for all values of $n$ and was confirmed by an analytic approximation for $\beta m \gg 1$ (see Eq.~\eqref{eq-bn-largemassv2}) and by numerical computations of $b_n$ (see Fig.~\ref{fig:bn5dLargeBetaMassNoCutoff}). The staggering due to the non-zero mass exists even when $n$ is large and thus the late-time behavior of Krylov complexity might be affected by the staggering effect. After studying the behavior of the Lanczos coefficients in several models (cf. App. \ref{app:OriginsStaggering}), we observe that staggering is absent if the following two conditions are satisfied: (I) the power spectrum is finite and positive at $\omega=0$, i.e., $0 <f^W(0) <\infty$; and (II) the derivative of the power spectrum $f^W(\omega)$ is a continuous function of $\omega$ for $-\Lambda < \omega < \Lambda$, where $\Lambda$ is a UV cutoff. In terms of the auto-correlation $C(t)$, the first condition implies that the integral $f_W(0)=\int C(t) dt$ is finite and positive, which means that the auto-correlation is positive most of the time. It is not clear to us what the second condition implies for $C(t)$, or the physical meaning of conditions I and II. We hope to come back to the physical interpretation of these mathematical statements in the future.

 The exponential growth rate $\tilde{\lambda}_{K}$ of $K_\mathcal{O}(t)$ computed for a finite range of $t$ decreases as a function of the mass, which is consistent with the conjectured bound~\eqref{eq:ImprovedChaosBound}. Introducing a hard UV cutoff for the momentum integral of a particle causes the saturation of the Lanczos coefficients and a transition from the exponential growth of $K_\mathcal{O}(t)$ to a linear growth (see Fig.~\ref{fig:KCUVcutoff}). We also considered a deformation of the power spectrum by a smooth cutoff (see Eq.~\eqref{wpssc}), which may provide a way to deform the UV structure of QFTs that arise as the continuum limits of lattice models.

This work on Lanczos coefficients and Krylov complexity in QFTs with momentum cutoffs was influenced by an important remark by Dymarsky and his collaborators related to the role that a UV cutoff would play in diagnosing quantum chaos in a discretized lattice model~\cite{DymarskyTalk, Avdoshkin:2022xuw}. From the perspective of the power spectrum, in Sec.~\ref{subsec:KrylovSpectral} and Sec.~\ref{subsec:MassScalarHeavyLimitHardUVCutoff}, we corroborated the observation that studying signatures of quantum chaos for spectral statistics at late times in QFTs on non-compact spaces with unbounded spectrum is an ill-posed question. The reason is that it is not possible to distinguish free from chaotic lattice models once the continuum limit is taken and their power spectrum is scrutinized. It is worth pointing out that similar statements regarding the study of quantum chaos in continuum theories from the perspective of their operator algebra have been made in~\cite{Radicevic:2021ykf}. Here it was argued that continuum theories cannot be expected to capture all aspects of spectral statistics, the cornerstone of quantum chaos. Instead, one needs to coarse-grain the continuum theory down to a classical theory with finite-dimensional state space and operator algebra of the same dimension.

In Sec.~\ref{sec:InterMassScalar}, we considered a real scalar field with cubic or quartic interaction terms propagating in 4-dimensional flat spacetime, and  numerically studied the effects of interactions on the Lanczos coefficients. The cubic (quartic) interaction term can be thought of as a relevant (marginally irrelevant) deformation of the free theory. In both cases, the interaction term breaks the smooth behavior of the Lanczos coefficients, producing staggering for all values of $n$. We observed that the presence of a mass gap in the power spectrum\footnote{We say that the spectral power has a mass gap if $f^W(\omega) = 0$ for $|\omega| < m$.} controls whether staggering is constant or not, which is consistent with the results of \cite{RecursionBook}.
The effect of quartic interactions can be accounted for by a simple shift in the mass ($m^2 \rightarrow m^2+m^2_\text{th}$), which produces a gap in the spectral power even when $m=0$.
The corresponding staggering effect does not depend on $n$. By contrast, in the case of cubic interactions, there is no mass gap, and the staggering decreases as we increase $n$. This suggests that the staggering produced by the cubic interaction term may not be present for sufficiently large values of $n$. Therefore, we do not expect this deformation to affect $\lambda_K$.

This study can be extended in several interesting directions, namely:
 \begin{itemize}
 \item We only calculated the exponential growth rate $\tilde{\lambda}_{K}$ for a finite range in $t$. It is important to semi-analytically determine the asymptotic rate $\lambda_{K}$ at $t\to\infty$. This could be accomplished by finding a compact closed-form expression of the probability amplitudes and computing the limit of the series~\eqref{eq:KrylovCDef}, provided it exists;
 
 \item In the $\phi^4$ theory, we observed that $\tilde{\lambda}_K$ decreases as we increase the coupling $g$. Since the staggering is constant in this case, there is a possibility that $\lambda_K$ also decreases in the presence of interactions. It would be interesting to check if this feature is also present in a matrix theory with quartic interactions \cite{Stanford:2015owe} because in this case the Lyapunov exponent $\lambda_L$ is well-defined, and one could study if the conjectured chaos bound~\eqref{eq:ImprovedChaosBound} is still satisfied;

 \item Also in the context of interacting QFTs, it would be interesting to investigate whether continuous tensor network techniques in the spirit of~\cite{Fernandez-Melgarejo:2020fzw} can be used to study Krylov complexity in generic interacting scalar field theories\footnote{We thank Adolfo del Campo for pointing out this interesting potential future direction.};

 \item Deformations of local QFTs due to interactions are represented by renormalization group flows. One may be able to classify the effects of the interactions on the Lanczos coefficient and Krylov complexity in terms of the renormalization group flows. In particular, we observed that the staggering produced by a relevant deformation decreases with $n$, while the staggering produced by  marginally irrelevant deformation does not depend on $n$.  Since the large-$n$ regime can be associated with UV physics, it is perhaps somewhat expected that the effects of the relevant deformation on the Lanczos coefficients decreases as we increase $n$. It would be interesting to further investigate these features, and check if they are also present in other models;

 \item In studies of K-complexity with the universal operator growth hypothesis, one typically considers a local scalar field operator having zero overlap with any conserved current. Studying K-complexity using conserved current operators instead of scalar operators might be an interesting direction of research since these operators have different scrambling properties which presumably affect the behavior of the Lanczos coefficients and K-complexity. It is perhaps more interesting to study this in the context of lattice models, where the K-complexity is known to provide a good diagnosis of quantum chaotic behavior;
 
 \end{itemize}

\section{Acknowledgements}

We thank the participants of the Asia Pacific Center for Theoretical Physics (APCTP) Focus Program ``Holography 2022: quantum matter and spacetime'' and ``Integrability, duality and related topics'' for valuable discussions on the subject. We are especially thankful to Anatoly Dymarsky for discussing his ongoing work on similar topics during the APCTP Focus Program ``Holography 2022: quantum matter and spacetime''~\cite{DymarskyTalk}, which motivated us to consider the relevance of ultra-violet (UV) cutoffs in our studies. We are also grateful to Adolfo del Campo, Anatoly Dymarsky, Song He, Pratik Nandy and Horacio M. Pastawski for comments and suggestions on the first version of this manuscript. H.~A.~Camargo also thanks Luis Apolo, Pawel Caputa, Christian Ferko, Chen-Te Ma and Pratik Nandy for discussions and correspondence. This work was supported by the Basic Science Research Program through the National Research Foundation of Korea (NRF) funded by the Ministry of Science, ICT \& Future Planning (NRF- 2021R1A2C1006791), the AI-based GIST Research Scientist Project grant funded by the GIST in 2023. This work was also supported by Creation of the Quantum Information Science R\&D Ecosystem (Grant No. 2022M3H3A106307411) through the National Research Foundation of Korea (NRF) funded by the Korean government (Ministry of Science and ICT). H.~A.~Camargo, V.~ Jahnke and M.~Nishida were supported by the Basic Science Research Program through the National Research Foundation of Korea (NRF) funded by the Ministry of Education (NRF-2022R1I1A1A01070589, NRF-2020R1I1A1A01073135, NRF-2020R1I1A1A01072726).

\section*{Appendices}

\appendix

\section{Analytic expressions for the Lanczos coefficients} \label{app:analyticbn}
In this appendix, we derive analytic results for the Lanczos coefficients for a 5-dimensional free QFT in the large mass regime. We start from the formula for the moments in the large mass limit~\eqref{eq:Moments5dLargeBetaMass}, which we reproduce here for convenience
\begin{equation}
\label{eq:Moments5dLargeBetaMassv3}
    \mu_{2n} =\frac{2^{-2} e^{\frac{m \beta}{2}} }{2+m \beta} \left( \frac{2}{\beta}\right)^{2n} \left[-m^2 \beta^2 \, \tilde{\Gamma} \left(2n+1,\frac{m \beta}{2}\right)+4 \tilde{\Gamma} \left( 2n+3,\frac{m \beta}{2}\right) \right]~.
\end{equation}
Expanding~\eqref{eq:Moments5dLargeBetaMass} in a power series in $1/(m \beta)$, we obtain
\begin{equation}
\label{eq:Moments5dLargeBetaMass2}
  \mu_{2n} =  m^{2 n} \left[1+  \frac{8 n}{\beta  m} -\frac{16 n}{\beta ^2 m^2}   ++\frac{48 n^2}{\beta ^2 m^2}+\frac{64 n}{\beta ^3 m^3}-\frac{288 n^2}{\beta ^3 m^3}+\frac{256 n^3}{\beta ^3 m^3} +\mathcal{O}\left( \frac{n^4}{\beta^4 m^4}\right)\right]\,.
\end{equation}
Using the non-linear recursion relation~\eqref{eq:Mu2nToBnRecursion}, we can obtain the Lanczos coefficients $b_{n}$ from the moments~\eqref{eq:Moments5dLargeBetaMass2} as a function of $\beta m$. Analyzing the first Lanczos coefficients allows us to determine how they behave as a function of $n$.
We derive the following analytic expression for the Lanczos coefficients
\begin{equation} \label{eq-bn-largemassvv3}
  b_n= m
\begin{cases} 
1+\frac{2(n+1)}{m \beta}+\frac{2(n+1)^2}{(m \beta)^2}-\frac{2(n+1)^3}{(m \beta)^3}+\mathcal{O}\left(\frac{n}{m \beta}\right)^4\,, \,\,\text{for}\,\,n\,\,\text{odd}~,\\
\frac{\sqrt{4 n (n+2)}}{m \beta}+\frac{\sqrt{4 n (n+1)^2(n+2)}}{(m \beta)^2}-\frac{(1+2n(n+2))\sqrt{n (n+2)}}{(m \beta)^3}+\mathcal{O}\left(\frac{n}{m \beta}\right)^4\,,\,\,\text{for}\,\, n\,\,\text{even}~.
\end{cases}  
\end{equation}
We checked that \eqref{eq-bn-largemassvv3} correctly reproduces the results for the Lanczos coefficients obtained numerically from~\eqref{eq:Moments5dLargeBetaMass} up to $n \approx 35$. See Fig.~\ref{fig:bn5dLargeBetaMassNoCutoff}.
By squaring \eqref{eq-bn-largemassvv3} and multiplying it by $\beta^2$, we obtain
\begin{equation} \label{eq-bn-largemassvv2}
  \beta^{2}b_{n}^{2}= m^{2}\beta^{2}
\begin{cases}
1+4\frac{1+n}{m\beta}+8\frac{(n+1)^2}{m^{2}\beta^{2}}+12\frac{(n+1)^3}{m^{3}\beta^{3}}+\cdots\,, \,\,\text{for}\,\,n\,\,\text{odd}~,\\
4\frac{n(n+2)}{m^{2}\beta^{2}}+8\frac{n(n+1)(n+2)}{m^{3}\beta^{3}}+\cdots\,,\,\,\text{for}\,\, n\,\,\text{even}~.
\end{cases}  
\end{equation}

\section{Auto-correlation Function in Odd-Dimensional Free QFTs}
\label{app:AutocOddd}

In this appendix we discuss the details of the thermal Wightman $2$-point function $C^{(d)}(t)$ for the field operator $\phi$ in free odd-dimensional QFTs in the large mass regime $m\beta \gg 1$~\eqref{eq:AutoCPhi0Lma}. The functions $\lbrace c^{(d)}_{i}(t)\rbrace$ for $d=5,7,9$ are given by
\begin{subequations}
\label{eq:AutoCPhi0Lma5d}
\begin{align}
    c_{1}^{(5)}(t)=&\frac{\beta ^3}{(\beta  m+2) \left(\beta ^2+4 t^2\right)^3}~,\\
    c_{2}^{(5)}(t)=&-4 t \left(\beta ^2 (\beta  m+3)+4 t^2 (\beta  m-1)\right)~,\\
    c_{3}^{(5)}(t)=&2 \beta ^3+m \left(\beta ^4-16 t^4\right)-24 \beta  t^2~,
\end{align}
\end{subequations}
\begin{subequations}
\label{eq:AutoCPhi0Lma7d}
\begin{align}
    c_{1}^{(7)}(t)=&\frac{\beta ^5}{\left(\beta ^2 m^2+6 \beta  m+12\right) \left(\beta ^2+4 t^2\right)^5}~,\\
    c_{2}^{(7)}(t)=&2t\Big(-3 \beta ^4 \left(\beta ^2 m^2+8 \beta  m+20\right)+64 m^2t^6\notag\\
   &-16 t^4 \left(\beta ^2 m^2-24 \beta  m+12\right)-20 \beta ^2 t^2 \left(\beta ^2
   m^2-24\right)\Big)~,\\
    c_{3}^{(7)}(t)=&\beta  m^2 \left(\beta ^2-12 t^2\right)
   \left(\beta ^2+4 t^2\right)^2\notag\\
   &+6 m \left(\beta ^6+64 t^6-80 \beta ^2 t^4-20 \beta ^4
   t^2\right)\notag\\
   &+12 \left(\beta ^5+80 \beta  t^4-40 \beta ^3 t^2\right)~,
\end{align}
\end{subequations}
\begin{subequations}
\label{eq:AutoCPhi0Lm9d}
\begin{align}
    c_{1}^{(9)}(t)=&\frac{\beta ^7}{\left(\beta ^3 m^3+12 \beta ^2 m^2+60 \beta  m+120\right) \left(\beta ^2+4
   t^2\right)^7}~,\\
    c_{2}^{(9)}(t)=&8t\Big(8 \beta ^4 t^2 \left(-\beta ^3m^3+105 \beta 
   m+525\right)+256 m^2 t^8 (\beta  m-3)\notag\\
   &+672 \beta ^2 t^4 \left(\beta ^2 m^2+5 \beta 
   m-15\right)\notag\\
   &-\beta ^6 \left(\beta ^3 m^3+15 \beta ^2 m^2+90 \beta  m+210\right)\notag\\
   &+128 t^6
   \left(\beta ^3 m^3+12 \beta ^2 m^2-45 \beta  m+15\right)\Big)~,\\
    c_{3}^{(9)}(t)=&m^3
   \left(\beta ^4+16 t^4-24 \beta ^2 t^2\right) \left(\beta ^2+4 t^2\right)^3\notag\\
   &+12 \beta 
   m^2 \left(\beta ^4+80 t^4-40 \beta ^2 t^2\right) \left(\beta ^2+4 t^2\right)^2\notag\\
   &-60 m
   \left(-\beta ^8+256 t^8-896 \beta ^2 t^6+56 \beta ^6 t^2\right)\notag\\
   &-120 \left(-\beta
   ^7+448 \beta  t^6-560 \beta ^3 t^4+84 \beta ^5 t^2\right)~.
\end{align}
\end{subequations}

\section{Krylov complexity in CFTs on a hyperbolic space}
\label{app:KcompHyperbolic}
By using a conformal map from $R^d$ to $S^{1}\times H^{d-1}$, we can study the Krylov complexity in CFTs whose spatial geometry is a hyperbolic space $H^{d-1}$. 
Due to the conformal map, the curvature scale of $H^{d-1}$ depends on the period $\beta$ of Euclidean time.
The scalar conformal two-point function on $S^{1}\times H^{d-1}$  is given by \cite{Haehl:2019eae}
\begin{align}
\frac{\mathcal{N}}{(-2\cos(\frac{2\pi}{\beta}(\tau_1-\tau_2))+2\cosh \mathbf{d}(1,2))^{\Delta}},\label{etfsh}
\end{align}
where $\Delta$ is a scaling dimension, $\tau_i$ is the Euclidean time, $\mathbf{d}(1,2)$ is the spatial distance in $H^{d-1}$, and $\mathcal{N}$ is a normalization factor. Substituting $\tau_1-\tau_2=it+\frac{\beta}{2}$ and $\mathbf{d}(1,2)=0$ to the Euclidean two-point function (\ref{etfsh}), we obtain the thermal Wightman two-point function
\begin{align}
C(t)=\frac{\mathcal{N}}{(4\cosh(\pi t/\beta))^{2\Delta}},\label{wtfh}
\end{align}
which agrees with (\ref{ess}) for $\alpha=\pi/\beta$ and $\eta=2\Delta$ up to the normalization factor. Therefore, the Krylov complexity in CFTs on the hyperbolic space with $\Delta=d/2-1$ calculated from Euclidean CFTs on $S^{1}\times H^{d-1}$ is \begin{align}
 K_{\mathcal{O}}(t)=(d-2)\sinh^2(\pi t/\beta).\label{kotcft}
 \end{align}

\section{Conditions for staggering}
\label{app:OriginsStaggering}
In this appendix, we study the properties of the power spectrum $f(\omega)$ that lead to staggering. We also comment on the implications of these conditions for the auto-correlation $C(t)$. 

Based on the study of the Lanczos coefficients for several models, we (tentatively) propose the following criteria which could help predict whether a given power spectrum could give rise to staggering. The conditions on the power spectrum $f(\omega)$ for the {\bf absence} of staggering appear to be:
\begin{align*}
    &\text{(I)} \,\,\,\,f(\omega) \,\,\text{is finite at $\omega=0$, i.e.,} \,\, 0< f(\omega)<\infty\,,\\
    &\text{(II)} \,\,\,\,  f'(\omega) \,\,\text{is a continuous function of}\,\, \omega \text{ for } -\Lambda < \omega < \Lambda.\,
\end{align*}
where $\Lambda$ is a UV cutoff. In all the examples we studied, the presence of staggering can be attributed to the violation of conditions (I) and/or (II).

\subsection*{Example 1}
\begin{figure}[h!]
     \centering
     \begin{subfigure}[b]{0.45\textwidth}
         \centering
         \includegraphics[width=\textwidth]{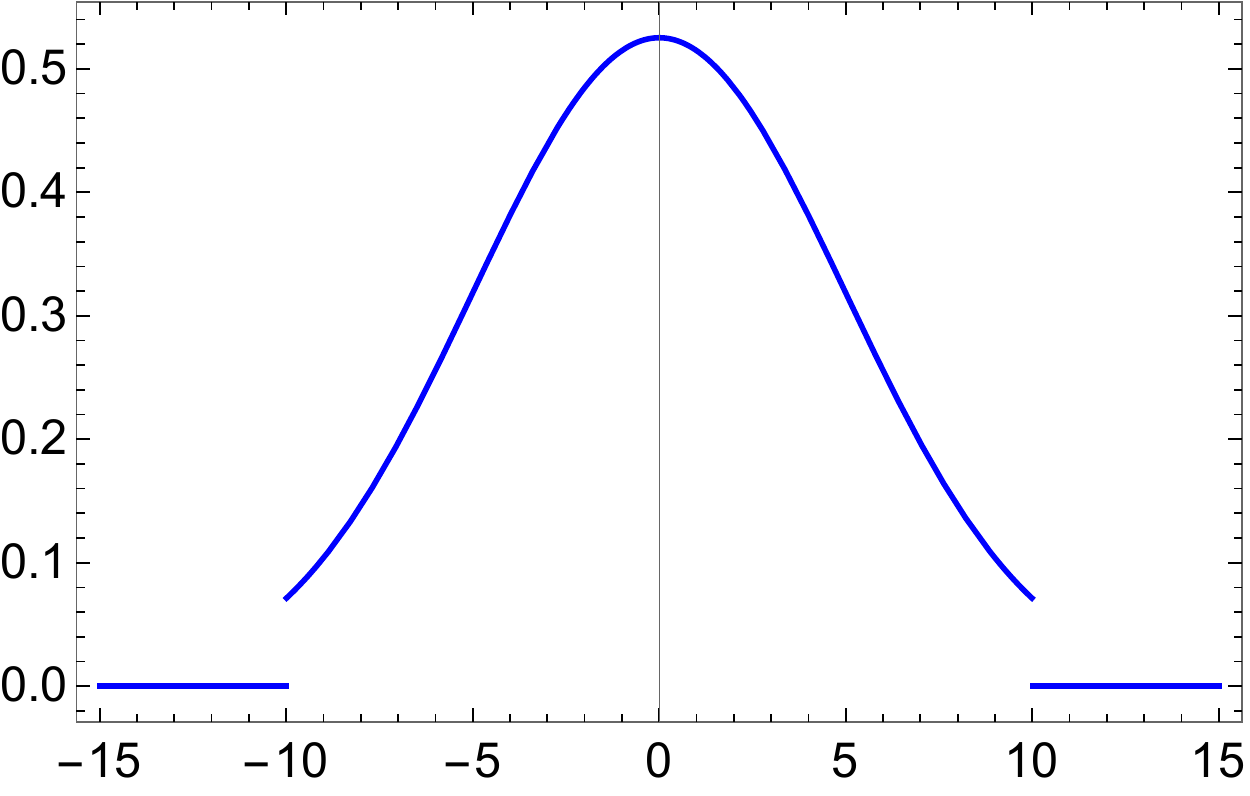}
         \caption{}
     \end{subfigure}
     \hfill
     \begin{subfigure}[b]{0.45\textwidth}
         \centering
         \includegraphics[width=\textwidth]{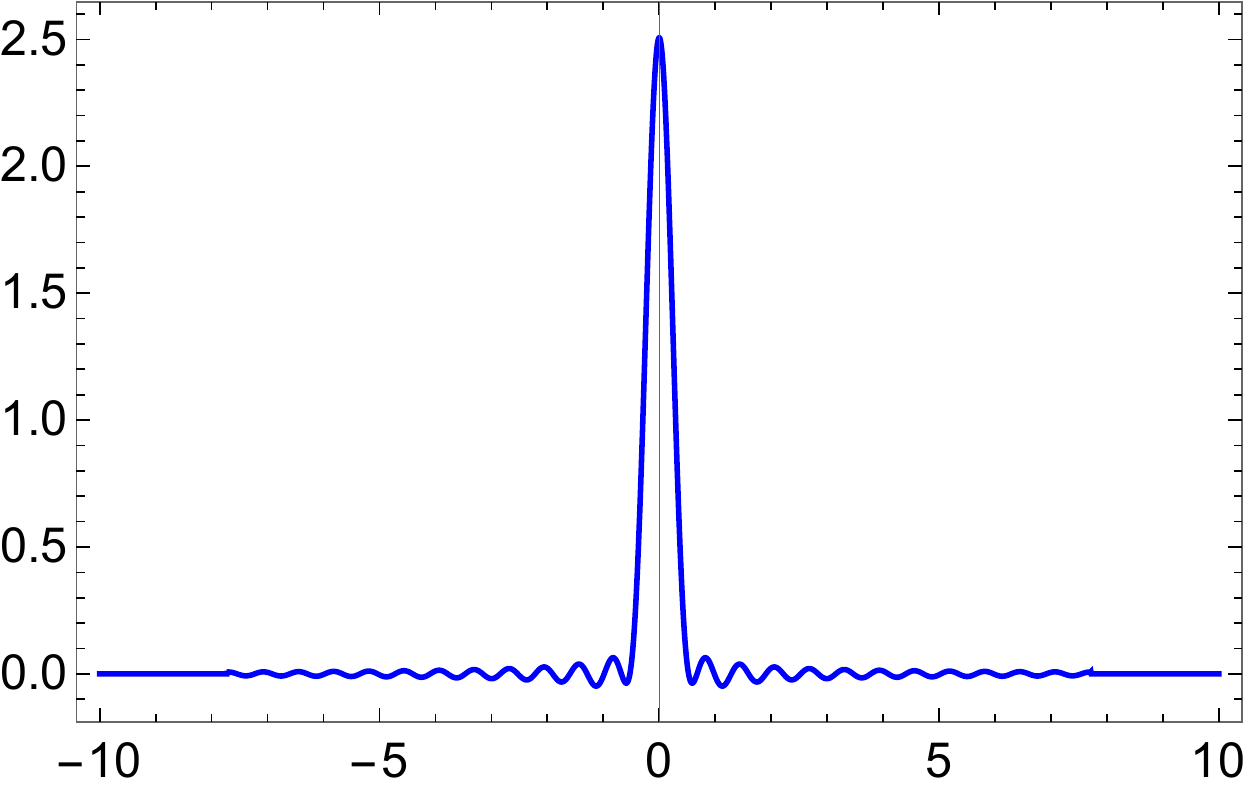}
         \caption{}
     \end{subfigure}
     \put(-240,15){\small $\omega$}
     \put(-5,15){\small $t$}
     \put(-195,148){\small $C(t)$}
     \put(-430,148){\small $f(\omega)$}
     \hfill
           \caption{(a) Power spectrum of the Example 1 for $\sigma=5$ and $\Lambda=10$. In (b), we show the corresponding auto-correlation.}
        \label{fig:example1}
\end{figure}

\begin{figure}
    \centering
    \includegraphics[width=0.45\textwidth]{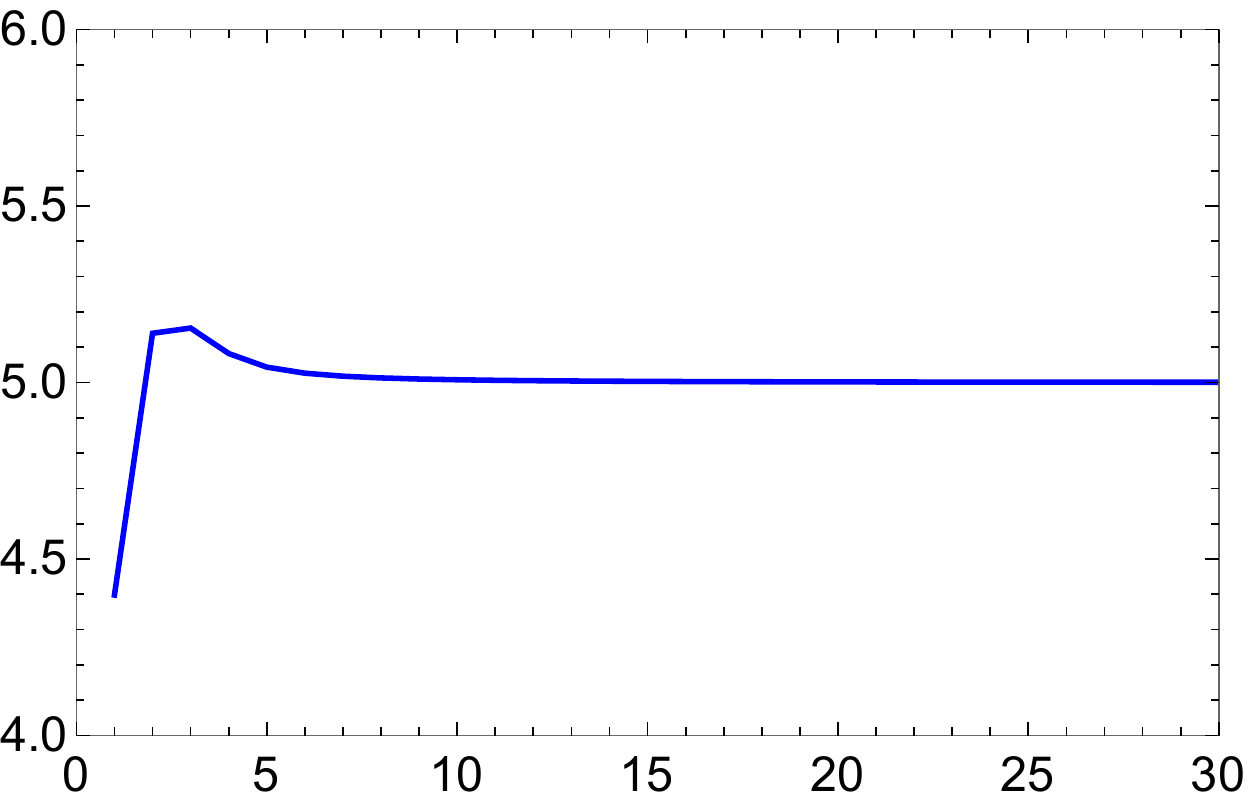}
    \put(5,3){\small $n$}
    \put(-195,130){\small $b_n$}
    \caption{Lanczos coefficients of the Example 1 for $\sigma=5$ and $\Lambda=10$. Here we connected the dots to make the absence of staggering more evident.}
    \label{fig:example1-bn}
\end{figure}

Let us first consider a case where there is no staggering, namely:
\[
 f(\omega) = \frac{\sqrt{2 \pi }}{\sigma \, \text{Erf}\left(\frac{\Lambda }{\sqrt{2} \sigma }\right)}
  \begin{cases} 
   e^{-\frac{\omega^2}{2 \sigma^2}} & \text{if } |\omega| \leq \Lambda \\
   0       & \text{if } |\omega| > \Lambda
  \end{cases}
\]
where $\Lambda$ is UV cutoff that makes the Lanczos coefficients to saturate to a constant value. It convenient to introduce such a cutoff because it is visually easier to see staggering when the Lanczos coefficients are not growing. The plots for the above power spectrum and the corresponding auto-correlation are shown in Fig.~\ref{fig:example1}, while the related results for the Lanczos coefficients are shown in Fig.~\ref{fig:example1-bn}.


\subsection*{Example 2}
\begin{figure}[h!]
     \centering
     \begin{subfigure}[b]{0.45\textwidth}
         \centering
         \includegraphics[width=\textwidth]{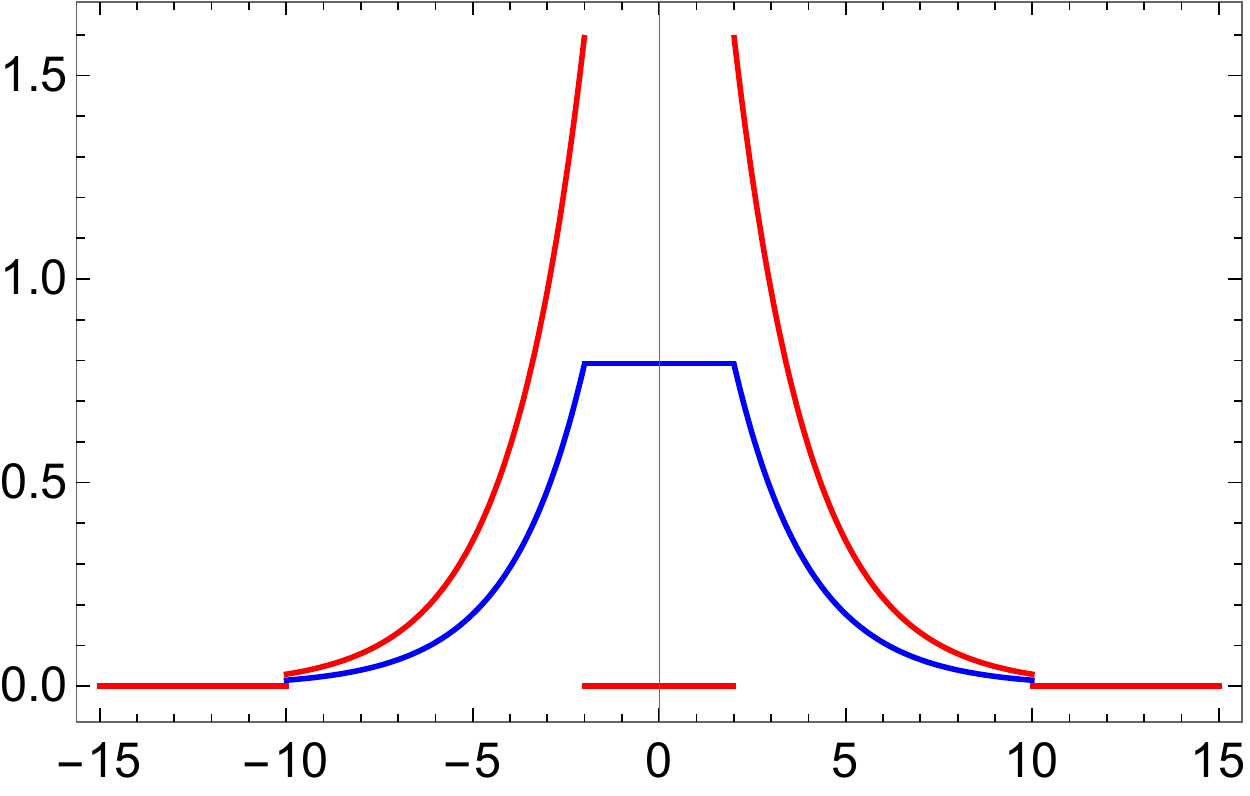}
         \caption{}
     \end{subfigure}
     \hfill
     \begin{subfigure}[b]{0.45\textwidth}
         \centering
         \includegraphics[width=\textwidth]{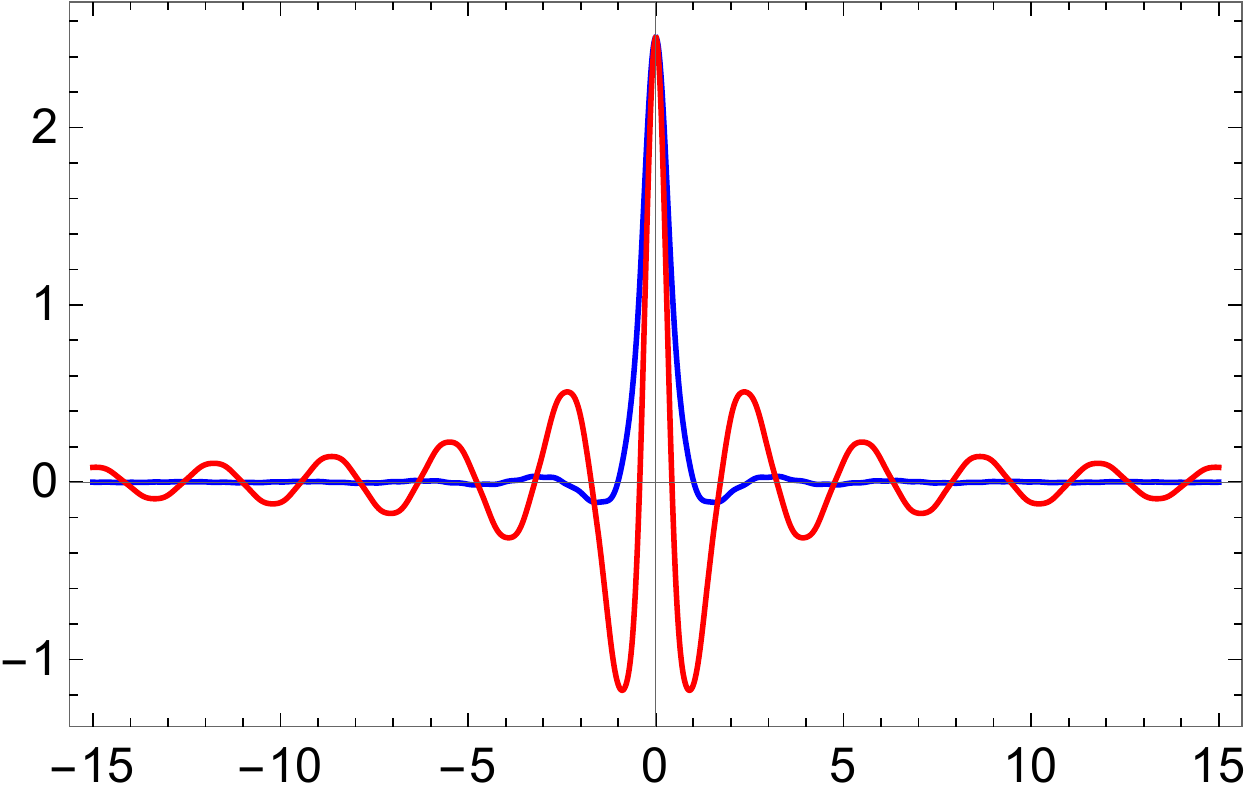}
         \caption{}
     \end{subfigure}
     \put(-240,15){\small $\omega$}
     \put(-5,15){\small $t$}
     \put(-195,148){\small $C(t)$}
     \put(-430,148){\small $f(\omega)$}
     \hfill
           \caption{(a) Power spectrum of the Example 2 with $m=2$, $a=1/2$, and $\Lambda=10$. The blue curves correspond to the case where $h=1$, where $f(\omega)$ is continuous and there is no mass gap. The red curves correspond to the case where $h=0$, where $f(\omega)$ is discontinuous at $\omega=m$ and vanishes for $\omega <m$. In (b), we show the corresponding auto-correlations. }
        \label{fig:example2}
\end{figure}

\begin{figure}[h!]
    \centering
    \includegraphics[width=0.45\textwidth]{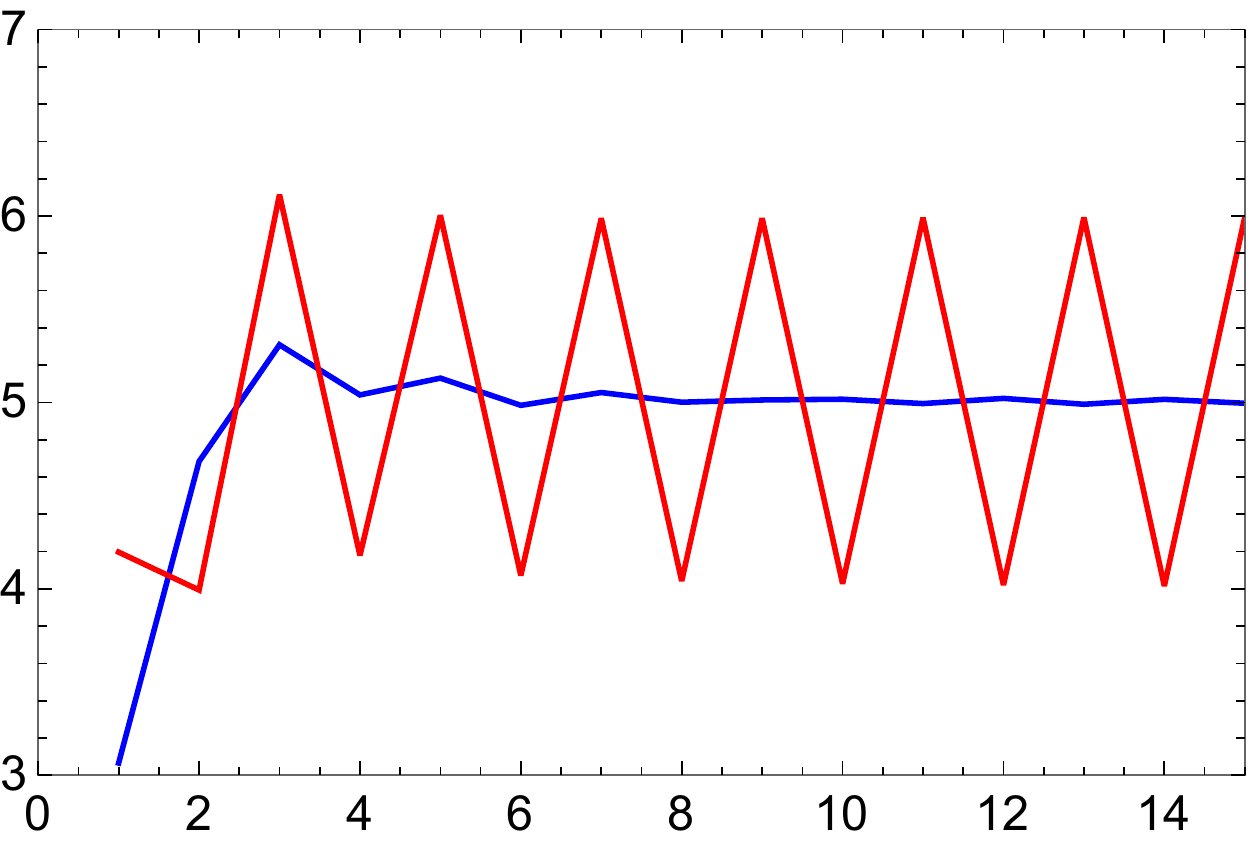}
    \put(5,3){\small $n$}
    \put(-195,133){\small $b_n$}
    \caption{Lanczos coefficients of the Example 2 for $h=0$ (red curve) and $h=1$ (blue curve). Here we connected the dots to make the staggering more evident.}
    \label{fig:example2-bn}
\end{figure}

We now consider two examples where there is staggering. Let us consider the case where the power spectrum is given by
\[
 f(\omega) = \frac{\pi  a e^{a (\Lambda +m)}}{e^{a \Lambda } (a h m+1)-e^{a m}}
  \begin{cases} 
   h\, e^{-a m} & \text{if } |\omega| \leq m \\
   e^{- a |\omega|}       & \text{if } m <|\omega| < \Lambda\\
   0       & \text{if } |\omega| \geq\Lambda
  \end{cases}
\]
Fig.~\ref{fig:example2} show plots of the above spectral function as well as its corresponding auto-correlation for $h=1$ and $h=0$. Fig.~\ref{fig:example2-bn} shows the corresponding Lanczos coefficients.

Both cases lead to staggering, however, in the case where $h=1$, staggering is smaller and decreases and we increase $n$, as opposed to the $f=0$ case, where staggering remains constant as we increase $n$. The $h=0$ case does not satisfy condition (I), while the $h=1$ case satisfies (I), but it does not satisfy condition (II), since the derivative of $f(w)$ is not continuous at $\omega=m$.

\subsection*{Example 3}
\begin{figure}[h!]
     \centering
     \begin{subfigure}[b]{0.45\textwidth}
         \centering
         \includegraphics[width=\textwidth]{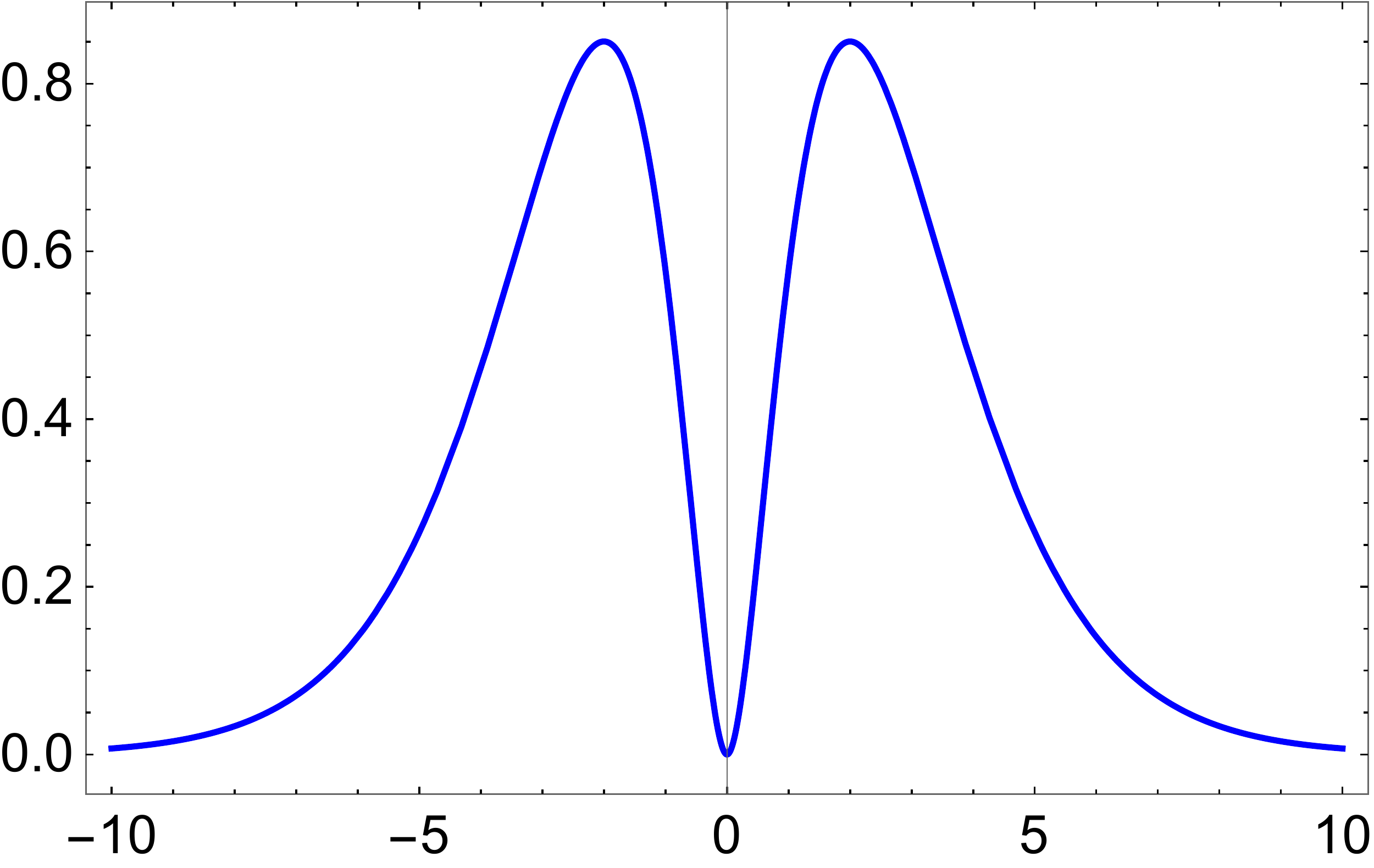}
         \caption{}
     \end{subfigure}
     \hfill
     \begin{subfigure}[b]{0.45\textwidth}
         \centering
         \includegraphics[width=\textwidth]{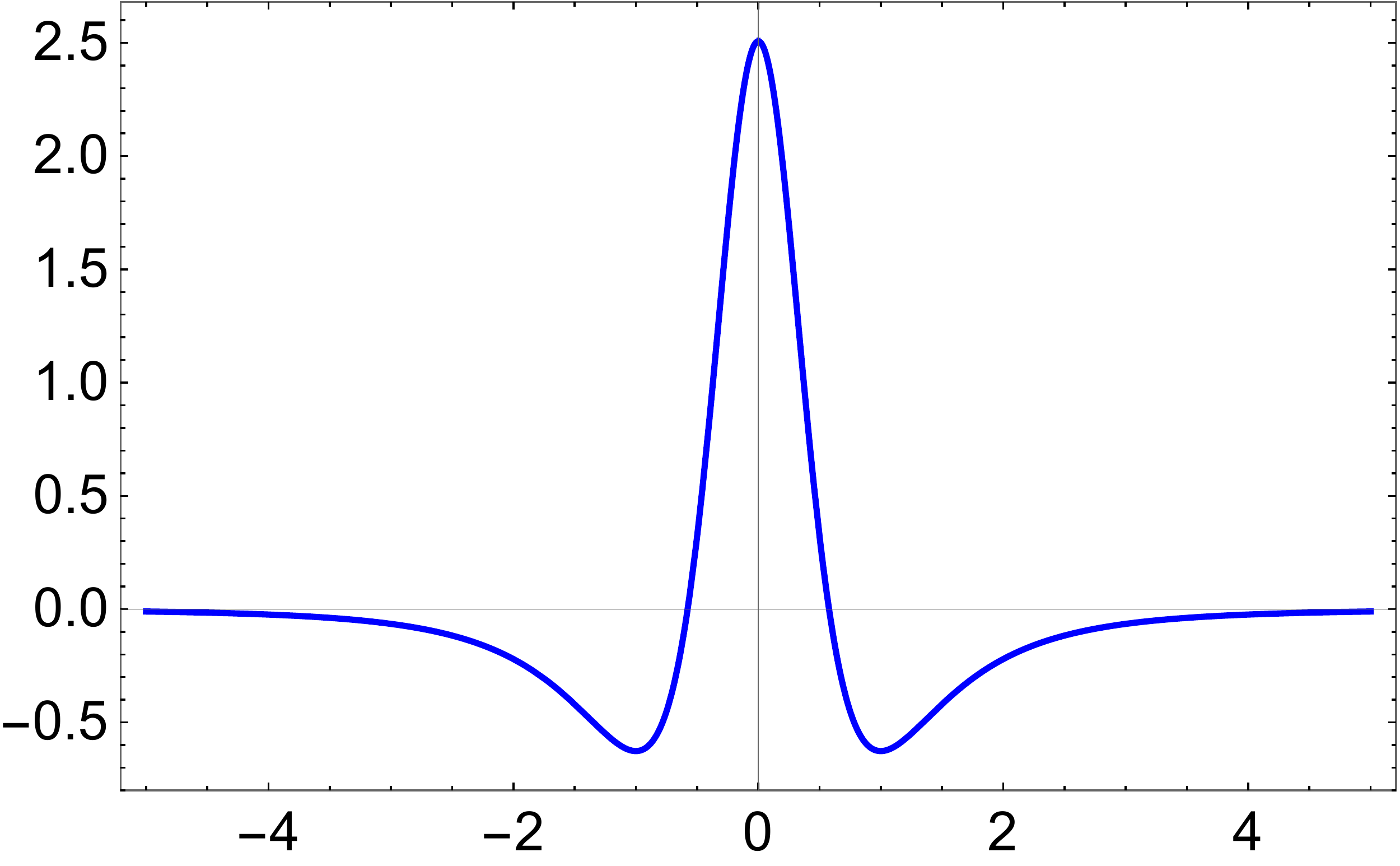}
         \caption{}
     \end{subfigure}
     \put(-240,15){\small $\omega$}
     \put(-5,15){\small $t$}
     \put(-195,148){\small $C(t)$}
     \put(-430,148){\small $f(\omega)$}
     \hfill
           \caption{(a) Power spectrum of the Example 3 for $\omega_0=1$, $\delta=2$, and $\lambda=2$.  In (b), we show the corresponding auto-correlation. }
        \label{fig:example3}
\end{figure}

\begin{figure}[h!]
    \centering
    \includegraphics[width=0.45\textwidth]{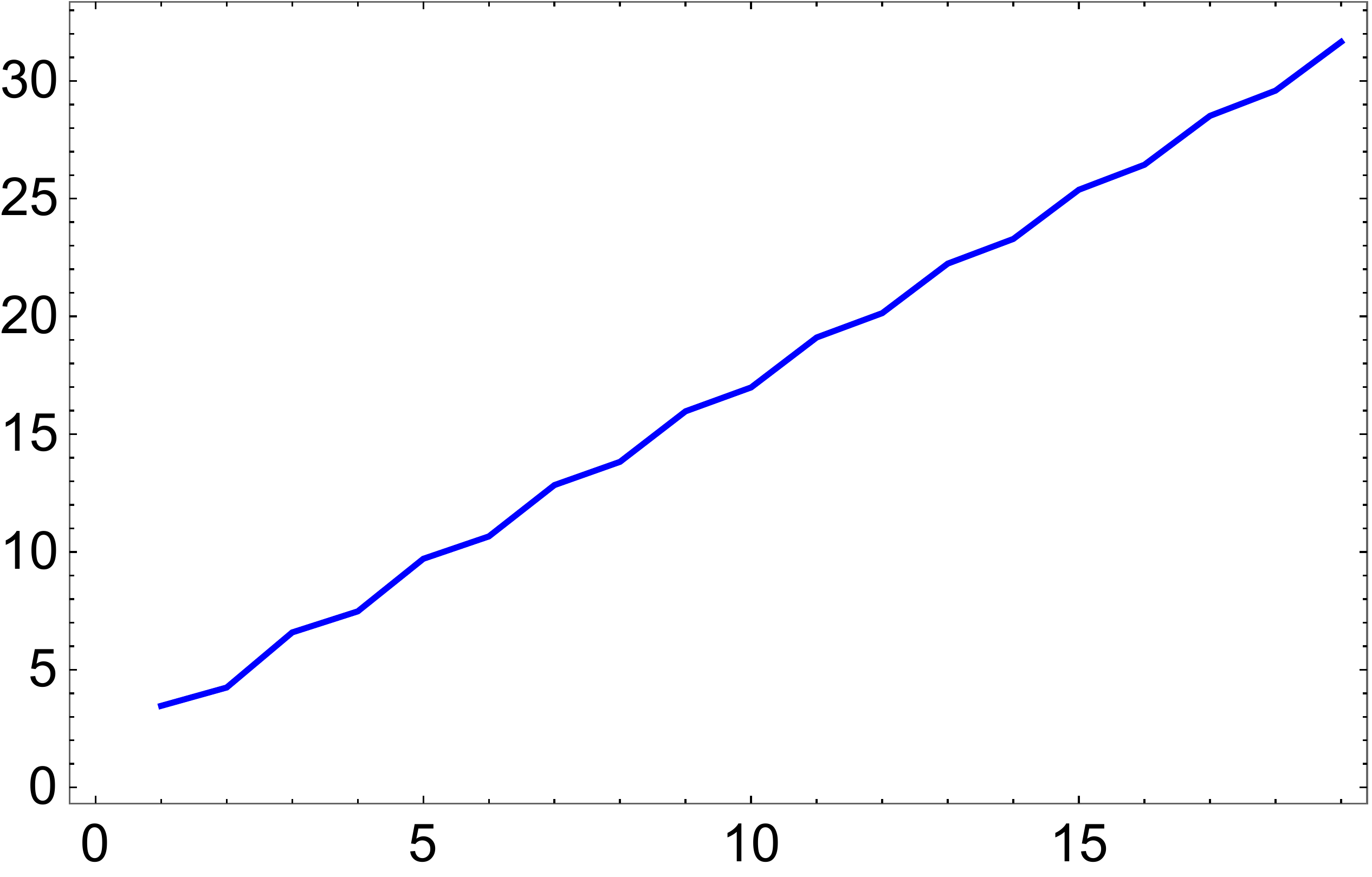}
    \put(5,3){\small $n$}
    \put(-195,130){\small $b_n$}
    \caption{Lanczos coefficients of the Example 3 for $\omega_0=1$, $\delta=2$, $\lambda=2$. Here we connected the dots to make the staggering more evident.}
    \label{fig:example3-bn}
\end{figure}

We now consider the textbook example of Eq. (\ref{eq:PowerSpectrumStaggering}), namely:
\begin{equation*}
f(\omega)=N(\omega_{0},\delta,\lambda)\left\vert\frac{\omega}{\omega_{0}}\right\vert^{\lambda}e^{-\left\vert\frac{\omega}{\omega_{0}}\right \vert^{\frac{2}{\delta}}}~\,.
\end{equation*}
Fig.~\ref{fig:example3} shows the plot of the above power spectrum and the corresponding auto-correlation, for $\omega_0=1$, $\delta=2$, and $\lambda=2$. Fig.~\ref{fig:example3-bn} shows the corresponding results for the Lanczos coefficients. The presence of staggering for $\lambda=2$ is probably because condition (I) is not satisfied in this case, namely, $f(0)=0$. 

\subsection*{Example 4}
\begin{figure}[h!]
     \centering
     \begin{subfigure}[b]{0.45\textwidth}
         \centering
         \includegraphics[width=\textwidth]{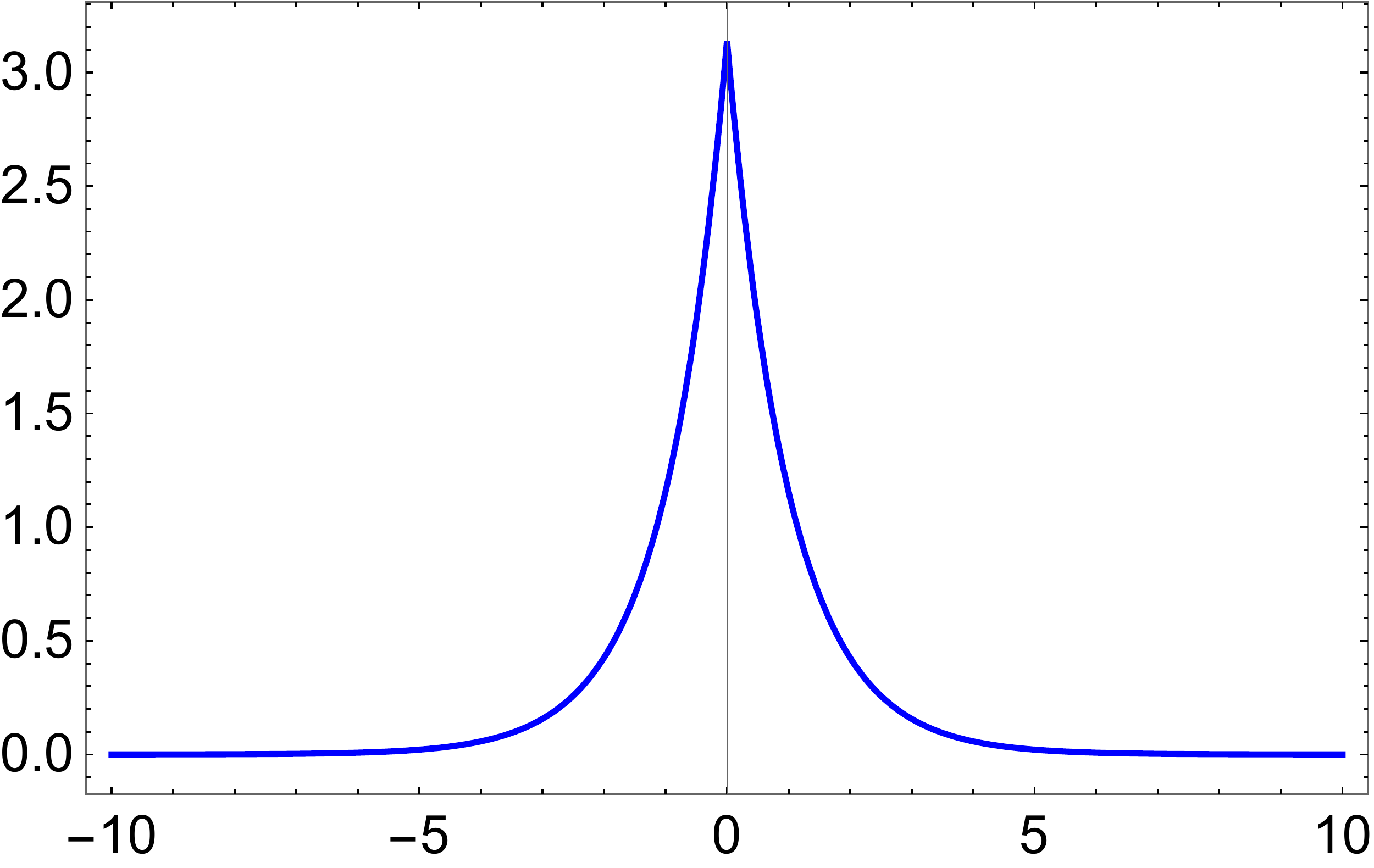}
         \caption{}
     \end{subfigure}
     \hfill
     \begin{subfigure}[b]{0.45\textwidth}
         \centering
         \includegraphics[width=\textwidth]{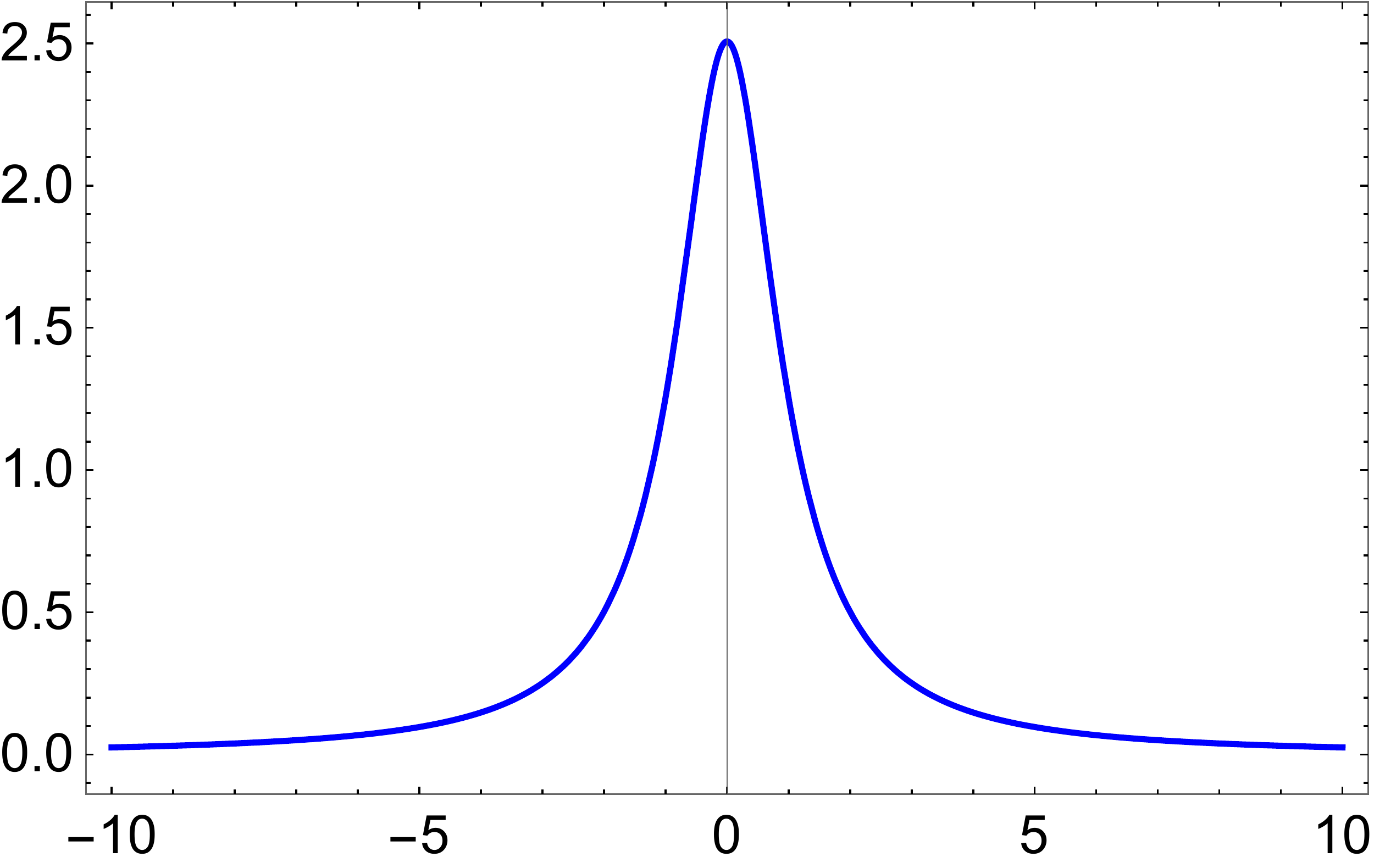}
         \caption{}
     \end{subfigure}
     \put(-240,15){\small $\omega$}
     \put(-5,15){\small $t$}
     \put(-195,148){\small $C(t)$}
     \put(-430,148){\small $f(\omega)$}
     \hfill
           \caption{(a) Power spectrum of the Example 4 for $\omega_0=1$,and $\lambda=30$.  In (b), we show the corresponding auto-correlation. }
        \label{fig:example4}
\end{figure}

\begin{figure}[h!]
    \centering
    \includegraphics[width=0.45\textwidth]{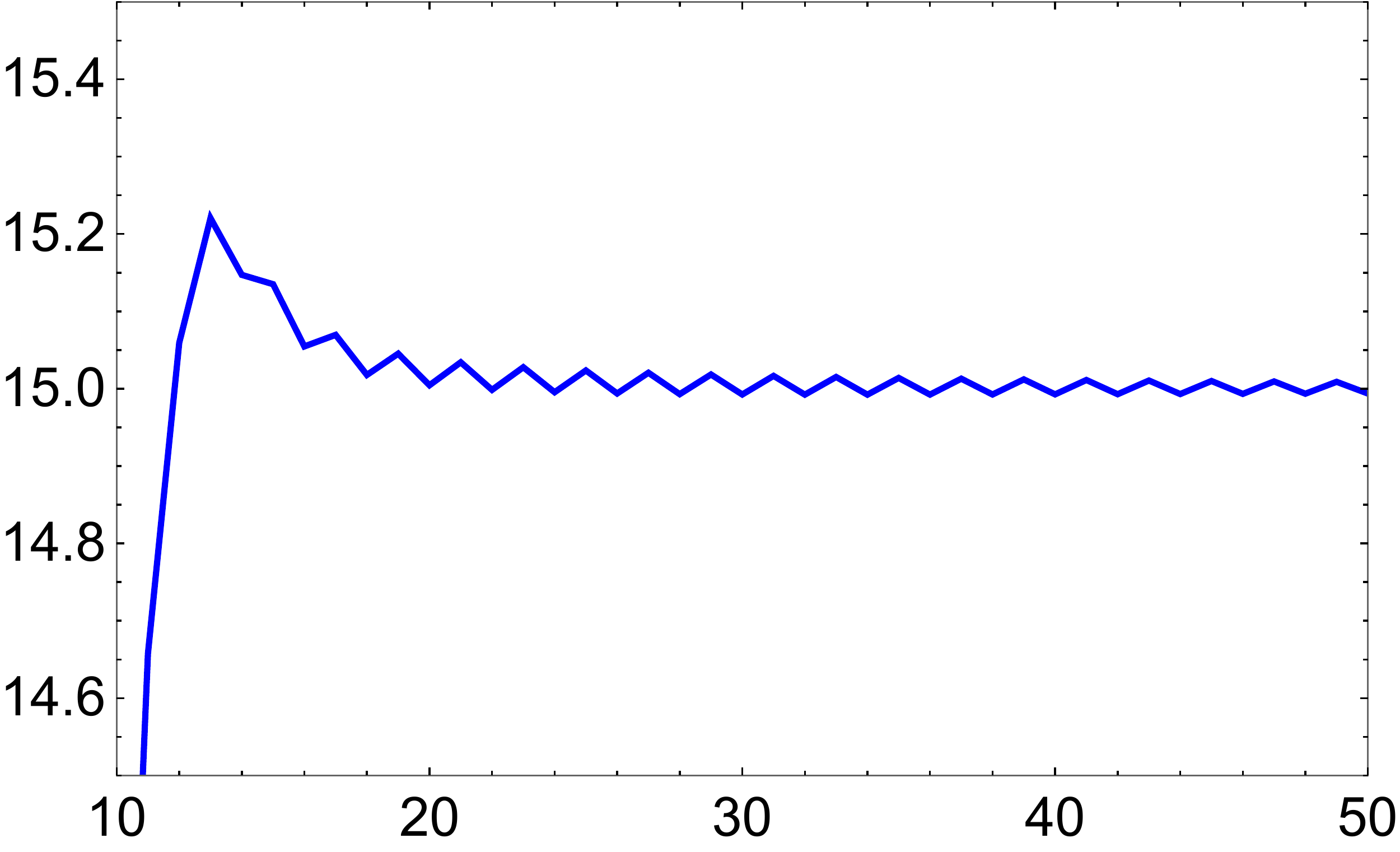}
    \put(5,3){\small $n$}
    \put(-195,130){\small $b_n$}
    \caption{Lanczos coefficients of the Example 4 for $\omega_0=1$, $\Lambda=30$. Here we connected the dots to make the staggering more evident.}
    \label{fig:example4-bn}
\end{figure}

We finally consider the following spectral function
\[
 f(\omega) = 
  \begin{cases} 
   N(\omega_{0},\Lambda) e^{-\left\vert\frac{\omega}{\omega_{0}}\right \vert}  & \text{if } |\omega| \leq \Lambda \\
   0       & \text{if } |\omega| > \Lambda
  \end{cases}
\]
which is based on the textbook example given in Eq. (\ref{eq:PowerSpectrumStaggering}) with $\lambda=0$, and $\delta=2$, but with the introduction of a UV cutoff $\Lambda$ (whose purpose is to make the presence of staggering visible). Fig.~\ref{fig:example4} shows the plot of the above power spectrum and the corresponding auto-correlation, for $\Lambda=30$ and $\omega_0=1$. Fig.~\ref{fig:example4-bn} shows the corresponding results for the Lanczos coefficients. The presence of staggering is probably due to the violation of condition II, namely, the derivative of $f(\omega)$ is not continuous at $\omega=0$.

\subsection*{Implications for the auto-correlation function}
In terms of the auto-correlation $C(t)$, condition (I) implies that the integral $f(\omega=0)=\int_{-\infty}^{\infty} C(t) dt$ is finite, which means that $C(t)$ is positive most of the time.  This feature of $C(t)$ can be observed in the examples above where $f(0)$ is finite.  At the moment we lack a concrete physical interpretation of this property. Additionally, it is not clear to us what condition (II) implies for $C(t)$.

However, we note that in the case of the (free) massive scalar field, the mass gap produces oscillations in the auto-correlation function, given by Eq.~\eqref{eq:AutoCPhi0Lma}, such that $\int_{-\infty}^{\infty} C(t) dt=0$, violating condition (I) similarly to Example 2 with $h=0$. Furthermore, in this case condition (II) is violated due also to the mass gap and the dimensionality of the spacetime in $d\neq 4$. Thus, in this case there seems to be two different origins to the staggering for the massive scalar field. We also remark that for the massless scalar field, staggering was observed for $d>4$. This is consistent with our analysis, given that in this case the power spectra have similar features to Example 3\footnote{In $d=5$, both conditions (I) and (II) are violated, whereas for $d>5$ only condition (I) is violated. We did not explore the cases $d<4$ in detail, although we may expect staggering in $d=3$ due to a violation of condition (I).}. 

\bibliographystyle{JHEP}
\bibliography{references}  
\end{document}